\documentclass[
    aps,
    prd,
    nofootinbib,
    superscriptaddress,
    10pt,
    notitlepage,
    preprintnumbers,
    twocolumn]{revtex4-1}

\usepackage{amssymb,amsmath,verbatim,mathtools,needspace,enumitem,etoolbox,graphicx}
\usepackage[dvipsnames, usenames]{xcolor}

\definecolor{linkcolor}{rgb}{0.0,0.3,0.5}
\usepackage[unicode, colorlinks=true, linkcolor=linkcolor, citecolor=linkcolor, filecolor=linkcolor,urlcolor=linkcolor, pdfusetitle]{hyperref}
\usepackage{epstopdf}

\usepackage{bm}
\usepackage{dcolumn}
\usepackage[utf8]{inputenc} 
\usepackage{latexsym}
\usepackage{rotating}
\usepackage{xcolor}
\usepackage{longtable}
\usepackage{enumerate}
\usepackage{mathtools}
\usepackage{url}
\usepackage{rotating}
\usepackage{tabularx}

\usepackage{tabularx}
\newcommand{\ben}{\begin{enumerate}}
\newcommand{\een}{\end{enumerate}}

\def\be{\begin{equation}}
\def\ee{\end{equation}}
\def\bea{\begin{eqnarray}}
\def\eea{\end{eqnarray}}
\newcommand{\beq}{\begin{eqnarray}}
\newcommand{\eeq}{\end{eqnarray}} 
\newcommand{\ba}{\begin{align}}
\newcommand{\ea}{\end{align}}

\def\ba{\bar{a}}

\newcommand{\tn}{\textnormal}

\definecolor{rb4}{HTML}{27408B}

\begin{document}
\title{Distinguishing double neutron star from neutron star-black hole binary populations with gravitational wave observations}
\author{Margherita Fasano}
\email{margherita.fasano@roma1.infn.it}
\affiliation{Dipartimento di Fisica, Sapienza Università di Roma \& Sezione INFN Roma1, P.A. Moro 5, 00185, Roma, Italy}

\author{Kaze W.K. Wong}
\email{kazewong@jhu.edu}
\affiliation{Department of Physics and Astronomy, Johns Hopkins University, 3400 N. Charles Street, Baltimore, Maryland 21218, USA}

\author{Andrea Maselli}
\email{andrea.maselli@uniroma1.it}
\affiliation{Dipartimento di Fisica, Sapienza Università di Roma \& Sezione INFN Roma1, P.A. Moro 5, 00185, Roma, Italy}

\author{Emanuele Berti}
\email{berti@jhu.edu}
\affiliation{Department of Physics and Astronomy, Johns Hopkins University, 3400 N. Charles Street, Baltimore, Maryland 21218, USA}

\author{Valeria Ferrari}
\email{valeria.ferrari@uniroma1.it}
\affiliation{Dipartimento di Fisica, Sapienza Università di Roma \& Sezione INFN Roma1, P.A. Moro 5, 00185, Roma, Italy}

\author{B.~S.~Sathyaprakash}
\email{bss25@psu.edu}
\affiliation{Institute for Gravitation and the Cosmos, Department of Physics,  Pennsylvania State University, University Park, PA, 16802, USA}
\affiliation{Department of Astronomy \& Astrophysics, Pennsylvania State University, University Park, PA, 16802, USA}
\affiliation{School of Physics and Astronomy, Cardiff University, Cardiff, CF24 3AA, GB}

\begin{abstract}
Gravitational waves from the merger of two neutron stars cannot be easily distinguished from those produced by a comparable-mass mixed binary in which one of the companions is a black hole.  Low-mass black holes are interesting because they could form in the aftermath of the coalescence of two neutron stars, from the collapse of massive stars, from matter overdensities in the primordial Universe, or as the outcome of the interaction between neutron stars and dark matter. Gravitational waves carry the imprint of the internal composition of neutron stars via the so-called tidal deformability parameter, which depends on the neutron star equation of state and is equal to zero for black holes.  We present a new data analysis strategy powered by Bayesian inference and machine learning to identify mixed binaries, hence low-mass black holes, using the distribution of the tidal deformability parameter inferred from gravitational-wave observations.
\end{abstract}
\maketitle

\section{Introduction}
The past few years have seen remarkable advances in gravitational-wave (GW) astronomy. The ground-breaking discovery of merging binary black holes (BHs)~\cite{Abbott:2016blz, Abbott:2016nmj, Abbott:2017vtc, Abbott:2017oio} was soon followed by the spectacular observation by the Laser Interferometer Gravitational-Wave Observatory (LIGO)~\cite{TheLIGOScientific:2014jea} and Virgo~\cite{TheVirgo:2014hva} of the coalescence of binary neutron stars (BNSs)~\cite{TheLIGOScientific:2017qsa,Abbott:2018wiz}, whose counterpart and afterglow was also witnessed in the entire electromagnetic spectrum by dozens of telescopes and detectors around the world and in space~\cite{GBM:2017lvd}. This latter observation has already shed light on a number of unsolved problems in physics and astronomy: it provided the first direct evidence that BNSs power the central engines of short gamma ray bursts \cite{Monitor:2017mdv}, identified the merger debris of such systems as prolific sites of the formation of r-process elements \cite{Lattimer:1974slx, GBM:2017lvd}, and confirmed that GWs travel essentially at the speed of light \cite{Monitor:2017mdv}. Most importantly for our present purposes, the discovery of GW170817 has helped demonstrate that GW observations can infer the tidal deformability of neutron stars (NSs)~\cite{De:2018uhw, Abbott:2018wiz, Abbott:2018exr, LIGOScientific:2019eut, Landry:2020vaw} (but see~\cite{Kastaun:2019bxo} for caveats).

LIGO and Virgo observations have so far firmly confirmed GWs from two classes of ultra-compact binaries: binary BHs and BNSs. With the recent discovery of GW190425, they have potentially also observed the first example of a mixed system containing a BH and a NS (BHNS)~\cite{Abbott:2020uma}, although GW190425 could well be a BNS merger \cite{Han:2020qmn}. When the masses of BHs in such systems are similar to those of NSs, how can one tell them apart? 

The presence of NSs in a binary can leave behind relativistic ejecta that predominantly contain energetic neutrons, which source r-process heavy elements and kilonovae \cite{Metzger:2016pju}.Indeed several studies used
electromagnetic information to understand whether the low-mass compact binary mergers detected so far are BHNS or BNS systems~\cite{Hinderer:2018pei,Coughlin:2019kqf,Siegel:2019mlp,Kyutoku:2020xka,Barbieri:2019bdq}. However, if the primary companion is a massive BH (where the precise mass threshold depends on the BH spin~\cite{Pannarale:2015jia,Foucart:2018rjc,Zappa:2019ntl}) then no ejecta might be left behind, as tidal forces will be small. If instead the BH mass is comparable to the NS mass, the electromagnetic afterglow might be similar to that of BNS mergers. Simulations suggest that the disk mass in this case may be small, so that the electromagnetic counterpart may be hard to detect~\cite{Foucart:2018rjc,Foucart:2019bxj}.

Besides, not all binaries detected by LIGO and Virgo might be accessible for electromagnetic follow-ups for various reasons, including their large distance, the line-of-sight dependence of the ejecta (see e.g.~\cite{Kawaguchi:2020osi}), and large uncertainties in the sky position of the source as determined by LIGO and Virgo.

Even so, discriminating the BNS population from the BHNS population is an important science goal for GW detectors, as this could shed light on the origin of the two populations, testing astrophysical models of the formation and evolution of such systems.  The presence of a NS in a binary can, in principle, be inferred by GW observations as the tidal field of the companion (BH or NS) can induce quadrupole deformation in the NS. This deformation is measured in terms of a dimensionless ``tidal deformability'' parameter $\Lambda,$ which is related to the quadrupolar $\ell=2$ tidal Love number $k_2$ and the radius $R$ and mass $M$ of the NS via $\Lambda=(2/3) k_2 (c^2R/GM)^5$~\cite{Hinderer:2007mb,Binnington:2009bb,Damour:2009vw}. LIGO/Virgo GW observations have direct access to this parameter, as the quadrupole deformation of the star leads to a faster rate of inspiral of the orbit. This is captured in the observed waveform as a fifth post-Newtonian order (i.e., ${\cal O}(v/c)^{10}$) correction to the orbital phase evolution of the system. At this order the deformability parameters $\Lambda_i$ ($i=1,\,2$) do not appear separately, but as a dimensionless combination called the \emph{effective tidal deformability}, which also depends on the mass ratio $q=M_2/M_1$ of the system:
\begin{equation}
  \tilde{\Lambda}\equiv \frac{16}{13}\frac{(1+12q)\Lambda_1+(q+12)q^4\Lambda_2}{(1+q)^5}.
  \label{eqn:Lambdone}
\end{equation}

While the primary goal behind measuring the tidal deformability is to determine the equation of state (EOS) of dense hadronic or quark matter in NS cores, in this paper we wish to exploit this measurement to distinguish BNS from BHNS systems. In particular, our goal is to develop a new statistic to discriminate between the two populations and measure a population hyperparameter that gives the fraction of BNS and BHNS systems in the observed population. To this end, we exploit the fact that according to our current understanding BHs have zero tidal deformability (see Refs.\,\cite{Damour:2009vw,Landry:2015zfa,Pani:2015hfa,Gralla:2017djj} for further details), while NSs, depending on the stiffness of the EOS, could have a large tidal deformability~\cite{Hinderer:2007mb,LIGOScientific:2019eut}. 

While it has long been known that mass measurements are not sufficient to distinguish between BNS and NSBH systems~\cite{Hannam:2013uu,Tsokaros:2019lnx}, our work differs from similar recent proposals. Measurements of the tidal deformabilities $\Lambda_{1}$ and $\Lambda_2$ of the individual binary components could be consistent with a NSBH system even for large-SNR signals and large tidal effects if at least one of the two tidal deformabilities is consistent with zero at the 50\% confidence level~\cite{Chatziioannou:2018vzf}, therefore it is hard to distinguish BNS from NSBH systems with GWs alone\footnote{Distinguishing low-mass binary BHs from BNSs is easier, as both tidal deformabilities vanish in the case of binary BHs.} if we assume that $\Lambda_1$ and $\Lambda_2$ are independent~\cite{Yang:2017gfb}. 
However, certain NS properties that can be measured via GWs can be expected to be similar for all NSs. This ``universality'' can be used to distinguish between the two classes of binaries, as described in Ref.~\cite{Chen:2019aiw}. The main caveat of this method is the requirement that the NS radius must be approximately constant for all NSs in binary systems, at least within statistical errors. This is reasonable when the EOS is hadronic, but it is not expected to hold if the EOS allows for phase transitions to quark matter~\cite{Han:2018mtj,Chen:2019rja}. Conversely, the method we propose can be applied to any EOS model. We consider two ``extreme'' EOS models (see Sec.~\ref{sec:massDist}), one of which (the ALF2 EOS) indeed leads to hybrid stars. 
Previous work developed a method to distinguish BNSs and low-mass binary BHs solely from their GW signals, considering the imprint of the tidal deformability of the NSs on the GW signal for systems undergoing prompt BH formation after merger~\cite{Chen:2020fzm}. More recently, tidal heating of BH horizons has been suggested as a way of distinguishing BNS from BHNS systems~\cite{Datta:2020gem}.

The rest of this paper is organized as follows. In Sec.~\ref{sec:massDist} we describe our assumptions on the mass distribution and the EOS, and their implications for the distribution of the effective tidal deformability parameter. In Sec.~\ref{sec:setup} we use hierarchical Bayesian inference to reconstruct the fraction of BHNS (BNS) systems from simulated observations. Finally, in Sec.~\ref{sec:conc} we discuss our results and point out possible directions for future work.
Appendix~\ref{app:lowmassBHs} lists some of the proposed formation scenarios that could produce BHs in the mass range $\sim 1$--$3\ M_\odot$. Appendix~\ref{app:EOS} shows that EOS uncertainties can affect the inference in the worst-case scenario where we use the stiffest EOS to recover astrophysical systems that correspond to the softest EOS (or vice versa). Appendix~\ref{app:TD} shows that our results are largely insensitive to the inclusion of tidal disruption effects in the waveform models.

\section{The mass distribution of compact binaries}\label{sec:massDist}

In this section we discuss and motivate our assumptions on the mass distribution of BNSs and BHNSs, which is an important ingredient to distinguish between the two families of compact objects.

Stellar evolution theory suggests that the minimum mass of isolated, nonrotating NSs should be $\sim 1M_\odot$ (see e.g.~\cite{Hector2016,Shibata2018minM} and references therein), and there is a growing body of experimental and theoretical constraints on the upper end of the mass spectrum. The timing of radio pulsars recently established a new observational lower limit on the maximum mass of $\sim 2.14^{+0.20}_{-0.18}M_\odot$ at 95.4\% confidence level~\cite{Cromartie:2019kug}.  Bayesian inference based on the electromagnetic observations of pulsars, nuclear physics calculations of the EOS and the recent observation of GW170817 together imply values of the maximum mass of stable NSs clustering around $\sim2 \, M_\odot$~\cite{Antoniadis:2016hxz,Alsing2018,Kalogera:1996ci,Bombaci1996,Srinivasan2002,Chamel2013,Shibata2019maxM,Rezzolla:2017aly,Ai:2019rre,LIGOScientific:2019eut}, although there are claimed observations of even more massive NSs~\cite{Freire:2007jd}, and theoretically the maximum mass can be as large as $\sim 3\,M_\odot$ \cite{Rhoades:1974fn}. In general, the mass spectrum of isolated BHs can span several orders of magnitude ranging from sub-solar mass objects to the supermassive BHs of mass $\gtrsim10^{6}M_\odot$ found in galactic centers. In this work we are interested in BHs with masses comparable to NSs, and therefore we will focus on the range $1M_\odot\lesssim M_\tn{BH}\lesssim 3M_\odot$. 

\begin{figure*}[t]
  \centering
  \includegraphics[width=0.9\textwidth]{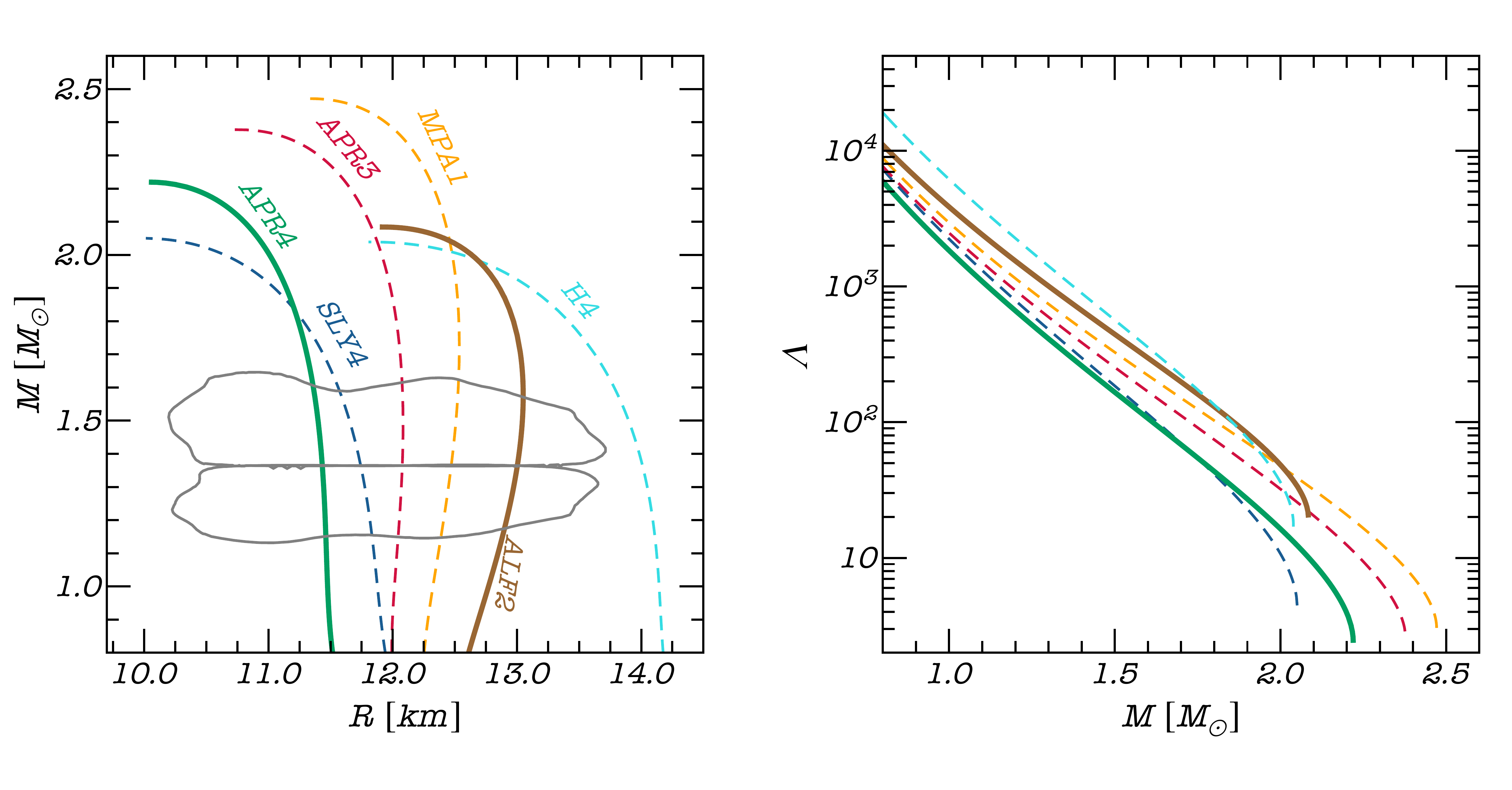} \\
  \caption{Left panel: Mass-radius relations for selected EOS models. Left to right: APR4~\cite{Akmal:1998cf} (thick green), SLY4~\cite{sly4} (dashed blue), APR3~\cite{Akmal:1998cf} (dashed red), MPA1~\cite{mpa1} (dashed orange), ALF2~\cite{Alford:2004pf} (thick brown), and H4~\cite{Lackey:2005tk} (dashed cyan). The gray lines represent the 90\% confidence regions for the companion masses and their radii for the LIGO/Virgo event GW170817, assuming a parametrized EOS and imposing a lower limit on the maximum mass of $1.97\,M_{\odot}$ (cf. Fig.~3 of Ref.~\cite{Abbott:2018exr}). Right panel: Dimensionless tidal deformability $\Lambda$ (on a log scale) as a function of the NS mass for the same EOS models considered in the left panel.} \label{fig:EOS} 
\end{figure*}

\subsection{Neutron star binaries}

We consider two BNS mass distribution models.

The first model (``double Gaussian,'' henceforth \texttt{BNS-DG}) is based on the electromagnetic observations of Galactic radio pulsars, whose evolutionary path is described in Refs.~\cite{Tauris:2017omb,Andrews:2019vou,Vigna-Gomez:2020bgo}.  In the standard isolated binary formation channel, the primary NS is spun up to $\sim 10$--$100$\,ms through accretion, whereas the secondary star spins down to a typical period of $\sim 1$\,s after birth. A recent Bayesian analysis using a sample of 17 Galactic BNSs~\cite{Farrow:2019xnc} indicates that the nonrecycled (secondary) NS mass is uniformly distributed within the range $M_\tn{NS} \in [1.14,1.46]M_{\odot}$, while the recycled (primary) NS follows a double Gaussian distribution
\begin{align}
  G(m|\theta) = &\frac{\alpha}{\sqrt{2 \pi}\sigma_1}
                  e^{\tfrac{(m-\mu_1)^2}{2\sigma_1^2}}+\frac{1-\alpha}{ \sqrt{2 \pi}\sigma_2} e^{\tfrac{(m-\mu_2)^2}{2\sigma_2^2}}\,,
  \label{eqn:DG}
\end{align}
where we introduced the four-dimensional parameter vector $\theta=(\mu_1,\mu_2,\sigma_1,\sigma_2)$ with $\mu_1=1.34 \ M_{\odot}$, $\mu_2=1.47 M_{\odot}$, $\sigma_1=0.02 M_{\odot}$, $\sigma_2=0.15 M_{\odot}$, and the ``mixing parameter'' $\alpha=0.68$. The \texttt{BNS-DG} prescription is completed by setting an EOS-dependent threshold $M^\tn{max}_\tn{EOS}$ for the maximum stellar mass (see Sec.~\ref{sec:setup}). This model is based on observations of galactic NSs, and therefore it should be viewed with some caution if we consider third-generation (3G) interferometers such as the Einstein Telescope~\cite{Punturo:2010zz,Sathyaprakash:2012jk,Maggiore:2019uih} or Cosmic Explorer~\cite{Evans:2016mbw}, which are expected to detect binary systems out to large redshifts~\cite{Evans:2016mbw,Reitze:2019iox}.

In the second, more agnostic model (\texttt{BNS-U}), both NS masses are extracted from a uniform distribution with $M_\tn{NS}\in [1M_{\odot}, M^\tn{max}_\tn{EOS}]$. This model is less physically motivated, but we use it to bracket uncertainties and to take into account the recent detection of GW190425~\cite{Abbott:2020uma}, which seems to suggest that the formation and evolution of the BNS population observed in GWs may be different from the Galactic population~\cite{Romero-Shaw:2020aaj}.

\subsection{Black hole-neutron star binaries}

The formation and evolution of BHNS binaries are arguably even more uncertain. ``Low-mass'' BHs can form from the gravitational collapse of stars of mass $\gtrsim 8 \ M_{\odot}$ or from overdensities in the early Universe (``primordial BHs'', henceforth PBHs~\cite{Hawking:1971ei,Carr:1974nx,Green:2014faa,Cai:2019igo,Escriva:2019phb,Gow:2019pok,DeLuca:2020ioi,Young:2020xmk,Lin:2020goi,Liu:2020cds}).  If the BH mass $m>M^\tn{max}_\tn{EOS}$ (where no stable NS configurations are allowed) the BH could have either primordial or stellar origin \cite{Roncadelli:2009qj}. There are several (more or less exotic) formation scenarios that could produce BHs in the mass range $\sim 1$--$3\ M_\odot$. To improve readability, we briefly review them in the Appendix~\ref{app:lowmassBHs}.

There are large uncertainties in current estimates of BNS and stellar BHNS merger rates (see e.g.~\cite{Baibhav:2019gxm}) and in key parameters of some of the more ``exotic'' formation scenarios, such as the fraction of dark matter in PBHs $f_\tn{PBH}$ (see e.g.~\cite{Sasaki:2018dmp,Raidal:2018bbj,Albert:2019qxd,Hertzberg:2020hsz,DeLuca:2020fpg,Carr:2020gox}), but it is reasonable to expect that BNS merger rates should be larger than BHNS merger rates in the mass range of interest here. LIGO-Virgo observations have measured a 90\% credible rate (to the nearest significant figure) of 100--4000\,yr$^{-1}$\,Gpc$^{-3}$ for BNS mergers, while the upper limit (in the absence of any candidates) on BHNS binaries is 600\,yr$^{-1}$\,Gpc$^{-3}$~\cite{LIGOScientific:2018mvr,TheLIGOScientific:2016pea}. However, we will be agnostic and allow for the possibility that BHNS rates may dominate over BNS rates.  We adopt a flat distribution for the BH mass $M_\tn{BH}$ in the range $[1, 3]M_{\odot}$, and (just as we did for BNSs) we consider either the double Gaussian distribution of Eq.~(\ref{eqn:DG}) or a uniform NS mass distribution in the range $M_\tn{NS} \in [1, M^\tn{max}_\tn{EOS}]M_{\odot}$. In the following we will refer to these models as \texttt{BHNS-DG} and \texttt{BHNS-U}, respectively.

\begin{table}[t]	
  \centering
  \caption{Radius and dimensionless tidal deformability $\Lambda \equiv \lambda/m^5$ for a prototype $1.4M_\odot$ NS modelled with two examples of theoretical EOSs, namely APR4~\cite{Akmal:1998cf} and ALF2~\cite{Alford:2004pf}, which represent to cases of soft and stiff nuclear matter, respectively.}
  \begin{tabular}{cccc}
     \hline
    \hline
\texttt{EOS} &  $R_\tn{NS} $ [km] & $\Lambda$ & $ M^\tn{max}_\tn{EOS}$ $[M_{\odot}]$\\
     \hline
APR4 &  11.43  & 260.35&2.21 \\ 
ALF2 & 13.02 & 666.23& 2.08\\
   \hline
  \hline
\end{tabular}
  \label{table:massradius}
\end{table}

\begin{figure*}[t]
\centering
\includegraphics[width=\textwidth]{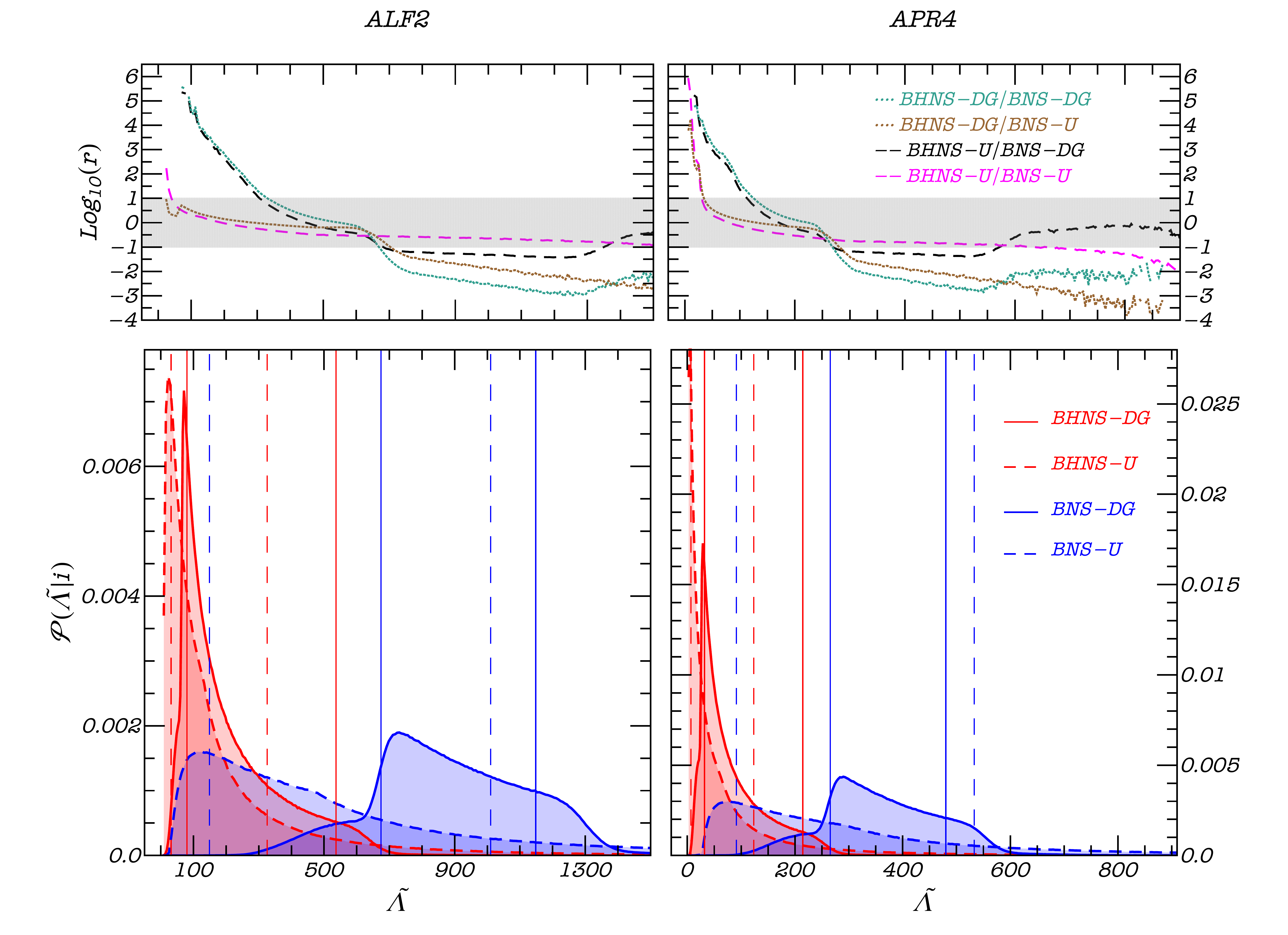}
\caption{Bottom: Conditional probability distributions $\mathcal{P}(\widetilde{\Lambda}|i)$, where $i=$ \texttt{BHNS-DG} (solid red), \texttt{BHNS-U} (dashed red), \texttt{BNS-DG} (solid blue) or \texttt{BNS-U} (dashed blue), for EOS ALF2 (left panel) and APR4 (right panel). Vertical lines identify the 68\% confidence intervals for the three distributions (cf. Table~\ref{table:massradius}). Top: $\log_{10} (r)$, where $r$ is the probability ratio defined in Eq.~(\ref{eq:r}) for different combinations of mass distribution models, as indicated in the legend. When this ratio is above the shaded region the binary is likely to be a BHNS. Below the shaded region it is likely to be a BNS. In the grey shaded region, the binary's origin is uncertain.}
\label{fig:p} 
\end{figure*}

\subsection{Choice of EOS}

In our analysis we consider two EOS models, APR4~\cite{Akmal:1998cf} and ALF2~\cite{Alford:2004pf}, as prototypes for ``soft'' and ``stiff'' nuclear matter. Soft and stiff EOSs lead to more and less compact stellar configurations, respectively. The APR4 EOS is computed from a nonrelativistic model which includes relativistic boost corrections to the two- and three-nucleon interactions using variational methods~\cite{Akmal:1998cf}.  The ALF2 EOS is a nuclear and quark matter EOS based on the so called MIT bag model, with a phase transition from nucleons to deconfined quarks at density $\rho \sim 8 \cdot 10^{14} \ \mathrm{g \cdot cm^{-3}}$ \cite{Alford:2004pf}.  As shown in Table~\ref{table:massradius}, for a given mass (here chosen to be the ``canonical'' $M=1.4M_\odot$) the APR4 EOS yields a NS with smaller radius and tidal deformability than the ALF2 EOS.
  
Both of these models are compatible with the LIGO/Virgo events GW170817~\cite{De:2018uhw, Abbott:2018wiz,Abbott:2018exr} and GW190425 \cite{Abbott:2020uma} and with recent electromagnetic observations~\cite{Raaijmakers:2019qny,Riley:2019yda,Miller:2019cac,Bogdanov:2019ixe,Bogdanov:2019qjb}.  Furthermore, as shown in the left panel of Fig.~\ref{fig:EOS}, APR4 and ALF2 span a wide range of mass-radius configurations.  The difference in stiffness between the two models has a large impact on the tidal deformability parameter $\Lambda$, which differs by a factor $\gtrsim2$ between the two models for a given mass. In the right panel of Fig.~\ref{fig:EOS} we plot the tidal deformability as a function of mass for the same equations of state. The soft EOS APR4 yields smaller values of $\Lambda$ than the stiff EOS ALF2 for all NS masses.  This is important for our purposes, because large values of $\Lambda$ enhance matter effects in the signal, and, therefore, lead to tighter constraints \cite{Hinderer:2018mrj}.

\subsection{Tidal deformability probability distribution}

We populate the models \texttt{BNS-DG}, \texttt{BNS-U}, \texttt{BHNS-DG} and \texttt{BHNS-U} with $n\sim 8 \cdot 10^6$ samples, each representing a binary with component masses randomly selected according to the criteria described in Sec.~\ref{sec:massDist}. For each system (BNS and BHNS), we compute the NS tidal deformability by solving the relativistic stellar structure equations for a given EOS~\cite{Hinderer:2007mb}, while the tidal deformability of the BH is assumed to be zero~\cite{Poisson:2014gka,Landry:2015zfa,Pani:2015hfa,Pani:2015hfa}. From these data sets we compute the dimensionless tidal deformability $\widetilde{\Lambda}$ defined in Eq.~\eqref{eqn:Lambdone}, and the corresponding conditional probability distributions $\mathcal{P}(\widetilde{\Lambda}|\tn{BNS})$ and $\mathcal{P}(\widetilde{\Lambda}|\tn{BHNS})$.

The blue and red histograms in Fig.~\ref{fig:p} show the probability distributions $\mathcal{P}(\widetilde{\Lambda}|\tn{BHNS})$ (red) and $\mathcal{P}(\widetilde{\Lambda}|\tn{BNS})$ (blue) for EOS ALF2 (left panel) and APR4 (right panel). Within each panel, solid (dashed) lines correspond to a double Gaussian (uniform) mass distribution for NSs. The EOS has a small effect on the qualitative shape of the probability functions for both BNS and BHNS systems. However, the stiffness of the EOS does change the median and the 68\% confidence intervals, as shown in Table~\ref{table:median}.
To guide the eye, in Fig.~\ref{fig:p} we mark all 68\% confidence intervals by vertical lines.

Compare for example the probability functions $\mathcal{P}(\widetilde{\Lambda}|\tn{BHNS-DG})$ and $\mathcal{P}(\widetilde{\Lambda}|\tn{BNS-DG})$. The left panel of Fig.~\ref{fig:p} shows that they have some overlap when $300\lesssim\widetilde{\Lambda}\lesssim 700$ for ALF2, while the right panel implies that they overlap for $100\lesssim\widetilde{\Lambda}\lesssim300$ for APR4. However, for both EOS models the 68\% confidence levels of the two distributions are disjoint. These qualitative considerations provide a first visual indication that it may indeed be possible to disentangle the nature of low-mass binaries from these probability distributions, with small and large values of $\widetilde{\Lambda}$ corresponding to BHNS and BNS systems, respectively, while intermediate values would suggest a mixture of the two populations.

The mass function of NSs does not significantly affect $\mathcal{P}(\widetilde{\Lambda}|\tn{BHNS})$, which remains sharply peaked at small values of $\widetilde{\Lambda}$, but it does change the qualitative behavior of $\mathcal{P}(\widetilde{\Lambda}|\tn{BNS})$.  Table~\ref{table:median} shows that the median value of $\widetilde{\Lambda}$ is significantly lower for $\mathcal{P}(\widetilde{\Lambda}|\tn{BNS-U})$ than for $\mathcal{P}(\widetilde{\Lambda}|\tn{BNS-DG})$, so the region in which the tidal deformability of BHNS and of BNS overlap increases significantly, and we can expect that our ability to distinguish BHNSs from BNSs will degrade significantly.  Note that this is a blessing and a curse: reconstructing the form of the probability distributions with future GW observations may allow us to reconstruct the mass distribution and the formation scenarios of BNS components.

\begin{table}[t]
  \centering
  \caption{Median and 68\% confidence intervals of the tidal deformability $\widetilde{\Lambda}$ (cf. Fig.~\ref{fig:p}.)}
  \begin{tabular}{ccc}
     \hline
    \hline
\texttt{model}  & \multicolumn{2}{c}{\texttt{EOS}} \\
     \hline
     &ALF2&APR4\\
\texttt{BHNS-U}  &  $100^{+225}_{-70}$  & $30^{+90}_{-24} $\\
\texttt{BHNS-DG} &  $160^{+380}_{-78}$  & $60^{+150}_{-31}$\\
\texttt{BNS-U}   &  $410^{+600}_{-260}$ &$230^{+300}_{-140}$ \\
\texttt{BNS-DG}  & $ 870^{+280}_{-200}$ &$350^{+130}_{-87}$ \\
   \hline
  \hline
\end{tabular}
  \label{table:median}
\end{table}

The distributions of the tidal deformability shown in the bottom panels of Fig.~\ref{fig:p} can be exploited to identify the specific type of binary. Let us introduce the ratio
\begin{equation}
r(\widetilde{\Lambda}) \equiv 
\frac{{\cal P}(\widetilde{\Lambda}|\tn{BHNS})}
{{\cal P}(\widetilde{\Lambda}|\tn{BNS})}\, .
\label{eq:r}
\end{equation}
Large (small) values of $r(\widetilde{\Lambda})$ indicate that $\widetilde{\Lambda}$ is more likely to come from a BHNS (BNS, respectively).  In the top panels of Fig.~\ref{fig:p} we plot $r(\widetilde{\Lambda})$ for the four possible combinations of EOS models (either ALF2 or APR4) and NS mass distributions (either double Gaussian or uniform).  When this ratio is above the shaded region, the binary is likely to be a BHNS. Below the shaded region, it is likely to be a BNS. In the grey shaded region, the binary's origin is uncertain.

The range of $\widetilde{\Lambda}$ corresponding to an uncertain binary origin (i.e., to the ratio $r$ being in the shaded region) depends sensitively on the mass distribution of NSs in BNS systems, being large when the mass distribution is flat. In general, the mixing fraction between BHNS and BNS systems will be hard to measure when the tidal deformability distributions for BNSs and BHNSs have a large overlap, i.e. when $r$ is in the shaded region over a broad range of values of $\Lambda$.

\begin{figure*}[t]
\centering
\includegraphics[width=0.48\textwidth]{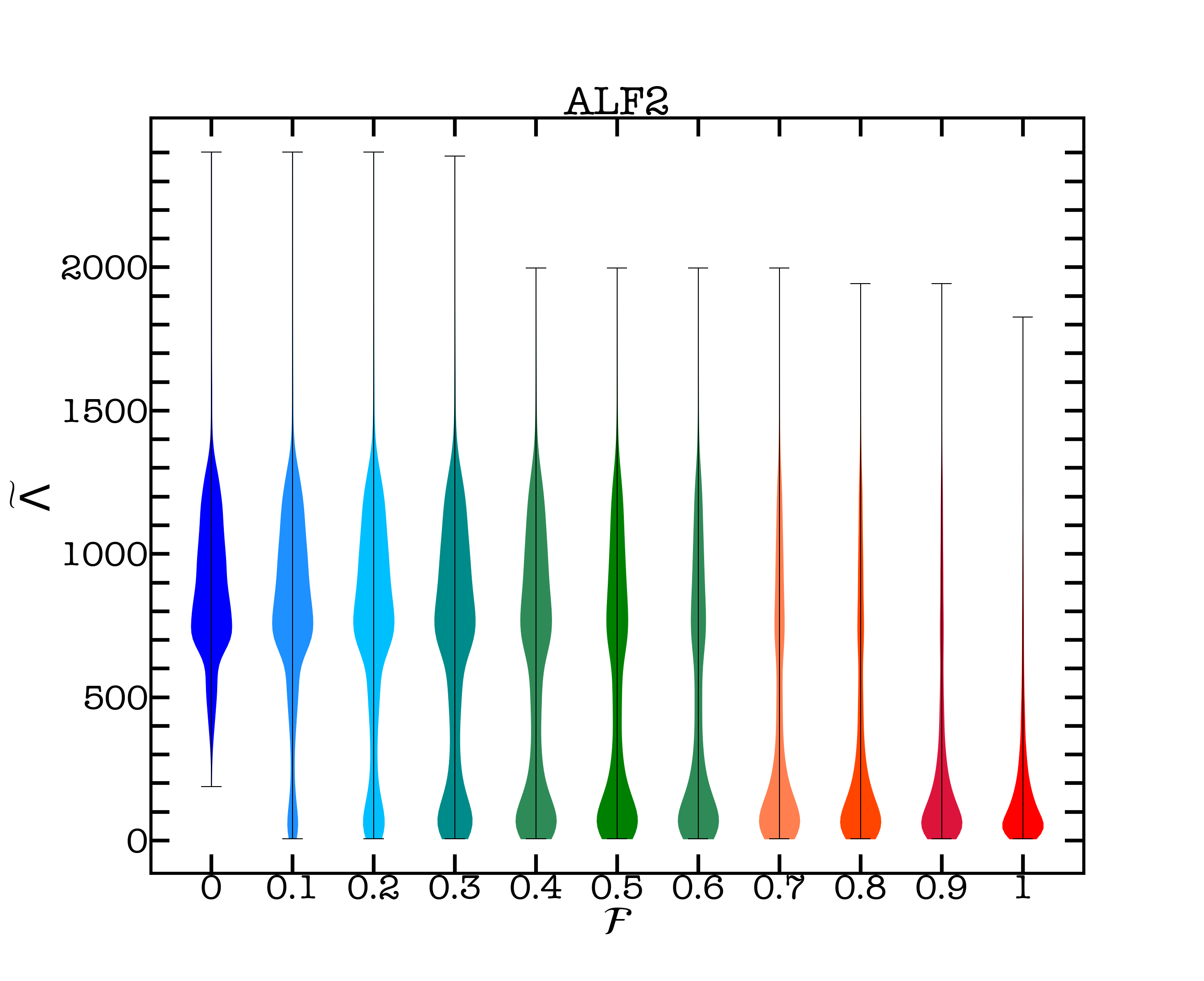}
\includegraphics[width=0.48\textwidth]{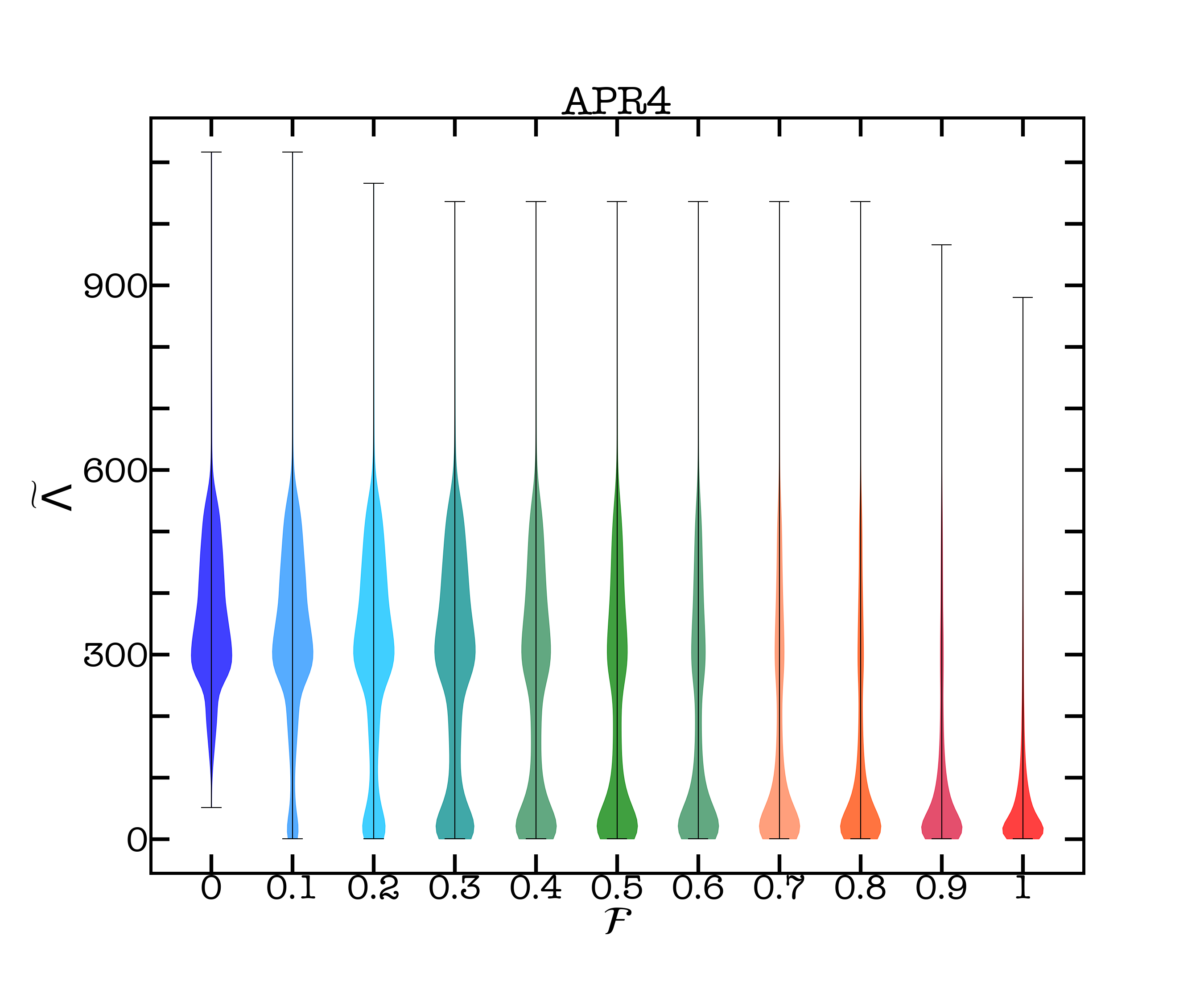}
\caption{Violin plot showing the probability distribution of $\mathcal{P}(\widetilde{\Lambda})$ defined in Eq.~\eqref{eq:plambda} for the ALF2 (left) and APR4 (right) EOS and for selected values of ${\cal F}=[0,\,0.1,\,0.2,\,\dots 1]$. The population is dominated by BNSs when ${\cal F}\to 0$, and by BHNSs when ${\cal F}\to 1$.}
\label{fig:es} 
\end{figure*}

\begin{figure*}[t]
\centering
\includegraphics[width=1\textwidth]{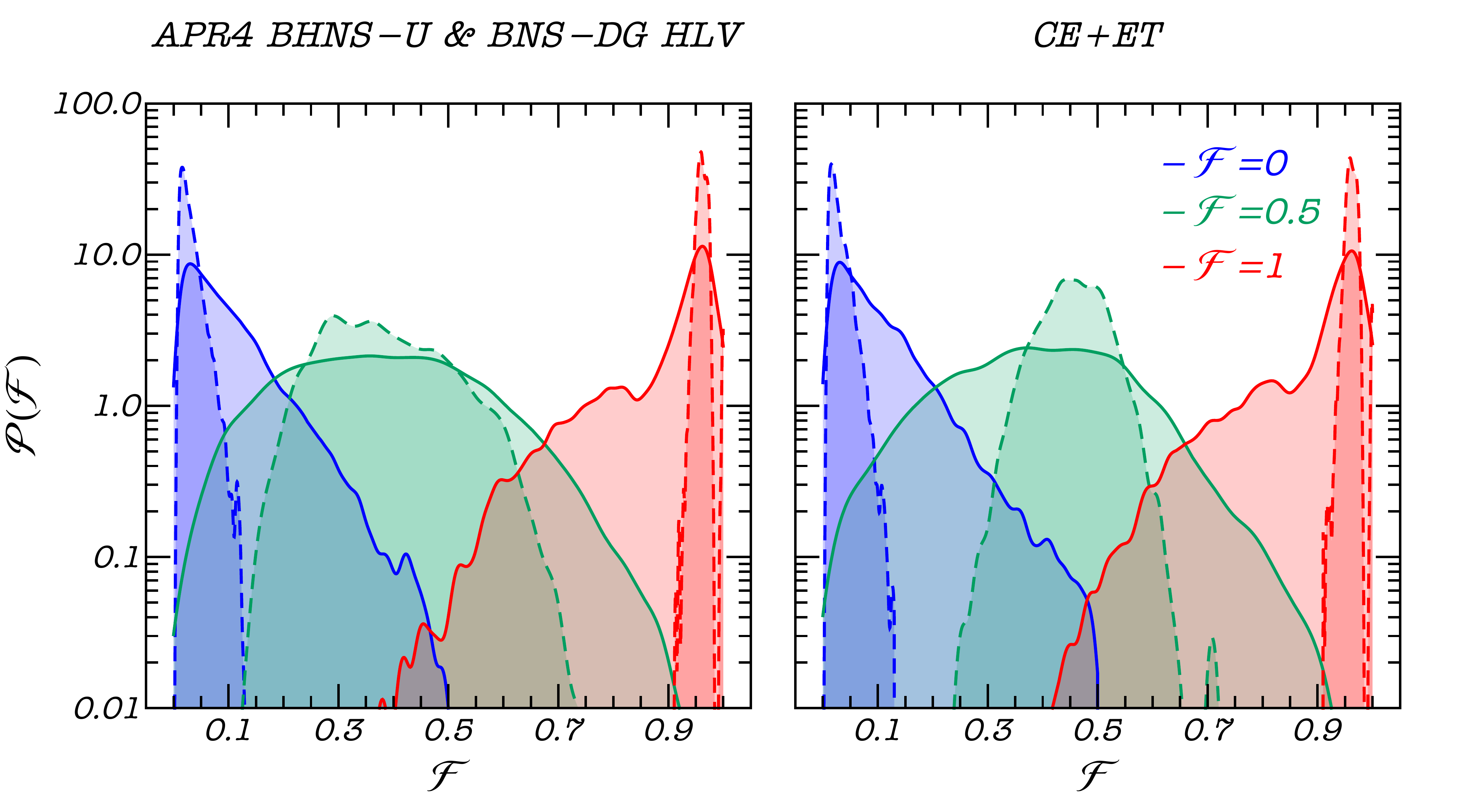}
\caption{Reconstructed probability density functions of the parameter $\mathcal{F}$ assuming $N_{\rm obs}=10$ observations (solid lines) or $N_{\rm obs}=60$ observations (dashed lines). The left panel refers to a second-generation detector network (HLV), and the right panel to a third-generation network composed of two CEs and ET. We focus on three extreme cases: a pure BNS population ($\mathcal{F}=0$, blue), a ``perfectly mixed'' population ($\mathcal{F}=0.5$, green), and a pure BHNS population ($\mathcal{F}=1$, red). For concreteness here we focus on EOS APR4 and we compare the mass distribution models \texttt{BHNS-U} and \texttt{BNS-DG}, but results are qualitatively similar for other EOS models and mass distributions.}
\label{fig:samplef} 
\end{figure*}

\section{Bayesian inference}\label{sec:setup}

The previous qualitative considerations can be put on a more solid footing by a Bayesian analysis. First of all, we can combine the probability distributions of the two compact binary families to obtain the observable probability distribution of $\widetilde{\Lambda}$:
\begin{align}
  \mathcal{P}(\widetilde{\Lambda})
  &=\mathcal{P}(\widetilde{\Lambda}|\tn{BHNS})\mathcal{P}^{\rm obs}(\tn{BHNS})\nonumber\\
  &+\mathcal{P}(\widetilde{\Lambda}|\tn{BNS})\mathcal{P}^{\rm obs}(\tn{BNS})\ ,
  \label{eq:plambda}
\end{align}
where $\mathcal{P}^{\rm obs}(\tn{BNS})$ and $\mathcal{P}^{\rm obs}(\tn{BHNS})$ are 
the probabilities to observe a BNS and a BHNS system, respectively.

Equation~\eqref{eq:plambda} can be used to infer the relative abundance of BNSs and BHNSs as follows. We define a ``mixing fraction'' parameter $\mathcal{F} = \mathcal{P}^{\rm obs}(\tn{BHNS})$ such that $0 \leq \mathcal{F}\leq1$ and $\mathcal{P}^{\rm obs}(\tn{BNS}) = 1- \cal{F}$.  Then Eq.~\eqref{eq:plambda} reads
\begin{equation}
  \mathcal{P}(\widetilde{\Lambda})
  =\mathcal{P}(\widetilde{\Lambda}|\tn{BHNS})\mathcal{F} +\mathcal{P}(\widetilde{\Lambda}|\tn{BNS})(1-\mathcal{F})\ .
  \label{eq:plambda1}
\end{equation}

\begin{table*}[ht]
  \centering
  \caption{Median and 68\% confidence intervals of $\cal{F}$ for the two EOS models ALF2 and APR4, a  number of observations $N_{\rm obs}=10$ or $N_{\rm obs}=60$, and two detector networks:  (i) a second-generation HLV network and (ii) a third-generation network composed of two CEs and ET. We focus on three extreme cases: a pure BNS population ($\mathcal{F}=0$), a ``perfectly mixed'' population ($\mathcal{F}=0.5$), and a pure BHNS population ($\mathcal{F}=1$).}
  \begin{tabular}{ccc|ccc}
    \hline
    \hline
\texttt{EOS} & \multicolumn{2}{c}{HLV}  & \multicolumn{2}{c}{CE+ET}\\
      \hline
                  & $N_{\rm obs}=10$ & $N_{\rm obs}=60$ & $N_{\rm obs}=10$   & $N_{\rm obs}=60$  \\
     \hline
     \hline
    &  \multicolumn{4}{c}{ $\mathcal{F} = $ 0} \\
     \hline\\[-7pt]
ALF2 &  $0.074^{+0.097}_{-0.046}$   &   $0.024^{+0.022}_{-0.010}$ &  $0.072^{+0.094}_{-0.045}$  &  $0.023^{+0.022}_{-0.0093}$   \\[3pt]
APR4 &  $0.075^{+0.110}_{-0.053}$   &   $0.024^{+0.032}_{-0.014}$ &  $0.075^{+0.100}_{-0.050}$  &  $0.023^{+0.023}_{-0.0096}$\\[3pt]
  \hline
  \hline
  &  \multicolumn{4}{c}{ $ \mathcal{F} = $ 0.5} \\
     \hline\\[-7pt]
ALF2 &  $0.53^{+0.13}_{-0.16}$    & $0.54^{+0.062}_{-0.062}$  & $0.53^{+0.13}_{-0.16}$ & $0.54^{+0.059}_{-0.059}$ \\[3pt]
APR4 &  $0.38^{+0.18}_{-0.17}$    & $0.37^{+0.130}_{-0.091}$  & $0.40^{+0.15}_{-0.16}$ & $0.46^{+0.054}_{-0.059}$ \\[3pt]
   \hline
  \hline
  &  \multicolumn{4}{c}{ $ \mathcal{F} = $ 1 } \\
     \hline\\[-7pt]
ALF2 & $0.94^{+0.031}_{-0.15}$     & $0.96^{+0.011}_{-0.008}$  &$0.94^{+0.03}_{-0.15}$ &  $0.962^{+0.0096}_{-0.0078}$ \\[3pt]
APR4 & $0.89^{+0.078}_{-0.15}$     & $0.96^{+0.015}_{-0.009}$  &$0.89^{+0.08}_{-0.15}$ &  $0.961^{+0.0103}_{-0.0080}$ \\[3pt]
   \hline
  \hline
\end{tabular}
  \label{table:fmi}
\end{table*}

For simplicity, in the following we will compare the \texttt{BHNS-U} and \texttt{BNS-DG} models only.  In Fig.~\ref{fig:es} we show how the shape of $\mathcal{P}(\widetilde{\Lambda})$ changes with $\mathcal{F}$. The left panel refers to the ALF2 EOS, and the right panel to the APR4 EOS. The two plots are qualitatively similar, although the range of possible values for $\widetilde{\Lambda}$ is very different. When $\mathcal{F}$ is close to unity most binaries are BHNSs, and $\mathcal{P}(\widetilde{\Lambda})$ has a single peak at values of $\widetilde{\Lambda}\lesssim 300 (200)$ for ALF2 (APR4). As $\mathcal{F}$ decreases below $\simeq 0.7$ the distribution becomes bimodal, with a second broad peak between $300 \lesssim \widetilde{\Lambda}\lesssim 1400$ ($100 \lesssim \widetilde{\Lambda}\lesssim 600$) for ALF2 (APR4). This second peak becomes more and more dominant in the limit $\mathcal{F}\to 0$, when BNSs dominate the observed population.

We sample the probability distribution of the hyperparameter $\mathcal{F}$ using a machine-learning emulator trained on numerical predictions and inserted into a Bayesian hierarchical framework~\cite{Taylor:2018iat,Wong:2019uni,Wong:2020jdt}.  We train a Gaussian process regression interpolant on 100 values of $\mathcal{F}\in  [0, \, 1]$.  The resulting emulator slots into a hierarchical Bayesian analysis and is fed with simulated data from observations with second- and third-generation interferometer networks, providing different constraints on $\mathcal{F}$.  We consider two detector configurations: (i) a network consisting of LIGO Hanford, LIGO Livingston \cite{TheLIGOScientific:2014jea} and Virgo \cite{TheVirgo:2014hva} (HLV), all operating at design sensitivity~\cite{HVL}, and (ii) a third-generation network \cite{Sathyaprakash:2019rom} composed of two Cosmic Explorer (CE) detectors~\cite{Reitze:2019iox,Evans:2016mbw} and one Einstein Telescope (ET)~\cite{Maggiore:2019uih,Punturo:2010zz}.

The observations injected within the code are simulated using the publicly available code \texttt{BILBY}, a Bayesian inference library for GW astronomy~\cite{bilby1,bilby2,Ashton:2018jfp}. For each binary injected in the data analysis pipeline, we randomly draw the component masses according to the specific model\footnote{\texttt{BILBY} cannot handle very small (and rare) values of $\tilde \Lambda$. Therefore we discard the lowest values of $\tilde \Lambda$, and this effectively sets the upper mass limit for BHs in our catalogs at $2M_\odot$.}, while the luminosity distance $d_L$ is sampled from a uniform distribution between 10 and 120~Mpc.

For the GW signal we use the \texttt{IMRPhenomPv2\_NRTidal} model~\cite{Dietrich:2018uni,Dietrich:2019kaq}. Numerical relativity simulations have shown that tidal disruption may occur in BHNS mergers, affecting the merger dynamics and introducing a characteristic frequency cutoff in the waveform. In general, the occurrence of tidal disruption and the waveform morphology are sensitive to the BH spin, the binary mass ratio and the EOS~\cite{Pannarale:2013uoa,Pannarale:2015jka,Pannarale:2015jia,Kruger:2020gig,Thompson:2020nei}. We focus (conservatively) on systems where tidal disruption does not occur, because the frequency cutoff due to tidal disruption makes BHNS binaries easier to tell apart from BNS binaries~\cite{Ferrari:2009bw,Maselli:2013rza}. We also verified by an explicit calculation that using PhenomNSBH~\cite{Thompson:2020nei}, a waveform model which includes tidal disruption, would not affect our conclusions (see Appendix~\ref{app:TD}).

We assume an isotropic source position and orientation in the sky.  We neglect the spins of both NSs (in which case the dimensionless spin parameter is expected to be $\lesssim0.3$~\cite{Miller:2014aaa}) and BHs (see e.g.~\cite{Baibhav:2020xdf,DeLuca:2020bjf}).  Although BHs, in principle, can have large spin, this should not significantly affect our results, because tidal deformability effects are expected to be dominant over spin effects at the relevant post-Newtonian order~\cite{Mora:2003wt}.
Moreover, we are focusing on black hole masses up to $2 \ M_{\odot}$. These may involve BHs of primordial origin or dark matter cores (see Appendix \ref{app:lowmassBHs}), and theoretical calculation in these scenarios suggest that BH spins should be negligible~\cite{DeLuca:2020bjf,Giddings:2008gr,East:2019dxt}.

For each posterior distribution of $\mathcal{F}$ we run three independent chains of $\sim 10^4$ samples, discarding the first $\sim10$ \% points as burn in.  The convergence of the Markov Chain Monte Carlo simulations is determined by cross-checking the chains through a standard Rubin test~\cite{Gilks:1996}.

We compute $\mathcal{P}(\widetilde{\Lambda})$, as defined in Eq.~\eqref{eq:plambda1}, for 100 values of $\mathcal{F}\in  [0, \, 1]$.  We have checked that the Gaussian process regression interpolant emulator reconstructs the probability distributions $\mathcal{P}(\widetilde{\Lambda})$ shown in Fig.~\ref{fig:es} with accuracy better than 10\% for any value of $\cal{F}$.

We choose a subset of binaries such that the inferred $\mathcal{P}(\widetilde{\Lambda})$ lies within the regions where we can correctly distinguish BNSs from BHNSs for both EOS models. For BHNSs (BNSs) we consider $\widetilde{\Lambda}$ in the range $\sim [50,\,200]$ ($[300,\,600]$) for APR4 and $[120,\,500]$ ($[400,\,1200]$) for ALF2. We inject these values into the machine learning emulator and reconstruct $\mathcal{P}(\mathcal{F)}$.

Figure~\ref{fig:samplef} shows the results of 10 and 60 simulated BNS and BHNS events assuming the ALF2 EOS in the HLV network (left panel) and for the the third-generation network of two CEs and one ET (right panel).  We reconstruct $\mathcal{P}(\cal{F})$ through a hierarchical Bayesian analysis under three assumptions: a ``pure BNS'' population ($\mathcal{F}=0$), a ``pure BHNS'' population ($\mathcal{F}=1$) and a ``perfectly mixed'' population ($\mathcal{F}=0.5$).  For the third-generation networks (right panel) the tidal deformability errors are roughly one order of magnitude smaller than for the second-generation network (left panel). This leads to slightly narrower probability distributions, but our results indicate that (quite remarkably) present detectors are sufficient to discriminate between the two populations, as long as the number of observations is large enough.  The median and the 68\% confidence intervals of the distributions are listed in Table~\ref{table:fmi}. Note that the comparison in Fig.~\ref{fig:samplef} (where we fix the number of observations) is somewhat unfair, because the higher sensitivity of third-generation detectors implies that event rates must increase with the cube of the sensitivity enhancement. As the number of events and detectors improve, the reconstruction of ${\cal P}(\mathcal{F})$ and our ability to determine $\mathcal{F}$ will get sensibly better.

This is one of the main conclusions of this work: current interferometers should already be able to determine the nature of low-mass compact binaries by measuring the tidal deformability distribution.  Roughly $\mathcal{O}(10)$ GW observations in the low-mass range can identify whether ${\cal F}$ favors double NSs or mixed binaries when one of the two families dominates the population, and a few tens of observations are sufficient to measure $\cal{F}$ with an accuracy $\sim 0.1$ even if both families contribute to the overall observed population.

\section{Conclusions}
\label{sec:conc}
A new era in astronomy has begun with the observation of compact binary coalescences by the LIGO and Virgo GW detectors. This complementary window to observe the Universe can inform our knowledge of fundamental physics and astrophysics. In particular, we can address the long-standing problem of how compact object binaries form and evolve by measuring their fundamental properties, such as the distribution of their masses and spins and their cosmological merger rates.

In this paper we have addressed how GW observations could be exploited to measure another key property of the population, namely the relative abundance of BNSs and BHNSs when the BHs masses are similar to those of NSs.  Delayed supernovae, the coalescence of NSs, certain models of dark matter and physical processes in the primordial Universe might produce such BHs. It is, therefore, critical to discriminate the two populations to test the different formation scenarios of BHs.

A crucial difference between BNS and BHNS systems arises because the dimensionless tidal deformability of NSs is $\Lambda \sim {\rm few} \times 100$, while it is predicted to be zero for BHs. Consequently, the effective tidal deformability $\widetilde{\Lambda}$ of a binary defined in Eq.~(\ref{eqn:Lambdone}), which depends on the tidal deformability of the binary components and their mass ratio, is significantly larger for BNSs ($\widetilde{\Lambda}_{\rm BNS} \sim 400$--1200 for the stiffer EOS, and $\sim 300$--600 for the softer EOS considered in this paper) than it is for BHNSs ($\widetilde{\Lambda}_{\rm BHNS} \sim 120$--500 for the stiffer EOS, and $\sim 50$--200 for the softer EOS). We exploit this asymmetry in the distribution of $\widetilde{\Lambda}$ to differentiate between the two populations.

To this end, we introduced a population hyperparameter $\cal F$ measuring the fraction of BHNS population relative to BNS population in the observed catalog of sources. We have shown that it is possible to infer the hyperparameter ${\cal F}$ from the measured distribution of $\widetilde{\Lambda}$. The distribution peaks at large (small) values of $\widetilde{\Lambda}$ if the population contains no BHNS (BNS) systems and ${\cal F} = 0$ (${\cal F} = 1$), while it will be bimodal if the population contains a significant population of BHNS systems, say $0.2 < {\cal F} < 0.8$.

The highlight of this investigation is that the network of GW detectors that are currently operational (LIGO Hanford, LIGO Livingston and Virgo) can constrain $\cal F$ at 68\% confidence level to the range $[0,\,0.2], [0.7,\,1]$ and $[0.3,\,0.7]$ with only 10 detections if the population is dominated by BNSs or BHNSs or an equal admixture of both, respectively. A larger number of observations, with 60 events, would increase our ability to reconstruct ${\cal F},$ pinning down the confidence intervals to $[0,\,0.05], [0.9,\,1]$ and $[0.4,\,0.6],$ for the same populations.

Our results are largely insensitive to the EOS of dense matter although stiffer equations of state do allow for a moderately better constraint on $\cal F.$ On the other hand, the mass ratio of the companion stars spreads the range of possible values of effective tidal deformability, limiting the accuracy with which the hyperparameter $\cal F$ can be inferred. If NS masses are confined to a narrower range than is assumed in this paper, then it will be possible to measure the relative fraction of BNSs and BHNSs more accurately. This is where CE and ET could make an impact: they will be able to provide us with a very precise distribution of NS masses by accurately measuring the masses of thousands of NSs.

\begin{acknowledgments}
We thank Omar Benhar, Stefania Marassi, Raffaella Schneider, Luca Graziani and Lucas Tonetto for useful discussions and advice on the methods used in this work and Nathan Johnson-McDaniel and Wolfgang Kastaun for comments on the manuscript.  E.B. and K.W. are supported by NSF Grants No. PHY-1912550 and AST-1841358, NASA ATP Grants No. 17-ATP17-0225 and 19-ATP19-0051, and NSF-XSEDE Grant No. PHY-090003 and B.S.S. in part by NSF Grant No. PHY-1836779, AST-1716394 and AST-1708146. This research was conducted using computational resources at the Maryland Advanced Research Computing Center (MARCC). The authors would like to acknowledge networking support by the GWverse COST Action CA16104, ``Black holes, gravitational waves and fundamental physics.''  We acknowledge support from the Amaldi Research Center funded by the MIUR program ``Dipartimento di Eccellenza'' (CUP: B81I18001170001).
\end{acknowledgments}

\appendix

\section{Low-mass black hole formation scenarios}
\label{app:lowmassBHs}

In this appendix we present a short overview of formation scenarios that could produce BHs in the mass range $\sim [1-3]\ M_\odot$.

\noindent {\bf \em Supernovae.}  One possibility to produce low-mass BHs is through supernova explosion. If the explosion is driven by rapidly growing instabilities with timescale of $10-20$~ms, it is expected to form BHs with masses $> 5\ M_{\odot}$. However, instabilities may develop over a longer ($>200$~ms) timescales and lead to lower mass remnants~\cite{Belczynski:2011bn}. In both cases, gravitational collapse could produce BHs compatible with the mass range considered in this paper.

\noindent {\bf \em Accretion-induced collapse.}  A second possibility is that NSs may gain mass through accretion and collapse to low-mass BHs~\cite{1983PThPh..70.1144N,Vietri:1999kj,MacFadyen:2005xm,Dermer:2006pw}.  Given current uncertainties on the maximum NS mass, it is not clear how to distinguish low-mass BHs formed in accretion-induced collapse from those formed in other channels.

\noindent {\bf \em Hierarchical mergers.} The hierarchical merger of BHs in dense environments is a possible channel to form the heaviest BHs observed by LIGO and Virgo~\cite{fishbach2017ligo,gerosa2017merging,antonini2016merging}.  Similarly, the remnant BHs produced by a BNS merger should often have masses below $3\ M_\odot$ and they could merge again in dense stellar environments, forming BHNS binaries with low-mass BHs via dynamical interactions~\cite{Gupta:2019nwj} (cf.~\cite{Ye:2019xvf} for caveats on the rates). An alternative scenario involves 2+2 quadruple systems, i.e.  wide binary systems in which each component is itself a binary~\cite{Fragione:2020aki}.

\noindent {\bf \em PBHs.} Current observational constraints on the PBH abundance from microlensing indicate that their mass fraction compared to dark matter may be as large as $f_\tn{PBH} \lesssim 10\%$~\cite{Carr:2020gox}.  If this bound is saturated, and we assume that the cross section for the dynamical capture of a NS and a BH of similar mass are comparable (this is reasonable, since the process is dominated by GW emission~\cite{East:2011xa}), then the merger rate of BHNSs may be even larger than the merger rate of dynamically formed BNSs~\cite{Yang:2017gfb}.

\noindent {\bf \em PBH captures.} Another possibility is that NSs, white dwarfs or even main sequence stars could capture mini PBHs with $M_\tn{BH}\ll 1M_\odot$. Efficient accretion from the star could then increase the PBH mass up to $\sim 1-3\ M_\odot$ \cite{Capela:2013yf,Fuller:2017uyd}. However, it is still not clear which fraction of NSs could survive this process to form a bound BHNS system \cite{Capela:2013yf}.

\noindent {\bf \em Dark matter cores.} It has been speculated that asymmetric dark matter could accumulate within the NS cores through nucleon scattering, and eventually form a BH seed~\cite{Goldman:1989nd,Bramante:2014zca,Bramante:2017ulk}, providing yet another possibility for converting a NS to a BH of similar mass.

\begin{figure}[t]
\centering
\includegraphics[width=0.5\textwidth]{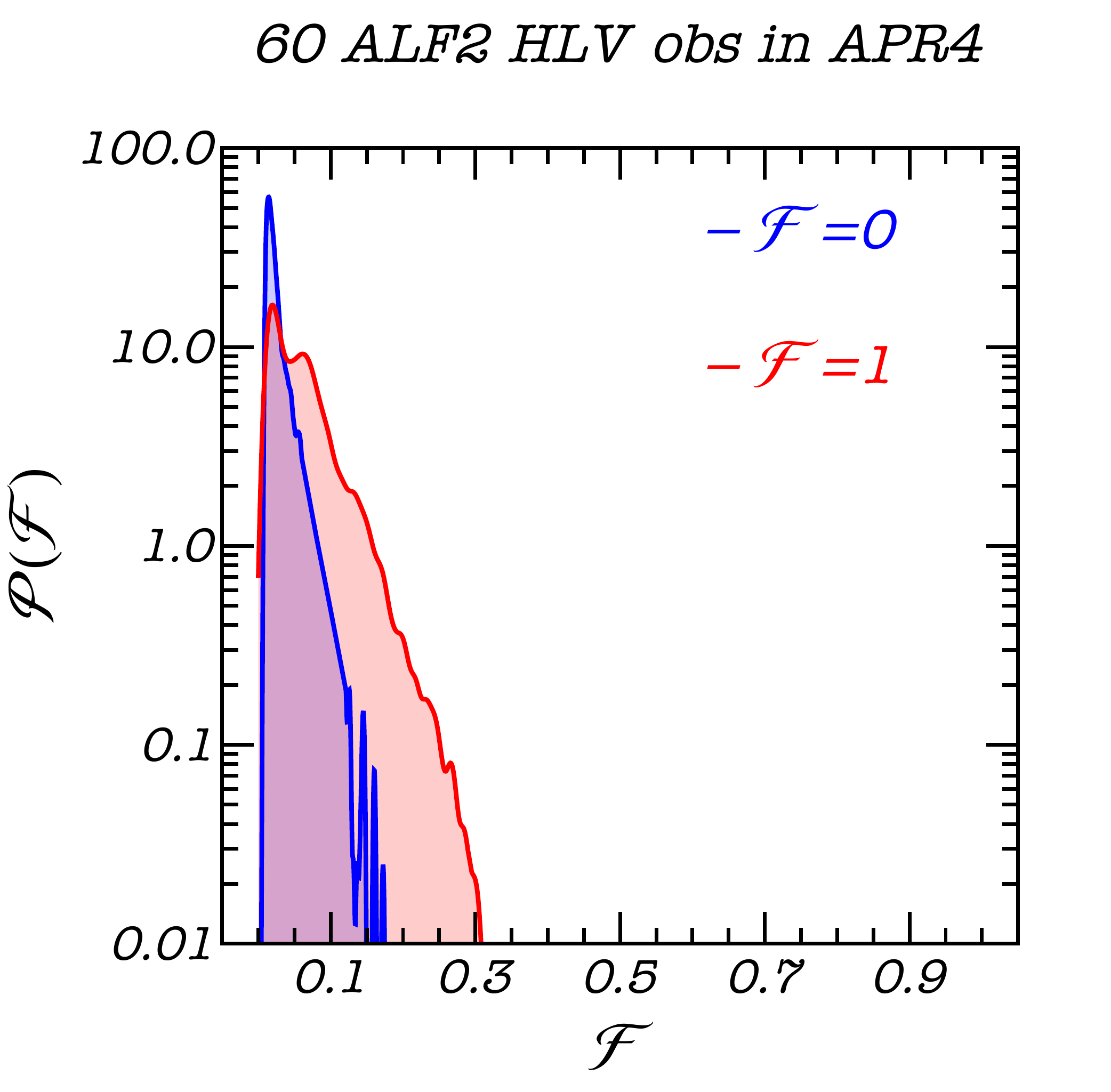}
\caption{Reconstructed probability density functions of the parameter $\mathcal{F}$ assuming $N_{\rm obs}=60$ observations with a second-generation detector network (HLV). Here we look at the most pessimistic scenario where we inject an ALF2 population into a Bayesian framework trained with the APR4 EOS, and we focus on the two extreme cases: a pure \texttt{BNS-DG} population ($\mathcal{F}=0$, blue) and a pure \texttt{BHNS-U} population ($\mathcal{F}=1$, red).  }
\label{fig:fig5} 
\end{figure}

\begin{figure}[t]
    \centering
    \includegraphics[width=0.5\textwidth]{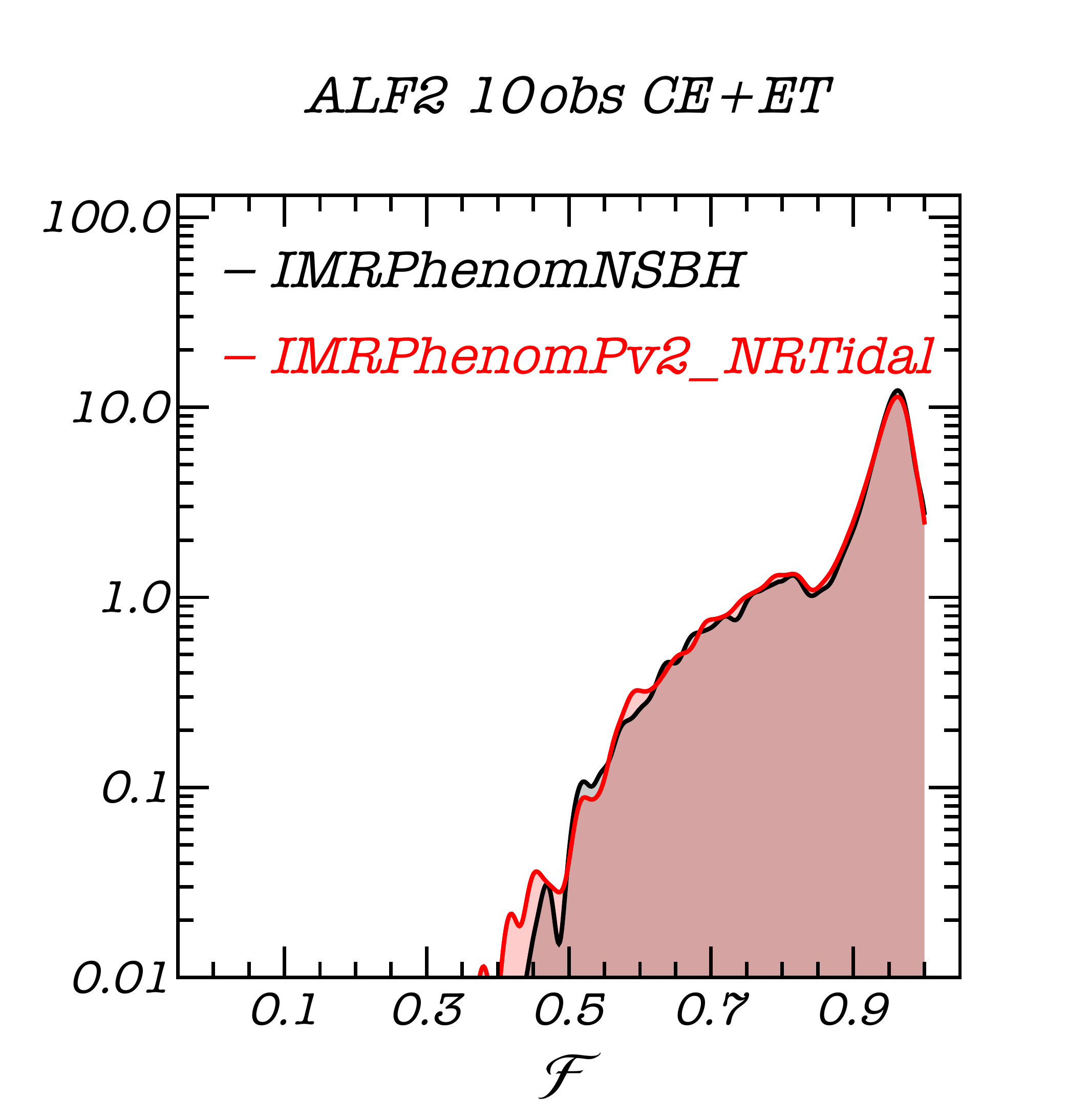}
    \caption{Probability distribution of $\mathcal{F}$ obtained by injecting 10 observations where tidal disruption occurs, as determined by imposing $q< Q_{D}(\mathcal{C},\mathcal{\chi})$ (cf. Eq.~(2) of Ref.~\cite{Pannarale:2015jia}). We simulate our binaries using either the \texttt{IMRPhenomPv2\_NRTidal} (red) or the \texttt{PhenomNSBH} (black) waveform models, using the ALF2 EOS and the CE+ET detector network. Allowing for tidal disruption does not significantly change the distributions inferred in the right panel of Fig.~\ref{fig:samplef}.}
    \label{fig:fig6}
\end{figure}

\section{EOS systematics}
\label{app:EOS}

Uncertainties in the NS EOS can affect our ability to distinguish BHNS from BNS systems. In this appendix we focus on the most pessimistic scenario compatible with current observations, and we analyze 60 detections with the HLV network to estimate the worst-case impact of EOS uncertainties on our results.

We inject GW observations of binaries modelled with the ALF2 EOS and different values of the mixing parameter ${\cal F}$ into the Bayesian framework trained with APR4. When we inject purely BNS binaries (blue curve in Fig.~\ref{fig:fig5}) we correctly recover the expected fraction of BNS/BHNS, i.e. ${\cal F}=0$. On the other hand, when we 
inject purely BHNS systems (red distribution in Fig.~\ref{fig:fig5}) we obtain inconsistent results, i.e. the posterior probability of ${\cal F}$ peaks around the wrong estimate. This is because the ALF2 EOS yields large values of the tidal deformability $\widetilde{\Lambda}$ relative to APR4, and therefore the observed events are misinterpreted as BNSs when they would be interpreted as BHNS with the ``correct'' assumption on the EOS.

Vice versa, we can inject observations of binaries modelled with APR4 into a Bayesian framework trained with ALF2. In this case, when we inject a pure-BHNS population we can correctly recover it as such, because APR4 yields values of $\widetilde{\Lambda}$ which are generally smaller than those derived using ALF2. However, when we inject BNS systems the approach leads to the wrong reconstruction.

In summary, EOS uncertainties can dramatically affect the inference in the worst-case scenario where we use the stiffest EOS to recover astrophysical systems that correspond to the softest EOS, or vice versa. We expect the uncertainty in the EOS to reduce significantly through electromagnetic and GW observations well before third-generation detectors become operational. A more realistic Bayesian analysis should reconstruct the EOS and the population fraction simultaneously. We plan to address this problem in future work.

\section{Tidal disruption}
\label{app:TD}

Tidal disruption can affect the merger dynamics. Investigations of BHNS mergers in numerical relativity have shown that the occurrence and nature of tidal disruption depends, in general, on the binary parameters, and in particular on the BH spin, the adopted EOS model, and the binary mass ratio (cf.~\cite{Pannarale:2015jia,Kruger:2020gig}). Tidal disruption is a characteristic signature of the presence of a BH in the binary, because it typically produces a sharp frequency cutoff in the GW signal~\cite{Vallisneri:1999nq}. From the presence of such a cutoff (if detectable) we can conclude that one of the merging objects is a BH~\cite{Ferrari:2009bw,Maselli:2013rza}.

In most of our work we have focused on systems where tidal disruption does not occur in the sensitive frequency band of  the detectors. This is a conservative choice, in the sense that our method does not rely on the additional information provided by tidal disruption. To check that tidal-disruption effects would not significantly affect our conclusions, we have computed the posterior probability distribution of $\mathcal{F}$ obtained using injections in which disruptive mergers can occur. Using the criterion in Eq.~(2) of \cite{Pannarale:2015jia}, this corresponds to binaries with $q< Q_{D}(\mathcal{C},\mathcal{\chi})$, where $q=M_{\rm BH}/M_{\rm NS}$ is the binary mass ratio, $\mathcal{C}=M_{\rm \rm NS}/R_{\rm NS}$ is the NS compactness, and $\mathcal{\chi}$ is the dimensionless BH spin. In Fig.~\ref{fig:fig6} we compare results obtained using the \texttt{IMRPhenomNSBH} model~\cite{Thompson:2020nei}  (black line), which allows for tidal disruption, against the results in right panel of Fig.~\ref{fig:samplef} (red line). Our main conclusions are clearly unaffected.

\bibliography{biblio}

\begin{thebibliography}{141}%
\makeatletter
\providecommand \@ifxundefined [1]{%
 \@ifx{#1\undefined}
}%
\providecommand \@ifnum [1]{%
 \ifnum #1\expandafter \@firstoftwo
 \else \expandafter \@secondoftwo
 \fi
}%
\providecommand \@ifx [1]{%
 \ifx #1\expandafter \@firstoftwo
 \else \expandafter \@secondoftwo
 \fi
}%
\providecommand \natexlab [1]{#1}%
\providecommand \enquote  [1]{``#1''}%
\providecommand \bibnamefont  [1]{#1}%
\providecommand \bibfnamefont [1]{#1}%
\providecommand \citenamefont [1]{#1}%
\providecommand \href@noop [0]{\@secondoftwo}%
\providecommand \href [0]{\begingroup \@sanitize@url \@href}%
\providecommand \@href[1]{\@@startlink{#1}\@@href}%
\providecommand \@@href[1]{\endgroup#1\@@endlink}%
\providecommand \@sanitize@url [0]{\catcode `\\12\catcode `\$12\catcode
  `\&12\catcode `\#12\catcode `\^12\catcode `\_12\catcode `\%12\relax}%
\providecommand \@@startlink[1]{}%
\providecommand \@@endlink[0]{}%
\providecommand \url  [0]{\begingroup\@sanitize@url \@url }%
\providecommand \@url [1]{\endgroup\@href {#1}{\urlprefix }}%
\providecommand \urlprefix  [0]{URL }%
\providecommand \Eprint [0]{\href }%
\providecommand \doibase [0]{http://dx.doi.org/}%
\providecommand \selectlanguage [0]{\@gobble}%
\providecommand \bibinfo  [0]{\@secondoftwo}%
\providecommand \bibfield  [0]{\@secondoftwo}%
\providecommand \translation [1]{[#1]}%
\providecommand \BibitemOpen [0]{}%
\providecommand \bibitemStop [0]{}%
\providecommand \bibitemNoStop [0]{.\EOS\space}%
\providecommand \EOS [0]{\spacefactor3000\relax}%
\providecommand \BibitemShut  [1]{\csname bibitem#1\endcsname}%
\let\auto@bib@innerbib\@empty
\bibitem [{\citenamefont {Abbott}\ \emph
  {et~al.}(2016{\natexlab{a}})\citenamefont {Abbott} \emph
  {et~al.}}]{Abbott:2016blz}%
  \BibitemOpen
  \bibfield  {author} {\bibinfo {author} {\bibfnamefont {B.~P.}\ \bibnamefont
  {Abbott}} \emph {et~al.} (\bibinfo {collaboration} {LIGO Scientific,
  Virgo}),\ }\href {\doibase 10.1103/PhysRevLett.116.061102} {\bibfield
  {journal} {\bibinfo  {journal} {Phys. Rev. Lett.}\ }\textbf {\bibinfo
  {volume} {116}},\ \bibinfo {pages} {061102} (\bibinfo {year}
  {2016}{\natexlab{a}})},\ \Eprint {http://arxiv.org/abs/1602.03837}
  {arXiv:1602.03837 [gr-qc]} \BibitemShut {NoStop}%
\bibitem [{\citenamefont {Abbott}\ \emph
  {et~al.}(2016{\natexlab{b}})\citenamefont {Abbott} \emph
  {et~al.}}]{Abbott:2016nmj}%
  \BibitemOpen
  \bibfield  {author} {\bibinfo {author} {\bibfnamefont {B.~P.}\ \bibnamefont
  {Abbott}} \emph {et~al.} (\bibinfo {collaboration} {LIGO Scientific,
  Virgo}),\ }\href {\doibase 10.1103/PhysRevLett.116.241103} {\bibfield
  {journal} {\bibinfo  {journal} {Phys. Rev. Lett.}\ }\textbf {\bibinfo
  {volume} {116}},\ \bibinfo {pages} {241103} (\bibinfo {year}
  {2016}{\natexlab{b}})},\ \Eprint {http://arxiv.org/abs/1606.04855}
  {arXiv:1606.04855 [gr-qc]} \BibitemShut {NoStop}%
\bibitem [{\citenamefont {Abbott}\ \emph
  {et~al.}(2017{\natexlab{a}})\citenamefont {Abbott} \emph
  {et~al.}}]{Abbott:2017vtc}%
  \BibitemOpen
  \bibfield  {author} {\bibinfo {author} {\bibfnamefont {B.~P.}\ \bibnamefont
  {Abbott}} \emph {et~al.} (\bibinfo {collaboration} {LIGO Scientific,
  VIRGO}),\ }\href {\doibase 10.1103/PhysRevLett.118.221101,
  10.1103/PhysRevLett.121.129901} {\bibfield  {journal} {\bibinfo  {journal}
  {Phys. Rev. Lett.}\ }\textbf {\bibinfo {volume} {118}},\ \bibinfo {pages}
  {221101} (\bibinfo {year} {2017}{\natexlab{a}})},\ \bibinfo {note} {[Erratum:
  Phys. Rev. Lett.121,no.12,129901(2018)]},\ \Eprint
  {http://arxiv.org/abs/1706.01812} {arXiv:1706.01812 [gr-qc]} \BibitemShut
  {NoStop}%
\bibitem [{\citenamefont {Abbott}\ \emph
  {et~al.}(2017{\natexlab{b}})\citenamefont {Abbott} \emph
  {et~al.}}]{Abbott:2017oio}%
  \BibitemOpen
  \bibfield  {author} {\bibinfo {author} {\bibfnamefont {B.~P.}\ \bibnamefont
  {Abbott}} \emph {et~al.} (\bibinfo {collaboration} {LIGO Scientific,
  Virgo}),\ }\href {\doibase 10.1103/PhysRevLett.119.141101} {\bibfield
  {journal} {\bibinfo  {journal} {Phys. Rev. Lett.}\ }\textbf {\bibinfo
  {volume} {119}},\ \bibinfo {pages} {141101} (\bibinfo {year}
  {2017}{\natexlab{b}})},\ \Eprint {http://arxiv.org/abs/1709.09660}
  {arXiv:1709.09660 [gr-qc]} \BibitemShut {NoStop}%
\bibitem [{\citenamefont {Aasi}\ \emph {et~al.}(2015)\citenamefont {Aasi} \emph
  {et~al.}}]{TheLIGOScientific:2014jea}%
  \BibitemOpen
  \bibfield  {author} {\bibinfo {author} {\bibfnamefont {J.}~\bibnamefont
  {Aasi}} \emph {et~al.} (\bibinfo {collaboration} {LIGO Scientific}),\ }\href
  {\doibase 10.1088/0264-9381/32/7/074001} {\bibfield  {journal} {\bibinfo
  {journal} {Class. Quant. Grav.}\ }\textbf {\bibinfo {volume} {32}},\ \bibinfo
  {pages} {074001} (\bibinfo {year} {2015})},\ \Eprint
  {http://arxiv.org/abs/1411.4547} {arXiv:1411.4547 [gr-qc]} \BibitemShut
  {NoStop}%
\bibitem [{\citenamefont {Acernese}\ \emph {et~al.}(2015)\citenamefont
  {Acernese} \emph {et~al.}}]{TheVirgo:2014hva}%
  \BibitemOpen
  \bibfield  {author} {\bibinfo {author} {\bibfnamefont {F.}~\bibnamefont
  {Acernese}} \emph {et~al.} (\bibinfo {collaboration} {VIRGO}),\ }\href
  {\doibase 10.1088/0264-9381/32/2/024001} {\bibfield  {journal} {\bibinfo
  {journal} {Class. Quant. Grav.}\ }\textbf {\bibinfo {volume} {32}},\ \bibinfo
  {pages} {024001} (\bibinfo {year} {2015})},\ \Eprint
  {http://arxiv.org/abs/1408.3978} {arXiv:1408.3978 [gr-qc]} \BibitemShut
  {NoStop}%
\bibitem [{\citenamefont {Abbott}\ \emph
  {et~al.}(2017{\natexlab{c}})\citenamefont {Abbott} \emph
  {et~al.}}]{TheLIGOScientific:2017qsa}%
  \BibitemOpen
  \bibfield  {author} {\bibinfo {author} {\bibfnamefont {B.~P.}\ \bibnamefont
  {Abbott}} \emph {et~al.} (\bibinfo {collaboration} {LIGO Scientific,
  Virgo}),\ }\href {\doibase 10.1103/PhysRevLett.119.161101} {\bibfield
  {journal} {\bibinfo  {journal} {Phys. Rev. Lett.}\ }\textbf {\bibinfo
  {volume} {119}},\ \bibinfo {pages} {161101} (\bibinfo {year}
  {2017}{\natexlab{c}})},\ \Eprint {http://arxiv.org/abs/1710.05832}
  {arXiv:1710.05832 [gr-qc]} \BibitemShut {NoStop}%
\bibitem [{\citenamefont {Abbott}\ \emph
  {et~al.}(2019{\natexlab{a}})\citenamefont {Abbott} \emph
  {et~al.}}]{Abbott:2018wiz}%
  \BibitemOpen
  \bibfield  {author} {\bibinfo {author} {\bibfnamefont {B.~P.}\ \bibnamefont
  {Abbott}} \emph {et~al.} (\bibinfo {collaboration} {LIGO Scientific,
  Virgo}),\ }\href {\doibase 10.1103/PhysRevX.9.011001} {\bibfield  {journal}
  {\bibinfo  {journal} {Phys. Rev.}\ }\textbf {\bibinfo {volume} {X9}},\
  \bibinfo {pages} {011001} (\bibinfo {year} {2019}{\natexlab{a}})},\ \Eprint
  {http://arxiv.org/abs/1805.11579} {arXiv:1805.11579 [gr-qc]} \BibitemShut
  {NoStop}%
\bibitem [{\citenamefont {Abbott}\ \emph
  {et~al.}(2017{\natexlab{d}})\citenamefont {Abbott} \emph
  {et~al.}}]{GBM:2017lvd}%
  \BibitemOpen
  \bibfield  {author} {\bibinfo {author} {\bibfnamefont {B.}~\bibnamefont
  {Abbott}} \emph {et~al.} (\bibinfo {collaboration} {LIGO Scientific, Virgo,
  Fermi GBM, INTEGRAL, IceCube, AstroSat Cadmium Zinc Telluride Imager Team,
  IPN, Insight-Hxmt, ANTARES, Swift, AGILE Team, 1M2H Team, Dark Energy Camera
  GW-EM, DES, DLT40, GRAWITA, Fermi-LAT, ATCA, ASKAP, Las Cumbres Observatory
  Group, OzGrav, DWF (Deeper Wider Faster Program), AST3, CAASTRO, VINROUGE,
  MASTER, J-GEM, GROWTH, JAGWAR, CaltechNRAO, TTU-NRAO, NuSTAR, Pan-STARRS,
  MAXI Team, TZAC Consortium, KU, Nordic Optical Telescope, ePESSTO, GROND,
  Texas Tech University, SALT Group, TOROS, BOOTES, MWA, CALET, IKI-GW
  Follow-up, H.E.S.S., LOFAR, LWA, HAWC, Pierre Auger, ALMA, Euro VLBI Team, Pi
  of Sky, Chandra Team at McGill University, DFN, ATLAS Telescopes, High Time
  Resolution Universe Survey, RIMAS, RATIR, SKA South Africa/MeerKAT}),\ }\href
  {\doibase 10.3847/2041-8213/aa91c9} {\bibfield  {journal} {\bibinfo
  {journal} {Astrophys. J.}\ }\textbf {\bibinfo {volume} {848}},\ \bibinfo
  {pages} {L12} (\bibinfo {year} {2017}{\natexlab{d}})},\ \Eprint
  {http://arxiv.org/abs/1710.05833} {arXiv:1710.05833 [astro-ph.HE]}
  \BibitemShut {NoStop}%
\bibitem [{\citenamefont {Abbott}\ \emph
  {et~al.}(2017{\natexlab{e}})\citenamefont {Abbott} \emph
  {et~al.}}]{Monitor:2017mdv}%
  \BibitemOpen
  \bibfield  {author} {\bibinfo {author} {\bibfnamefont {B.}~\bibnamefont
  {Abbott}} \emph {et~al.} (\bibinfo {collaboration} {LIGO Scientific, Virgo,
  Fermi-GBM, INTEGRAL}),\ }\href {\doibase 10.3847/2041-8213/aa920c} {\bibfield
   {journal} {\bibinfo  {journal} {Astrophys. J.}\ }\textbf {\bibinfo {volume}
  {848}},\ \bibinfo {pages} {L13} (\bibinfo {year} {2017}{\natexlab{e}})},\
  \Eprint {http://arxiv.org/abs/1710.05834} {arXiv:1710.05834 [astro-ph.HE]}
  \BibitemShut {NoStop}%
\bibitem [{\citenamefont {Lattimer}\ and\ \citenamefont
  {Schramm}(1974)}]{Lattimer:1974slx}%
  \BibitemOpen
  \bibfield  {author} {\bibinfo {author} {\bibfnamefont {J.}~\bibnamefont
  {Lattimer}}\ and\ \bibinfo {author} {\bibfnamefont {D.}~\bibnamefont
  {Schramm}},\ }\href {\doibase 10.1086/181612} {\bibfield  {journal} {\bibinfo
   {journal} {Astrophys. J.}\ }\textbf {\bibinfo {volume} {192}},\ \bibinfo
  {pages} {L145} (\bibinfo {year} {1974})}\BibitemShut {NoStop}%
\bibitem [{\citenamefont {De}\ \emph {et~al.}(2018)\citenamefont {De},
  \citenamefont {Finstad}, \citenamefont {Lattimer}, \citenamefont {Brown},
  \citenamefont {Berger},\ and\ \citenamefont {Biwer}}]{De:2018uhw}%
  \BibitemOpen
  \bibfield  {author} {\bibinfo {author} {\bibfnamefont {S.}~\bibnamefont
  {De}}, \bibinfo {author} {\bibfnamefont {D.}~\bibnamefont {Finstad}},
  \bibinfo {author} {\bibfnamefont {J.~M.}\ \bibnamefont {Lattimer}}, \bibinfo
  {author} {\bibfnamefont {D.~A.}\ \bibnamefont {Brown}}, \bibinfo {author}
  {\bibfnamefont {E.}~\bibnamefont {Berger}}, \ and\ \bibinfo {author}
  {\bibfnamefont {C.~M.}\ \bibnamefont {Biwer}},\ }\href {\doibase
  10.1103/PhysRevLett.121.091102} {\bibfield  {journal} {\bibinfo  {journal}
  {Phys. Rev. Lett.}\ }\textbf {\bibinfo {volume} {121}},\ \bibinfo {pages}
  {091102} (\bibinfo {year} {2018})},\ \bibinfo {note} {[Erratum:
  Phys.Rev.Lett. 121, 259902 (2018)]},\ \Eprint
  {http://arxiv.org/abs/1804.08583} {arXiv:1804.08583 [astro-ph.HE]}
  \BibitemShut {NoStop}%
\bibitem [{\citenamefont {Abbott}\ \emph {et~al.}(2018)\citenamefont {Abbott}
  \emph {et~al.}}]{Abbott:2018exr}%
  \BibitemOpen
  \bibfield  {author} {\bibinfo {author} {\bibfnamefont {B.~P.}\ \bibnamefont
  {Abbott}} \emph {et~al.} (\bibinfo {collaboration} {LIGO Scientific,
  Virgo}),\ }\href {\doibase 10.1103/PhysRevLett.121.161101} {\bibfield
  {journal} {\bibinfo  {journal} {Phys. Rev. Lett.}\ }\textbf {\bibinfo
  {volume} {121}},\ \bibinfo {pages} {161101} (\bibinfo {year} {2018})},\
  \Eprint {http://arxiv.org/abs/1805.11581} {arXiv:1805.11581 [gr-qc]}
  \BibitemShut {NoStop}%
\bibitem [{\citenamefont {Abbott}\ \emph
  {et~al.}(2020{\natexlab{a}})\citenamefont {Abbott} \emph
  {et~al.}}]{LIGOScientific:2019eut}%
  \BibitemOpen
  \bibfield  {author} {\bibinfo {author} {\bibfnamefont {B.~P.}\ \bibnamefont
  {Abbott}} \emph {et~al.} (\bibinfo {collaboration} {LIGO Scientific,
  Virgo}),\ }\href {\doibase 10.1088/1361-6382/ab5f7c} {\bibfield  {journal}
  {\bibinfo  {journal} {Class. Quant. Grav.}\ }\textbf {\bibinfo {volume}
  {37}},\ \bibinfo {pages} {045006} (\bibinfo {year} {2020}{\natexlab{a}})},\
  \Eprint {http://arxiv.org/abs/1908.01012} {arXiv:1908.01012 [gr-qc]}
  \BibitemShut {NoStop}%
\bibitem [{\citenamefont {Landry}\ \emph {et~al.}(2020)\citenamefont {Landry},
  \citenamefont {Essick},\ and\ \citenamefont
  {Chatziioannou}}]{Landry:2020vaw}%
  \BibitemOpen
  \bibfield  {author} {\bibinfo {author} {\bibfnamefont {P.}~\bibnamefont
  {Landry}}, \bibinfo {author} {\bibfnamefont {R.}~\bibnamefont {Essick}}, \
  and\ \bibinfo {author} {\bibfnamefont {K.}~\bibnamefont {Chatziioannou}},\
  }\href@noop {} {\  (\bibinfo {year} {2020})},\ \Eprint
  {http://arxiv.org/abs/2003.04880} {arXiv:2003.04880 [astro-ph.HE]}
  \BibitemShut {NoStop}%
\bibitem [{\citenamefont {Kastaun}\ and\ \citenamefont
  {Ohme}(2019)}]{Kastaun:2019bxo}%
  \BibitemOpen
  \bibfield  {author} {\bibinfo {author} {\bibfnamefont {W.}~\bibnamefont
  {Kastaun}}\ and\ \bibinfo {author} {\bibfnamefont {F.}~\bibnamefont {Ohme}},\
  }\href {\doibase 10.1103/PhysRevD.100.103023} {\bibfield  {journal} {\bibinfo
   {journal} {Phys. Rev. D}\ }\textbf {\bibinfo {volume} {100}},\ \bibinfo
  {pages} {103023} (\bibinfo {year} {2019})},\ \Eprint
  {http://arxiv.org/abs/1909.12718} {arXiv:1909.12718 [gr-qc]} \BibitemShut
  {NoStop}%
\bibitem [{\citenamefont {Abbott}\ \emph
  {et~al.}(2020{\natexlab{b}})\citenamefont {Abbott} \emph
  {et~al.}}]{Abbott:2020uma}%
  \BibitemOpen
  \bibfield  {author} {\bibinfo {author} {\bibfnamefont {B.~P.}\ \bibnamefont
  {Abbott}} \emph {et~al.} (\bibinfo {collaboration} {LIGO Scientific,
  Virgo}),\ }\href@noop {} {\  (\bibinfo {year} {2020}{\natexlab{b}})},\
  \Eprint {http://arxiv.org/abs/2001.01761} {arXiv:2001.01761 [astro-ph.HE]}
  \BibitemShut {NoStop}%
\bibitem [{\citenamefont {Han}\ \emph {et~al.}(2020)\citenamefont {Han},
  \citenamefont {Tang}, \citenamefont {Hu}, \citenamefont {Li}, \citenamefont
  {Jiang}, \citenamefont {Jin}, \citenamefont {Fan},\ and\ \citenamefont
  {Wei}}]{Han:2020qmn}%
  \BibitemOpen
  \bibfield  {author} {\bibinfo {author} {\bibfnamefont {M.-Z.}\ \bibnamefont
  {Han}}, \bibinfo {author} {\bibfnamefont {S.-P.}\ \bibnamefont {Tang}},
  \bibinfo {author} {\bibfnamefont {Y.-M.}\ \bibnamefont {Hu}}, \bibinfo
  {author} {\bibfnamefont {Y.-J.}\ \bibnamefont {Li}}, \bibinfo {author}
  {\bibfnamefont {J.-L.}\ \bibnamefont {Jiang}}, \bibinfo {author}
  {\bibfnamefont {Z.-P.}\ \bibnamefont {Jin}}, \bibinfo {author} {\bibfnamefont
  {Y.-Z.}\ \bibnamefont {Fan}}, \ and\ \bibinfo {author} {\bibfnamefont
  {D.-M.}\ \bibnamefont {Wei}},\ }\href {\doibase 10.3847/2041-8213/ab745a}
  {\bibfield  {journal} {\bibinfo  {journal} {Astrophys. J. Lett.}\ }\textbf
  {\bibinfo {volume} {891}},\ \bibinfo {pages} {L5} (\bibinfo {year} {2020})},\
  \Eprint {http://arxiv.org/abs/2001.07882} {arXiv:2001.07882 [astro-ph.HE]}
  \BibitemShut {NoStop}%
\bibitem [{\citenamefont {Metzger}(2017)}]{Metzger:2016pju}%
  \BibitemOpen
  \bibfield  {author} {\bibinfo {author} {\bibfnamefont {B.~D.}\ \bibnamefont
  {Metzger}},\ }\href {\doibase 10.1007/s41114-017-0006-z} {\bibfield
  {journal} {\bibinfo  {journal} {Living Rev. Rel.}\ }\textbf {\bibinfo
  {volume} {20}},\ \bibinfo {pages} {3} (\bibinfo {year} {2017})},\ \Eprint
  {http://arxiv.org/abs/1610.09381} {arXiv:1610.09381 [astro-ph.HE]}
  \BibitemShut {NoStop}%
\bibitem [{\citenamefont {Hinderer}\ \emph {et~al.}(2019)\citenamefont
  {Hinderer} \emph {et~al.}}]{Hinderer:2018pei}%
  \BibitemOpen
  \bibfield  {author} {\bibinfo {author} {\bibfnamefont {T.}~\bibnamefont
  {Hinderer}} \emph {et~al.},\ }\href {\doibase 10.1103/PhysRevD.100.063021}
  {\bibfield  {journal} {\bibinfo  {journal} {Phys. Rev. D}\ }\textbf {\bibinfo
  {volume} {100}},\ \bibinfo {pages} {06321} (\bibinfo {year} {2019})},\
  \Eprint {http://arxiv.org/abs/1808.03836} {arXiv:1808.03836 [astro-ph.HE]}
  \BibitemShut {NoStop}%
\bibitem [{\citenamefont {Coughlin}\ and\ \citenamefont
  {Dietrich}(2019)}]{Coughlin:2019kqf}%
  \BibitemOpen
  \bibfield  {author} {\bibinfo {author} {\bibfnamefont {M.~W.}\ \bibnamefont
  {Coughlin}}\ and\ \bibinfo {author} {\bibfnamefont {T.}~\bibnamefont
  {Dietrich}},\ }\href {\doibase 10.1103/PhysRevD.100.043011} {\bibfield
  {journal} {\bibinfo  {journal} {Phys. Rev. D}\ }\textbf {\bibinfo {volume}
  {100}},\ \bibinfo {pages} {043011} (\bibinfo {year} {2019})},\ \Eprint
  {http://arxiv.org/abs/1901.06052} {arXiv:1901.06052 [astro-ph.HE]}
  \BibitemShut {NoStop}%
\bibitem [{\citenamefont {Siegel}(2019)}]{Siegel:2019mlp}%
  \BibitemOpen
  \bibfield  {author} {\bibinfo {author} {\bibfnamefont {D.~M.}\ \bibnamefont
  {Siegel}},\ }\href {\doibase 10.1140/epja/i2019-12888-9} {\bibfield
  {journal} {\bibinfo  {journal} {Eur. Phys. J. A}\ }\textbf {\bibinfo {volume}
  {55}},\ \bibinfo {pages} {203} (\bibinfo {year} {2019})},\ \Eprint
  {http://arxiv.org/abs/1901.09044} {arXiv:1901.09044 [astro-ph.HE]}
  \BibitemShut {NoStop}%
\bibitem [{\citenamefont {Kyutoku}\ \emph {et~al.}(2020)\citenamefont
  {Kyutoku}, \citenamefont {Fujibayashi}, \citenamefont {Hayashi},
  \citenamefont {Kawaguchi}, \citenamefont {Kiuchi}, \citenamefont {Shibata},\
  and\ \citenamefont {Tanaka}}]{Kyutoku:2020xka}%
  \BibitemOpen
  \bibfield  {author} {\bibinfo {author} {\bibfnamefont {K.}~\bibnamefont
  {Kyutoku}}, \bibinfo {author} {\bibfnamefont {S.}~\bibnamefont
  {Fujibayashi}}, \bibinfo {author} {\bibfnamefont {K.}~\bibnamefont
  {Hayashi}}, \bibinfo {author} {\bibfnamefont {K.}~\bibnamefont {Kawaguchi}},
  \bibinfo {author} {\bibfnamefont {K.}~\bibnamefont {Kiuchi}}, \bibinfo
  {author} {\bibfnamefont {M.}~\bibnamefont {Shibata}}, \ and\ \bibinfo
  {author} {\bibfnamefont {M.}~\bibnamefont {Tanaka}},\ }\href {\doibase
  10.3847/2041-8213/ab6e70} {\bibfield  {journal} {\bibinfo  {journal}
  {Astrophys. J. Lett.}\ }\textbf {\bibinfo {volume} {890}},\ \bibinfo {pages}
  {L4} (\bibinfo {year} {2020})},\ \Eprint {http://arxiv.org/abs/2001.04474}
  {arXiv:2001.04474 [astro-ph.HE]} \BibitemShut {NoStop}%
\bibitem [{\citenamefont {Barbieri}\ \emph {et~al.}(2019)\citenamefont
  {Barbieri}, \citenamefont {Salafia}, \citenamefont {Colpi}, \citenamefont
  {Ghirlanda}, \citenamefont {Perego},\ and\ \citenamefont
  {Colombo}}]{Barbieri:2019bdq}%
  \BibitemOpen
  \bibfield  {author} {\bibinfo {author} {\bibfnamefont {C.}~\bibnamefont
  {Barbieri}}, \bibinfo {author} {\bibfnamefont {O.~S.}\ \bibnamefont
  {Salafia}}, \bibinfo {author} {\bibfnamefont {M.}~\bibnamefont {Colpi}},
  \bibinfo {author} {\bibfnamefont {G.}~\bibnamefont {Ghirlanda}}, \bibinfo
  {author} {\bibfnamefont {A.}~\bibnamefont {Perego}}, \ and\ \bibinfo {author}
  {\bibfnamefont {A.}~\bibnamefont {Colombo}},\ }\href {\doibase
  10.3847/2041-8213/ab5c1e} {\bibfield  {journal} {\bibinfo  {journal}
  {Astrophys. J. Lett.}\ }\textbf {\bibinfo {volume} {887}},\ \bibinfo {pages}
  {L35} (\bibinfo {year} {2019})},\ \Eprint {http://arxiv.org/abs/1912.03894}
  {arXiv:1912.03894 [astro-ph.HE]} \BibitemShut {NoStop}%
\bibitem [{\citenamefont {Pannarale}\ \emph
  {et~al.}(2015{\natexlab{a}})\citenamefont {Pannarale}, \citenamefont {Berti},
  \citenamefont {Kyutoku}, \citenamefont {Lackey},\ and\ \citenamefont
  {Shibata}}]{Pannarale:2015jia}%
  \BibitemOpen
  \bibfield  {author} {\bibinfo {author} {\bibfnamefont {F.}~\bibnamefont
  {Pannarale}}, \bibinfo {author} {\bibfnamefont {E.}~\bibnamefont {Berti}},
  \bibinfo {author} {\bibfnamefont {K.}~\bibnamefont {Kyutoku}}, \bibinfo
  {author} {\bibfnamefont {B.~D.}\ \bibnamefont {Lackey}}, \ and\ \bibinfo
  {author} {\bibfnamefont {M.}~\bibnamefont {Shibata}},\ }\href {\doibase
  10.1103/PhysRevD.92.081504} {\bibfield  {journal} {\bibinfo  {journal} {Phys.
  Rev. D}\ }\textbf {\bibinfo {volume} {92}},\ \bibinfo {pages} {081504}
  (\bibinfo {year} {2015}{\natexlab{a}})},\ \Eprint
  {http://arxiv.org/abs/1509.06209} {arXiv:1509.06209 [gr-qc]} \BibitemShut
  {NoStop}%
\bibitem [{\citenamefont {Foucart}\ \emph {et~al.}(2018)\citenamefont
  {Foucart}, \citenamefont {Hinderer},\ and\ \citenamefont
  {Nissanke}}]{Foucart:2018rjc}%
  \BibitemOpen
  \bibfield  {author} {\bibinfo {author} {\bibfnamefont {F.}~\bibnamefont
  {Foucart}}, \bibinfo {author} {\bibfnamefont {T.}~\bibnamefont {Hinderer}}, \
  and\ \bibinfo {author} {\bibfnamefont {S.}~\bibnamefont {Nissanke}},\ }\href
  {\doibase 10.1103/PhysRevD.98.081501} {\bibfield  {journal} {\bibinfo
  {journal} {Phys. Rev.}\ }\textbf {\bibinfo {volume} {D98}},\ \bibinfo {pages}
  {081501} (\bibinfo {year} {2018})},\ \Eprint
  {http://arxiv.org/abs/1807.00011} {arXiv:1807.00011 [astro-ph.HE]}
  \BibitemShut {NoStop}%
\bibitem [{\citenamefont {Zappa}\ \emph {et~al.}(2019)\citenamefont {Zappa},
  \citenamefont {Bernuzzi}, \citenamefont {Pannarale}, \citenamefont
  {Mapelli},\ and\ \citenamefont {Giacobbo}}]{Zappa:2019ntl}%
  \BibitemOpen
  \bibfield  {author} {\bibinfo {author} {\bibfnamefont {F.}~\bibnamefont
  {Zappa}}, \bibinfo {author} {\bibfnamefont {S.}~\bibnamefont {Bernuzzi}},
  \bibinfo {author} {\bibfnamefont {F.}~\bibnamefont {Pannarale}}, \bibinfo
  {author} {\bibfnamefont {M.}~\bibnamefont {Mapelli}}, \ and\ \bibinfo
  {author} {\bibfnamefont {N.}~\bibnamefont {Giacobbo}},\ }\href {\doibase
  10.1103/PhysRevLett.123.041102} {\bibfield  {journal} {\bibinfo  {journal}
  {Phys. Rev. Lett.}\ }\textbf {\bibinfo {volume} {123}},\ \bibinfo {pages}
  {041102} (\bibinfo {year} {2019})},\ \Eprint
  {http://arxiv.org/abs/1903.11622} {arXiv:1903.11622 [gr-qc]} \BibitemShut
  {NoStop}%
\bibitem [{\citenamefont {Foucart}\ \emph {et~al.}(2019)\citenamefont
  {Foucart}, \citenamefont {Duez}, \citenamefont {Kidder}, \citenamefont
  {Nissanke}, \citenamefont {Pfeiffer},\ and\ \citenamefont
  {Scheel}}]{Foucart:2019bxj}%
  \BibitemOpen
  \bibfield  {author} {\bibinfo {author} {\bibfnamefont {F.}~\bibnamefont
  {Foucart}}, \bibinfo {author} {\bibfnamefont {M.}~\bibnamefont {Duez}},
  \bibinfo {author} {\bibfnamefont {L.}~\bibnamefont {Kidder}}, \bibinfo
  {author} {\bibfnamefont {S.}~\bibnamefont {Nissanke}}, \bibinfo {author}
  {\bibfnamefont {H.}~\bibnamefont {Pfeiffer}}, \ and\ \bibinfo {author}
  {\bibfnamefont {M.}~\bibnamefont {Scheel}},\ }\href {\doibase
  10.1103/PhysRevD.99.103025} {\bibfield  {journal} {\bibinfo  {journal} {Phys.
  Rev. D}\ }\textbf {\bibinfo {volume} {99}},\ \bibinfo {pages} {103025}
  (\bibinfo {year} {2019})},\ \Eprint {http://arxiv.org/abs/1903.09166}
  {arXiv:1903.09166 [astro-ph.HE]} \BibitemShut {NoStop}%
\bibitem [{\citenamefont {Kawaguchi}\ \emph {et~al.}(2020)\citenamefont
  {Kawaguchi}, \citenamefont {Shibata},\ and\ \citenamefont
  {Tanaka}}]{Kawaguchi:2020osi}%
  \BibitemOpen
  \bibfield  {author} {\bibinfo {author} {\bibfnamefont {K.}~\bibnamefont
  {Kawaguchi}}, \bibinfo {author} {\bibfnamefont {M.}~\bibnamefont {Shibata}},
  \ and\ \bibinfo {author} {\bibfnamefont {M.}~\bibnamefont {Tanaka}},\
  }\href@noop {} {\  (\bibinfo {year} {2020})},\ \Eprint
  {http://arxiv.org/abs/2002.01662} {arXiv:2002.01662 [astro-ph.HE]}
  \BibitemShut {NoStop}%
\bibitem [{\citenamefont {Hinderer}(2008)}]{Hinderer:2007mb}%
  \BibitemOpen
  \bibfield  {author} {\bibinfo {author} {\bibfnamefont {T.}~\bibnamefont
  {Hinderer}},\ }\href {\doibase 10.1086/533487} {\bibfield  {journal}
  {\bibinfo  {journal} {Astrophys. J.}\ }\textbf {\bibinfo {volume} {677}},\
  \bibinfo {pages} {1216} (\bibinfo {year} {2008})},\ \Eprint
  {http://arxiv.org/abs/0711.2420} {arXiv:0711.2420 [astro-ph]} \BibitemShut
  {NoStop}%
\bibitem [{\citenamefont {Binnington}\ and\ \citenamefont
  {Poisson}(2009)}]{Binnington:2009bb}%
  \BibitemOpen
  \bibfield  {author} {\bibinfo {author} {\bibfnamefont {T.}~\bibnamefont
  {Binnington}}\ and\ \bibinfo {author} {\bibfnamefont {E.}~\bibnamefont
  {Poisson}},\ }\href {\doibase 10.1103/PhysRevD.80.084018} {\bibfield
  {journal} {\bibinfo  {journal} {Phys. Rev.}\ }\textbf {\bibinfo {volume}
  {D80}},\ \bibinfo {pages} {084018} (\bibinfo {year} {2009})},\ \Eprint
  {http://arxiv.org/abs/0906.1366} {arXiv:0906.1366 [gr-qc]} \BibitemShut
  {NoStop}%
\bibitem [{\citenamefont {Damour}\ and\ \citenamefont
  {Nagar}(2009)}]{Damour:2009vw}%
  \BibitemOpen
  \bibfield  {author} {\bibinfo {author} {\bibfnamefont {T.}~\bibnamefont
  {Damour}}\ and\ \bibinfo {author} {\bibfnamefont {A.}~\bibnamefont {Nagar}},\
  }\href {\doibase 10.1103/PhysRevD.80.084035} {\bibfield  {journal} {\bibinfo
  {journal} {Phys. Rev. D}\ }\textbf {\bibinfo {volume} {80}},\ \bibinfo
  {pages} {084035} (\bibinfo {year} {2009})},\ \Eprint
  {http://arxiv.org/abs/0906.0096} {arXiv:0906.0096 [gr-qc]} \BibitemShut
  {NoStop}%
\bibitem [{\citenamefont {Landry}\ and\ \citenamefont
  {Poisson}(2015)}]{Landry:2015zfa}%
  \BibitemOpen
  \bibfield  {author} {\bibinfo {author} {\bibfnamefont {P.}~\bibnamefont
  {Landry}}\ and\ \bibinfo {author} {\bibfnamefont {E.}~\bibnamefont
  {Poisson}},\ }\href {\doibase 10.1103/PhysRevD.91.104018} {\bibfield
  {journal} {\bibinfo  {journal} {Phys. Rev. D}\ }\textbf {\bibinfo {volume}
  {91}},\ \bibinfo {pages} {104018} (\bibinfo {year} {2015})},\ \Eprint
  {http://arxiv.org/abs/1503.07366} {arXiv:1503.07366 [gr-qc]} \BibitemShut
  {NoStop}%
\bibitem [{\citenamefont {Pani}\ \emph {et~al.}(2015)\citenamefont {Pani},
  \citenamefont {Gualtieri}, \citenamefont {Maselli},\ and\ \citenamefont
  {Ferrari}}]{Pani:2015hfa}%
  \BibitemOpen
  \bibfield  {author} {\bibinfo {author} {\bibfnamefont {P.}~\bibnamefont
  {Pani}}, \bibinfo {author} {\bibfnamefont {L.}~\bibnamefont {Gualtieri}},
  \bibinfo {author} {\bibfnamefont {A.}~\bibnamefont {Maselli}}, \ and\
  \bibinfo {author} {\bibfnamefont {V.}~\bibnamefont {Ferrari}},\ }\href
  {\doibase 10.1103/PhysRevD.92.024010} {\bibfield  {journal} {\bibinfo
  {journal} {Phys. Rev. D}\ }\textbf {\bibinfo {volume} {92}},\ \bibinfo
  {pages} {024010} (\bibinfo {year} {2015})},\ \Eprint
  {http://arxiv.org/abs/1503.07365} {arXiv:1503.07365 [gr-qc]} \BibitemShut
  {NoStop}%
\bibitem [{\citenamefont {Gralla}(2018)}]{Gralla:2017djj}%
  \BibitemOpen
  \bibfield  {author} {\bibinfo {author} {\bibfnamefont {S.~E.}\ \bibnamefont
  {Gralla}},\ }\href {\doibase 10.1088/1361-6382/aab186} {\bibfield  {journal}
  {\bibinfo  {journal} {Class. Quant. Grav.}\ }\textbf {\bibinfo {volume}
  {35}},\ \bibinfo {pages} {085002} (\bibinfo {year} {2018})},\ \Eprint
  {http://arxiv.org/abs/1710.11096} {arXiv:1710.11096 [gr-qc]} \BibitemShut
  {NoStop}%
\bibitem [{\citenamefont {Hannam}\ \emph {et~al.}(2013)\citenamefont {Hannam},
  \citenamefont {Brown}, \citenamefont {Fairhurst}, \citenamefont {Fryer},\
  and\ \citenamefont {Harry}}]{Hannam:2013uu}%
  \BibitemOpen
  \bibfield  {author} {\bibinfo {author} {\bibfnamefont {M.}~\bibnamefont
  {Hannam}}, \bibinfo {author} {\bibfnamefont {D.~A.}\ \bibnamefont {Brown}},
  \bibinfo {author} {\bibfnamefont {S.}~\bibnamefont {Fairhurst}}, \bibinfo
  {author} {\bibfnamefont {C.~L.}\ \bibnamefont {Fryer}}, \ and\ \bibinfo
  {author} {\bibfnamefont {I.~W.}\ \bibnamefont {Harry}},\ }\href {\doibase
  10.1088/2041-8205/766/1/L14} {\bibfield  {journal} {\bibinfo  {journal}
  {Astrophys. J.}\ }\textbf {\bibinfo {volume} {766}},\ \bibinfo {pages} {L14}
  (\bibinfo {year} {2013})},\ \Eprint {http://arxiv.org/abs/1301.5616}
  {arXiv:1301.5616 [gr-qc]} \BibitemShut {NoStop}%
\bibitem [{\citenamefont {Tsokaros}\ \emph {et~al.}(2020)\citenamefont
  {Tsokaros}, \citenamefont {Ruiz}, \citenamefont {Shapiro}, \citenamefont
  {Sun},\ and\ \citenamefont {Ury\=u}}]{Tsokaros:2019lnx}%
  \BibitemOpen
  \bibfield  {author} {\bibinfo {author} {\bibfnamefont {A.}~\bibnamefont
  {Tsokaros}}, \bibinfo {author} {\bibfnamefont {M.}~\bibnamefont {Ruiz}},
  \bibinfo {author} {\bibfnamefont {S.~L.}\ \bibnamefont {Shapiro}}, \bibinfo
  {author} {\bibfnamefont {L.}~\bibnamefont {Sun}}, \ and\ \bibinfo {author}
  {\bibfnamefont {K.}~\bibnamefont {Ury\=u}},\ }\href {\doibase
  10.1103/PhysRevLett.124.071101} {\bibfield  {journal} {\bibinfo  {journal}
  {Phys. Rev. Lett.}\ }\textbf {\bibinfo {volume} {124}},\ \bibinfo {pages}
  {071101} (\bibinfo {year} {2020})},\ \Eprint
  {http://arxiv.org/abs/1911.06865} {arXiv:1911.06865 [astro-ph.HE]}
  \BibitemShut {NoStop}%
\bibitem [{\citenamefont {Chatziioannou}\ \emph {et~al.}(2018)\citenamefont
  {Chatziioannou}, \citenamefont {Haster},\ and\ \citenamefont
  {Zimmerman}}]{Chatziioannou:2018vzf}%
  \BibitemOpen
  \bibfield  {author} {\bibinfo {author} {\bibfnamefont {K.}~\bibnamefont
  {Chatziioannou}}, \bibinfo {author} {\bibfnamefont {C.-J.}\ \bibnamefont
  {Haster}}, \ and\ \bibinfo {author} {\bibfnamefont {A.}~\bibnamefont
  {Zimmerman}},\ }\href {\doibase 10.1103/PhysRevD.97.104036} {\bibfield
  {journal} {\bibinfo  {journal} {Phys. Rev.}\ }\textbf {\bibinfo {volume}
  {D97}},\ \bibinfo {pages} {104036} (\bibinfo {year} {2018})},\ \Eprint
  {http://arxiv.org/abs/1804.03221} {arXiv:1804.03221 [gr-qc]} \BibitemShut
  {NoStop}%
\bibitem [{\citenamefont {Yang}\ \emph {et~al.}(2018)\citenamefont {Yang},
  \citenamefont {East},\ and\ \citenamefont {Lehner}}]{Yang:2017gfb}%
  \BibitemOpen
  \bibfield  {author} {\bibinfo {author} {\bibfnamefont {H.}~\bibnamefont
  {Yang}}, \bibinfo {author} {\bibfnamefont {W.~E.}\ \bibnamefont {East}}, \
  and\ \bibinfo {author} {\bibfnamefont {L.}~\bibnamefont {Lehner}},\ }\href
  {\doibase 10.3847/1538-4357/aaf723, 10.3847/1538-4357/aab2b0} {\bibfield
  {journal} {\bibinfo  {journal} {Astrophys. J.}\ }\textbf {\bibinfo {volume}
  {856}},\ \bibinfo {pages} {110} (\bibinfo {year} {2018})},\ \bibinfo {note}
  {[Erratum: Astrophys. J.870,no.2,139(2019)]},\ \Eprint
  {http://arxiv.org/abs/1710.05891} {arXiv:1710.05891 [gr-qc]} \BibitemShut
  {NoStop}%
\bibitem [{\citenamefont {Chen}\ and\ \citenamefont
  {Chatziioannou}(2019)}]{Chen:2019aiw}%
  \BibitemOpen
  \bibfield  {author} {\bibinfo {author} {\bibfnamefont {H.-Y.}\ \bibnamefont
  {Chen}}\ and\ \bibinfo {author} {\bibfnamefont {K.}~\bibnamefont
  {Chatziioannou}},\ }\href@noop {} {\  (\bibinfo {year} {2019})},\ \Eprint
  {http://arxiv.org/abs/1903.11197} {arXiv:1903.11197 [astro-ph.HE]}
  \BibitemShut {NoStop}%
\bibitem [{\citenamefont {Han}\ and\ \citenamefont
  {Steiner}(2019)}]{Han:2018mtj}%
  \BibitemOpen
  \bibfield  {author} {\bibinfo {author} {\bibfnamefont {S.}~\bibnamefont
  {Han}}\ and\ \bibinfo {author} {\bibfnamefont {A.~W.}\ \bibnamefont
  {Steiner}},\ }\href {\doibase 10.1103/PhysRevD.99.083014} {\bibfield
  {journal} {\bibinfo  {journal} {Phys. Rev. D}\ }\textbf {\bibinfo {volume}
  {99}},\ \bibinfo {pages} {083014} (\bibinfo {year} {2019})},\ \Eprint
  {http://arxiv.org/abs/1810.10967} {arXiv:1810.10967 [nucl-th]} \BibitemShut
  {NoStop}%
\bibitem [{\citenamefont {Chen}\ \emph
  {et~al.}(2020{\natexlab{a}})\citenamefont {Chen}, \citenamefont {Chesler},\
  and\ \citenamefont {Loeb}}]{Chen:2019rja}%
  \BibitemOpen
  \bibfield  {author} {\bibinfo {author} {\bibfnamefont {H.-Y.}\ \bibnamefont
  {Chen}}, \bibinfo {author} {\bibfnamefont {P.~M.}\ \bibnamefont {Chesler}}, \
  and\ \bibinfo {author} {\bibfnamefont {A.}~\bibnamefont {Loeb}},\ }\href
  {\doibase 10.3847/2041-8213/ab830f} {\bibfield  {journal} {\bibinfo
  {journal} {Astrophys. J.}\ }\textbf {\bibinfo {volume} {893}},\ \bibinfo
  {pages} {L4} (\bibinfo {year} {2020}{\natexlab{a}})},\ \Eprint
  {http://arxiv.org/abs/1909.04096} {arXiv:1909.04096 [astro-ph.HE]}
  \BibitemShut {NoStop}%
\bibitem [{\citenamefont {Chen}\ \emph
  {et~al.}(2020{\natexlab{b}})\citenamefont {Chen}, \citenamefont
  {Johnson-McDaniel}, \citenamefont {Dietrich},\ and\ \citenamefont
  {Dudi}}]{Chen:2020fzm}%
  \BibitemOpen
  \bibfield  {author} {\bibinfo {author} {\bibfnamefont {A.}~\bibnamefont
  {Chen}}, \bibinfo {author} {\bibfnamefont {N.~K.}\ \bibnamefont
  {Johnson-McDaniel}}, \bibinfo {author} {\bibfnamefont {T.}~\bibnamefont
  {Dietrich}}, \ and\ \bibinfo {author} {\bibfnamefont {R.}~\bibnamefont
  {Dudi}},\ }\href@noop {} {\  (\bibinfo {year} {2020}{\natexlab{b}})},\
  \Eprint {http://arxiv.org/abs/2001.11470} {arXiv:2001.11470 [astro-ph.HE]}
  \BibitemShut {NoStop}%
\bibitem [{\citenamefont {Datta}\ \emph {et~al.}(2020)\citenamefont {Datta},
  \citenamefont {Phukon},\ and\ \citenamefont {Bose}}]{Datta:2020gem}%
  \BibitemOpen
  \bibfield  {author} {\bibinfo {author} {\bibfnamefont {S.}~\bibnamefont
  {Datta}}, \bibinfo {author} {\bibfnamefont {K.~S.}\ \bibnamefont {Phukon}}, \
  and\ \bibinfo {author} {\bibfnamefont {S.}~\bibnamefont {Bose}},\ }\href@noop
  {} {\  (\bibinfo {year} {2020})},\ \Eprint {http://arxiv.org/abs/2004.05974}
  {arXiv:2004.05974 [gr-qc]} \BibitemShut {NoStop}%
\bibitem [{\citenamefont {Silva}\ \emph {et~al.}(2016)\citenamefont {Silva},
  \citenamefont {Sotani},\ and\ \citenamefont {Berti}}]{Hector2016}%
  \BibitemOpen
  \bibfield  {author} {\bibinfo {author} {\bibfnamefont {H.~O.}\ \bibnamefont
  {Silva}}, \bibinfo {author} {\bibfnamefont {H.}~\bibnamefont {Sotani}}, \
  and\ \bibinfo {author} {\bibfnamefont {E.}~\bibnamefont {Berti}},\ }\href
  {\doibase 10.1093/mnras/stw969} {\bibfield  {journal} {\bibinfo  {journal}
  {Mon. Not. Roy. Astron. Soc.}\ }\textbf {\bibinfo {volume} {459}},\ \bibinfo
  {pages} {4378} (\bibinfo {year} {2016})},\ \Eprint
  {http://arxiv.org/abs/1601.03407} {arXiv:1601.03407 [astro-ph.HE]}
  \BibitemShut {NoStop}%
\bibitem [{\citenamefont {Suwa}\ \emph {et~al.}(2018)\citenamefont {Suwa},
  \citenamefont {Yoshida}, \citenamefont {Shibata}, \citenamefont {Umeda},\
  and\ \citenamefont {Takahashi}}]{Shibata2018minM}%
  \BibitemOpen
  \bibfield  {author} {\bibinfo {author} {\bibfnamefont {Y.}~\bibnamefont
  {Suwa}}, \bibinfo {author} {\bibfnamefont {T.}~\bibnamefont {Yoshida}},
  \bibinfo {author} {\bibfnamefont {M.}~\bibnamefont {Shibata}}, \bibinfo
  {author} {\bibfnamefont {H.}~\bibnamefont {Umeda}}, \ and\ \bibinfo {author}
  {\bibfnamefont {K.}~\bibnamefont {Takahashi}},\ }\href {\doibase
  10.1093/mnras/sty2460} {\bibfield  {journal} {\bibinfo  {journal} {Mon. Not.
  Roy. Astron. Soc.}\ }\textbf {\bibinfo {volume} {481}},\ \bibinfo {pages}
  {3305} (\bibinfo {year} {2018})},\ \Eprint {http://arxiv.org/abs/1808.02328}
  {arXiv:1808.02328 [astro-ph.HE]} \BibitemShut {NoStop}%
\bibitem [{\citenamefont {Cromartie}\ \emph {et~al.}(2019)\citenamefont
  {Cromartie} \emph {et~al.}}]{Cromartie:2019kug}%
  \BibitemOpen
  \bibfield  {author} {\bibinfo {author} {\bibfnamefont {H.~T.}\ \bibnamefont
  {Cromartie}} \emph {et~al.},\ }\href {\doibase 10.1038/s41550-019-0880-2}
  {\bibfield  {journal} {\bibinfo  {journal} {Nat. Astron.}\ }\textbf {\bibinfo
  {volume} {4}},\ \bibinfo {pages} {72} (\bibinfo {year} {2019})},\ \Eprint
  {http://arxiv.org/abs/1904.06759} {arXiv:1904.06759 [astro-ph.HE]}
  \BibitemShut {NoStop}%
\bibitem [{\citenamefont {Antoniadis}\ \emph {et~al.}(2016)\citenamefont
  {Antoniadis}, \citenamefont {Tauris}, \citenamefont {Ozel}, \citenamefont
  {Barr}, \citenamefont {Champion},\ and\ \citenamefont
  {Freire}}]{Antoniadis:2016hxz}%
  \BibitemOpen
  \bibfield  {author} {\bibinfo {author} {\bibfnamefont {J.}~\bibnamefont
  {Antoniadis}}, \bibinfo {author} {\bibfnamefont {T.~M.}\ \bibnamefont
  {Tauris}}, \bibinfo {author} {\bibfnamefont {F.}~\bibnamefont {Ozel}},
  \bibinfo {author} {\bibfnamefont {E.}~\bibnamefont {Barr}}, \bibinfo {author}
  {\bibfnamefont {D.~J.}\ \bibnamefont {Champion}}, \ and\ \bibinfo {author}
  {\bibfnamefont {P.~C.~C.}\ \bibnamefont {Freire}},\ }\href@noop {} {\
  (\bibinfo {year} {2016})},\ \Eprint {http://arxiv.org/abs/1605.01665}
  {arXiv:1605.01665 [astro-ph.HE]} \BibitemShut {NoStop}%
\bibitem [{\citenamefont {{Alsing}}\ \emph {et~al.}(2018)\citenamefont
  {{Alsing}}, \citenamefont {{Silva}},\ and\ \citenamefont
  {{Berti}}}]{Alsing2018}%
  \BibitemOpen
  \bibfield  {author} {\bibinfo {author} {\bibfnamefont {J.}~\bibnamefont
  {{Alsing}}}, \bibinfo {author} {\bibfnamefont {H.~O.}\ \bibnamefont
  {{Silva}}}, \ and\ \bibinfo {author} {\bibfnamefont {E.}~\bibnamefont
  {{Berti}}},\ }\href {\doibase 10.1093/mnras/sty1065} {\bibfield  {journal}
  {\bibinfo  {journal} {Monthly Notices of the RAS}\ }\textbf {\bibinfo
  {volume} {478}},\ \bibinfo {pages} {1377} (\bibinfo {year} {2018})},\ \Eprint
  {http://arxiv.org/abs/1709.07889} {arXiv:1709.07889 [astro-ph.HE]}
  \BibitemShut {NoStop}%
\bibitem [{\citenamefont {Kalogera}\ and\ \citenamefont
  {Baym}(1996)}]{Kalogera:1996ci}%
  \BibitemOpen
  \bibfield  {author} {\bibinfo {author} {\bibfnamefont {V.}~\bibnamefont
  {Kalogera}}\ and\ \bibinfo {author} {\bibfnamefont {G.}~\bibnamefont
  {Baym}},\ }\href {\doibase 10.1086/310296} {\bibfield  {journal} {\bibinfo
  {journal} {Astrophys. J.}\ }\textbf {\bibinfo {volume} {470}},\ \bibinfo
  {pages} {L61} (\bibinfo {year} {1996})},\ \Eprint
  {http://arxiv.org/abs/astro-ph/9608059} {arXiv:astro-ph/9608059 [astro-ph]}
  \BibitemShut {NoStop}%
\bibitem [{\citenamefont {{Bombaci}}(1996)}]{Bombaci1996}%
  \BibitemOpen
  \bibfield  {author} {\bibinfo {author} {\bibfnamefont {I.}~\bibnamefont
  {{Bombaci}}},\ }\href@noop {} {\bibfield  {journal} {\bibinfo  {journal}
  {Astronomy and Astrophysics}\ }\textbf {\bibinfo {volume} {305}},\ \bibinfo
  {pages} {871} (\bibinfo {year} {1996})}\BibitemShut {NoStop}%
\bibitem [{\citenamefont {{Srinivasan}}(2002)}]{Srinivasan2002}%
  \BibitemOpen
  \bibfield  {author} {\bibinfo {author} {\bibfnamefont {G.}~\bibnamefont
  {{Srinivasan}}},\ }\href@noop {} {\bibfield  {journal} {\bibinfo  {journal}
  {Bulletin of the Astronomical Society of India}\ }\textbf {\bibinfo {volume}
  {30}},\ \bibinfo {pages} {523} (\bibinfo {year} {2002})}\BibitemShut
  {NoStop}%
\bibitem [{\citenamefont {Chamel}\ \emph {et~al.}(2013)\citenamefont {Chamel},
  \citenamefont {Haensel}, \citenamefont {Zdunik},\ and\ \citenamefont
  {Fantina}}]{Chamel2013}%
  \BibitemOpen
  \bibfield  {author} {\bibinfo {author} {\bibfnamefont {N.}~\bibnamefont
  {Chamel}}, \bibinfo {author} {\bibfnamefont {P.}~\bibnamefont {Haensel}},
  \bibinfo {author} {\bibfnamefont {J.~L.}\ \bibnamefont {Zdunik}}, \ and\
  \bibinfo {author} {\bibfnamefont {A.~F.}\ \bibnamefont {Fantina}},\ }\href
  {\doibase 10.1142/S021830131330018X} {\bibfield  {journal} {\bibinfo
  {journal} {Int. J. Mod. Phys.}\ }\textbf {\bibinfo {volume} {E22}},\ \bibinfo
  {pages} {1330018} (\bibinfo {year} {2013})},\ \Eprint
  {http://arxiv.org/abs/1307.3995} {arXiv:1307.3995 [astro-ph.HE]} \BibitemShut
  {NoStop}%
\bibitem [{\citenamefont {{Shibata}}\ \emph {et~al.}(2019)\citenamefont
  {{Shibata}}, \citenamefont {{Zhou}}, \citenamefont {{Kiuchi}},\ and\
  \citenamefont {{Fujibayashi}}}]{Shibata2019maxM}%
  \BibitemOpen
  \bibfield  {author} {\bibinfo {author} {\bibfnamefont {M.}~\bibnamefont
  {{Shibata}}}, \bibinfo {author} {\bibfnamefont {E.}~\bibnamefont {{Zhou}}},
  \bibinfo {author} {\bibfnamefont {K.}~\bibnamefont {{Kiuchi}}}, \ and\
  \bibinfo {author} {\bibfnamefont {S.}~\bibnamefont {{Fujibayashi}}},\ }\href
  {\doibase 10.1103/PhysRevD.100.023015} {\bibfield  {journal} {\bibinfo
  {journal} {\prd}\ }\textbf {\bibinfo {volume} {100}},\ \bibinfo {eid}
  {023015} (\bibinfo {year} {2019})},\ \Eprint
  {http://arxiv.org/abs/1905.03656} {arXiv:1905.03656 [astro-ph.HE]}
  \BibitemShut {NoStop}%
\bibitem [{\citenamefont {Rezzolla}\ \emph {et~al.}(2018)\citenamefont
  {Rezzolla}, \citenamefont {Most},\ and\ \citenamefont
  {Weih}}]{Rezzolla:2017aly}%
  \BibitemOpen
  \bibfield  {author} {\bibinfo {author} {\bibfnamefont {L.}~\bibnamefont
  {Rezzolla}}, \bibinfo {author} {\bibfnamefont {E.~R.}\ \bibnamefont {Most}},
  \ and\ \bibinfo {author} {\bibfnamefont {L.~R.}\ \bibnamefont {Weih}},\
  }\href {\doibase 10.3847/2041-8213/aaa401} {\bibfield  {journal} {\bibinfo
  {journal} {Astrophys. J.}\ }\textbf {\bibinfo {volume} {852}},\ \bibinfo
  {pages} {L25} (\bibinfo {year} {2018})},\ \bibinfo {note} {[Astrophys. J.
  Lett.852,L25(2018)]},\ \Eprint {http://arxiv.org/abs/1711.00314}
  {arXiv:1711.00314 [astro-ph.HE]} \BibitemShut {NoStop}%
\bibitem [{\citenamefont {Ai}\ \emph {et~al.}(2019)\citenamefont {Ai},
  \citenamefont {Gao},\ and\ \citenamefont {Zhang}}]{Ai:2019rre}%
  \BibitemOpen
  \bibfield  {author} {\bibinfo {author} {\bibfnamefont {S.}~\bibnamefont
  {Ai}}, \bibinfo {author} {\bibfnamefont {H.}~\bibnamefont {Gao}}, \ and\
  \bibinfo {author} {\bibfnamefont {B.}~\bibnamefont {Zhang}},\ }\href@noop {}
  {\  (\bibinfo {year} {2019})},\ \Eprint {http://arxiv.org/abs/1912.06369}
  {arXiv:1912.06369 [astro-ph.HE]} \BibitemShut {NoStop}%
\bibitem [{\citenamefont {Freire}\ \emph {et~al.}(2008)\citenamefont {Freire},
  \citenamefont {Ransom}, \citenamefont {Begin}, \citenamefont {Stairs},
  \citenamefont {Hessels}, \citenamefont {Frey},\ and\ \citenamefont
  {Camilo}}]{Freire:2007jd}%
  \BibitemOpen
  \bibfield  {author} {\bibinfo {author} {\bibfnamefont {P.~C.~C.}\
  \bibnamefont {Freire}}, \bibinfo {author} {\bibfnamefont {S.~M.}\
  \bibnamefont {Ransom}}, \bibinfo {author} {\bibfnamefont {S.}~\bibnamefont
  {Begin}}, \bibinfo {author} {\bibfnamefont {I.~H.}\ \bibnamefont {Stairs}},
  \bibinfo {author} {\bibfnamefont {J.~W.}\ \bibnamefont {Hessels}}, \bibinfo
  {author} {\bibfnamefont {L.~H.}\ \bibnamefont {Frey}}, \ and\ \bibinfo
  {author} {\bibfnamefont {F.}~\bibnamefont {Camilo}},\ }\href {\doibase
  10.1086/526338} {\bibfield  {journal} {\bibinfo  {journal} {Astrophys. J.}\
  }\textbf {\bibinfo {volume} {675}},\ \bibinfo {pages} {670} (\bibinfo {year}
  {2008})},\ \Eprint {http://arxiv.org/abs/0711.0925} {arXiv:0711.0925
  [astro-ph]} \BibitemShut {NoStop}%
\bibitem [{\citenamefont {Rhoades}\ and\ \citenamefont
  {Ruffini}(1974)}]{Rhoades:1974fn}%
  \BibitemOpen
  \bibfield  {author} {\bibinfo {author} {\bibfnamefont {C.~E.}\ \bibnamefont
  {Rhoades}, \bibfnamefont {Jr.}}\ and\ \bibinfo {author} {\bibfnamefont
  {R.}~\bibnamefont {Ruffini}},\ }\href {\doibase 10.1103/PhysRevLett.32.324}
  {\bibfield  {journal} {\bibinfo  {journal} {Phys. Rev. Lett.}\ }\textbf
  {\bibinfo {volume} {32}},\ \bibinfo {pages} {324} (\bibinfo {year}
  {1974})}\BibitemShut {NoStop}%
\bibitem [{\citenamefont {Akmal}\ \emph {et~al.}(1998)\citenamefont {Akmal},
  \citenamefont {Pandharipande},\ and\ \citenamefont
  {Ravenhall}}]{Akmal:1998cf}%
  \BibitemOpen
  \bibfield  {author} {\bibinfo {author} {\bibfnamefont {A.}~\bibnamefont
  {Akmal}}, \bibinfo {author} {\bibfnamefont {V.~R.}\ \bibnamefont
  {Pandharipande}}, \ and\ \bibinfo {author} {\bibfnamefont {D.~G.}\
  \bibnamefont {Ravenhall}},\ }\href {\doibase 10.1103/PhysRevC.58.1804}
  {\bibfield  {journal} {\bibinfo  {journal} {Phys. Rev.}\ }\textbf {\bibinfo
  {volume} {C58}},\ \bibinfo {pages} {1804} (\bibinfo {year} {1998})},\ \Eprint
  {http://arxiv.org/abs/nucl-th/9804027} {arXiv:nucl-th/9804027 [nucl-th]}
  \BibitemShut {NoStop}%
\bibitem [{\citenamefont {Douchin}\ and\ \citenamefont {Haensel}(2001)}]{sly4}%
  \BibitemOpen
  \bibfield  {author} {\bibinfo {author} {\bibfnamefont {F.}~\bibnamefont
  {Douchin}}\ and\ \bibinfo {author} {\bibfnamefont {P.}~\bibnamefont
  {Haensel}},\ }\href {\doibase 10.1051/0004-6361:20011402} {\bibfield
  {journal} {\bibinfo  {journal} {Astron. Astrophys.}\ }\textbf {\bibinfo
  {volume} {380}},\ \bibinfo {pages} {151} (\bibinfo {year} {2001})},\ \Eprint
  {http://arxiv.org/abs/astro-ph/0111092} {arXiv:astro-ph/0111092 [astro-ph]}
  \BibitemShut {NoStop}%
\bibitem [{\citenamefont {{M{\"u}ther}}\ \emph {et~al.}(1987)\citenamefont
  {{M{\"u}ther}}, \citenamefont {{Prakash}},\ and\ \citenamefont
  {{Ainsworth}}}]{mpa1}%
  \BibitemOpen
  \bibfield  {author} {\bibinfo {author} {\bibfnamefont {H.}~\bibnamefont
  {{M{\"u}ther}}}, \bibinfo {author} {\bibfnamefont {M.}~\bibnamefont
  {{Prakash}}}, \ and\ \bibinfo {author} {\bibfnamefont {T.~L.}\ \bibnamefont
  {{Ainsworth}}},\ }\href {\doibase 10.1016/0370-2693(87)91611-X} {\bibfield
  {journal} {\bibinfo  {journal} {Physics Letters B}\ }\textbf {\bibinfo
  {volume} {199}},\ \bibinfo {pages} {469} (\bibinfo {year}
  {1987})}\BibitemShut {NoStop}%
\bibitem [{\citenamefont {Alford}\ \emph {et~al.}(2005)\citenamefont {Alford},
  \citenamefont {Braby}, \citenamefont {Paris},\ and\ \citenamefont
  {Reddy}}]{Alford:2004pf}%
  \BibitemOpen
  \bibfield  {author} {\bibinfo {author} {\bibfnamefont {M.}~\bibnamefont
  {Alford}}, \bibinfo {author} {\bibfnamefont {M.}~\bibnamefont {Braby}},
  \bibinfo {author} {\bibfnamefont {M.~W.}\ \bibnamefont {Paris}}, \ and\
  \bibinfo {author} {\bibfnamefont {S.}~\bibnamefont {Reddy}},\ }\href
  {\doibase 10.1086/430902} {\bibfield  {journal} {\bibinfo  {journal}
  {Astrophys. J.}\ }\textbf {\bibinfo {volume} {629}},\ \bibinfo {pages} {969}
  (\bibinfo {year} {2005})},\ \Eprint {http://arxiv.org/abs/nucl-th/0411016}
  {arXiv:nucl-th/0411016 [nucl-th]} \BibitemShut {NoStop}%
\bibitem [{\citenamefont {Lackey}\ \emph {et~al.}(2006)\citenamefont {Lackey},
  \citenamefont {Nayyar},\ and\ \citenamefont {Owen}}]{Lackey:2005tk}%
  \BibitemOpen
  \bibfield  {author} {\bibinfo {author} {\bibfnamefont {B.~D.}\ \bibnamefont
  {Lackey}}, \bibinfo {author} {\bibfnamefont {M.}~\bibnamefont {Nayyar}}, \
  and\ \bibinfo {author} {\bibfnamefont {B.~J.}\ \bibnamefont {Owen}},\ }\href
  {\doibase 10.1103/PhysRevD.73.024021} {\bibfield  {journal} {\bibinfo
  {journal} {Phys. Rev.}\ }\textbf {\bibinfo {volume} {D73}},\ \bibinfo {pages}
  {024021} (\bibinfo {year} {2006})},\ \Eprint
  {http://arxiv.org/abs/astro-ph/0507312} {arXiv:astro-ph/0507312 [astro-ph]}
  \BibitemShut {NoStop}%
\bibitem [{\citenamefont {Tauris}\ \emph {et~al.}(2017)\citenamefont {Tauris}
  \emph {et~al.}}]{Tauris:2017omb}%
  \BibitemOpen
  \bibfield  {author} {\bibinfo {author} {\bibfnamefont {T.}~\bibnamefont
  {Tauris}} \emph {et~al.},\ }\href {\doibase 10.3847/1538-4357/aa7e89}
  {\bibfield  {journal} {\bibinfo  {journal} {Astrophys. J.}\ }\textbf
  {\bibinfo {volume} {846}},\ \bibinfo {pages} {170} (\bibinfo {year}
  {2017})},\ \Eprint {http://arxiv.org/abs/1706.09438} {arXiv:1706.09438
  [astro-ph.HE]} \BibitemShut {NoStop}%
\bibitem [{\citenamefont {Andrews}\ and\ \citenamefont
  {Mandel}(2019)}]{Andrews:2019vou}%
  \BibitemOpen
  \bibfield  {author} {\bibinfo {author} {\bibfnamefont {J.~J.}\ \bibnamefont
  {Andrews}}\ and\ \bibinfo {author} {\bibfnamefont {I.}~\bibnamefont
  {Mandel}},\ }\href {\doibase 10.3847/2041-8213/ab2ed1} {\bibfield  {journal}
  {\bibinfo  {journal} {Astrophys. J.}\ }\textbf {\bibinfo {volume} {880}},\
  \bibinfo {pages} {L8} (\bibinfo {year} {2019})},\ \Eprint
  {http://arxiv.org/abs/1904.12745} {arXiv:1904.12745 [astro-ph.HE]}
  \BibitemShut {NoStop}%
\bibitem [{\citenamefont {Vigna-Gómez}\ \emph {et~al.}(2020)\citenamefont
  {Vigna-Gómez}, \citenamefont {MacLeod}, \citenamefont {Neijssel},
  \citenamefont {Broekgaarden}, \citenamefont {Justham}, \citenamefont
  {Howitt}, \citenamefont {de~Mink},\ and\ \citenamefont
  {Mandel}}]{Vigna-Gomez:2020bgo}%
  \BibitemOpen
  \bibfield  {author} {\bibinfo {author} {\bibfnamefont {A.}~\bibnamefont
  {Vigna-Gómez}}, \bibinfo {author} {\bibfnamefont {M.}~\bibnamefont
  {MacLeod}}, \bibinfo {author} {\bibfnamefont {C.~J.}\ \bibnamefont
  {Neijssel}}, \bibinfo {author} {\bibfnamefont {F.~S.}\ \bibnamefont
  {Broekgaarden}}, \bibinfo {author} {\bibfnamefont {S.}~\bibnamefont
  {Justham}}, \bibinfo {author} {\bibfnamefont {G.}~\bibnamefont {Howitt}},
  \bibinfo {author} {\bibfnamefont {S.~E.}\ \bibnamefont {de~Mink}}, \ and\
  \bibinfo {author} {\bibfnamefont {I.}~\bibnamefont {Mandel}},\ }\href@noop {}
  {\  (\bibinfo {year} {2020})},\ \Eprint {http://arxiv.org/abs/2001.09829}
  {arXiv:2001.09829 [astro-ph.SR]} \BibitemShut {NoStop}%
\bibitem [{\citenamefont {Farrow}\ \emph {et~al.}(2019)\citenamefont {Farrow},
  \citenamefont {Zhu},\ and\ \citenamefont {Thrane}}]{Farrow:2019xnc}%
  \BibitemOpen
  \bibfield  {author} {\bibinfo {author} {\bibfnamefont {N.}~\bibnamefont
  {Farrow}}, \bibinfo {author} {\bibfnamefont {X.-J.}\ \bibnamefont {Zhu}}, \
  and\ \bibinfo {author} {\bibfnamefont {E.}~\bibnamefont {Thrane}},\ }\href
  {\doibase 10.3847/1538-4357/ab12e3} {\bibfield  {journal} {\bibinfo
  {journal} {Astrophys. J.}\ }\textbf {\bibinfo {volume} {876}},\ \bibinfo
  {pages} {18} (\bibinfo {year} {2019})},\ \Eprint
  {http://arxiv.org/abs/1902.03300} {arXiv:1902.03300 [astro-ph.HE]}
  \BibitemShut {NoStop}%
\bibitem [{\citenamefont {Punturo}\ \emph {et~al.}(2010)\citenamefont {Punturo}
  \emph {et~al.}}]{Punturo:2010zz}%
  \BibitemOpen
  \bibfield  {author} {\bibinfo {author} {\bibfnamefont {M.}~\bibnamefont
  {Punturo}} \emph {et~al.},\ }\href {\doibase 10.1088/0264-9381/27/19/194002}
  {\bibfield  {journal} {\bibinfo  {journal} {Class. Quant. Grav.}\ }\textbf
  {\bibinfo {volume} {27}},\ \bibinfo {pages} {194002} (\bibinfo {year}
  {2010})}\BibitemShut {NoStop}%
\bibitem [{\citenamefont {Sathyaprakash}\ \emph {et~al.}(2012)\citenamefont
  {Sathyaprakash} \emph {et~al.}}]{Sathyaprakash:2012jk}%
  \BibitemOpen
  \bibfield  {author} {\bibinfo {author} {\bibfnamefont {B.}~\bibnamefont
  {Sathyaprakash}} \emph {et~al.},\ }\href {\doibase
  10.1088/0264-9381/29/12/124013} {\bibfield  {journal} {\bibinfo  {journal}
  {Class. Quant. Grav.}\ }\textbf {\bibinfo {volume} {29}},\ \bibinfo {pages}
  {124013} (\bibinfo {year} {2012})},\ \bibinfo {note} {[Erratum:
  Class.Quant.Grav. 30, 079501 (2013)]},\ \Eprint
  {http://arxiv.org/abs/1206.0331} {arXiv:1206.0331 [gr-qc]} \BibitemShut
  {NoStop}%
\bibitem [{\citenamefont {Maggiore}\ \emph {et~al.}(2020)\citenamefont
  {Maggiore} \emph {et~al.}}]{Maggiore:2019uih}%
  \BibitemOpen
  \bibfield  {author} {\bibinfo {author} {\bibfnamefont {M.}~\bibnamefont
  {Maggiore}} \emph {et~al.},\ }\href {\doibase 10.1088/1475-7516/2020/03/050}
  {\bibfield  {journal} {\bibinfo  {journal} {JCAP}\ }\textbf {\bibinfo
  {volume} {03}},\ \bibinfo {pages} {050} (\bibinfo {year} {2020})},\ \Eprint
  {http://arxiv.org/abs/1912.02622} {arXiv:1912.02622 [astro-ph.CO]}
  \BibitemShut {NoStop}%
\bibitem [{\citenamefont {Abbott}\ \emph
  {et~al.}(2017{\natexlab{f}})\citenamefont {Abbott} \emph
  {et~al.}}]{Evans:2016mbw}%
  \BibitemOpen
  \bibfield  {author} {\bibinfo {author} {\bibfnamefont {B.~P.}\ \bibnamefont
  {Abbott}} \emph {et~al.} (\bibinfo {collaboration} {LIGO Scientific}),\
  }\href {\doibase 10.1088/1361-6382/aa51f4} {\bibfield  {journal} {\bibinfo
  {journal} {Class. Quant. Grav.}\ }\textbf {\bibinfo {volume} {34}},\ \bibinfo
  {pages} {044001} (\bibinfo {year} {2017}{\natexlab{f}})},\ \Eprint
  {http://arxiv.org/abs/1607.08697} {arXiv:1607.08697 [astro-ph.IM]}
  \BibitemShut {NoStop}%
\bibitem [{\citenamefont {Reitze}\ \emph {et~al.}(2019)\citenamefont {Reitze}
  \emph {et~al.}}]{Reitze:2019iox}%
  \BibitemOpen
  \bibfield  {author} {\bibinfo {author} {\bibfnamefont {D.}~\bibnamefont
  {Reitze}} \emph {et~al.},\ }\href@noop {} {\bibfield  {journal} {\bibinfo
  {journal} {Bull. Am. Astron. Soc.}\ }\textbf {\bibinfo {volume} {51}},\
  \bibinfo {pages} {035} (\bibinfo {year} {2019})},\ \Eprint
  {http://arxiv.org/abs/1907.04833} {arXiv:1907.04833 [astro-ph.IM]}
  \BibitemShut {NoStop}%
\bibitem [{\citenamefont {Romero-Shaw}\ \emph {et~al.}(2020)\citenamefont
  {Romero-Shaw}, \citenamefont {Farrow}, \citenamefont {Stevenson},
  \citenamefont {Thrane},\ and\ \citenamefont {Zhu}}]{Romero-Shaw:2020aaj}%
  \BibitemOpen
  \bibfield  {author} {\bibinfo {author} {\bibfnamefont {I.~M.}\ \bibnamefont
  {Romero-Shaw}}, \bibinfo {author} {\bibfnamefont {N.}~\bibnamefont {Farrow}},
  \bibinfo {author} {\bibfnamefont {S.}~\bibnamefont {Stevenson}}, \bibinfo
  {author} {\bibfnamefont {E.}~\bibnamefont {Thrane}}, \ and\ \bibinfo {author}
  {\bibfnamefont {X.-J.}\ \bibnamefont {Zhu}},\ }\href@noop {} {\  (\bibinfo
  {year} {2020})},\ \Eprint {http://arxiv.org/abs/2001.06492} {arXiv:2001.06492
  [astro-ph.HE]} \BibitemShut {NoStop}%
\bibitem [{\citenamefont {Hawking}(1971)}]{Hawking:1971ei}%
  \BibitemOpen
  \bibfield  {author} {\bibinfo {author} {\bibfnamefont {S.}~\bibnamefont
  {Hawking}},\ }\href@noop {} {\bibfield  {journal} {\bibinfo  {journal} {Mon.
  Not. Roy. Astron. Soc.}\ }\textbf {\bibinfo {volume} {152}},\ \bibinfo
  {pages} {75} (\bibinfo {year} {1971})}\BibitemShut {NoStop}%
\bibitem [{\citenamefont {Carr}\ and\ \citenamefont
  {Hawking}(1974)}]{Carr:1974nx}%
  \BibitemOpen
  \bibfield  {author} {\bibinfo {author} {\bibfnamefont {B.~J.}\ \bibnamefont
  {Carr}}\ and\ \bibinfo {author} {\bibfnamefont {S.~W.}\ \bibnamefont
  {Hawking}},\ }\href@noop {} {\bibfield  {journal} {\bibinfo  {journal} {Mon.
  Not. Roy. Astron. Soc.}\ }\textbf {\bibinfo {volume} {168}},\ \bibinfo
  {pages} {399} (\bibinfo {year} {1974})}\BibitemShut {NoStop}%
\bibitem [{\citenamefont {Green}(2015)}]{Green:2014faa}%
  \BibitemOpen
  \bibfield  {author} {\bibinfo {author} {\bibfnamefont {A.~M.}\ \bibnamefont
  {Green}},\ }\href {\doibase 10.1007/978-3-319-10852-0_5} {\bibfield
  {journal} {\bibinfo  {journal} {Fundam. Theor. Phys.}\ }\textbf {\bibinfo
  {volume} {178}},\ \bibinfo {pages} {129} (\bibinfo {year} {2015})},\ \Eprint
  {http://arxiv.org/abs/1403.1198} {arXiv:1403.1198 [gr-qc]} \BibitemShut
  {NoStop}%
\bibitem [{\citenamefont {Cai}\ and\ \citenamefont {Wang}(2020)}]{Cai:2019igo}%
  \BibitemOpen
  \bibfield  {author} {\bibinfo {author} {\bibfnamefont {R.-G.}\ \bibnamefont
  {Cai}}\ and\ \bibinfo {author} {\bibfnamefont {S.-J.}\ \bibnamefont {Wang}},\
  }\href {\doibase 10.1103/PhysRevD.101.043508} {\bibfield  {journal} {\bibinfo
   {journal} {Phys. Rev.}\ }\textbf {\bibinfo {volume} {D101}},\ \bibinfo
  {pages} {043508} (\bibinfo {year} {2020})},\ \Eprint
  {http://arxiv.org/abs/1910.07981} {arXiv:1910.07981 [astro-ph.CO]}
  \BibitemShut {NoStop}%
\bibitem [{\citenamefont {Escrivà}\ \emph {et~al.}(2019)\citenamefont
  {Escrivà}, \citenamefont {Germani},\ and\ \citenamefont
  {Sheth}}]{Escriva:2019phb}%
  \BibitemOpen
  \bibfield  {author} {\bibinfo {author} {\bibfnamefont {A.}~\bibnamefont
  {Escrivà}}, \bibinfo {author} {\bibfnamefont {C.}~\bibnamefont {Germani}}, \
  and\ \bibinfo {author} {\bibfnamefont {R.~K.}\ \bibnamefont {Sheth}},\
  }\href@noop {} {\  (\bibinfo {year} {2019})},\ \Eprint
  {http://arxiv.org/abs/1907.13311} {arXiv:1907.13311 [gr-qc]} \BibitemShut
  {NoStop}%
\bibitem [{\citenamefont {Gow}\ \emph {et~al.}(2020)\citenamefont {Gow},
  \citenamefont {Byrnes}, \citenamefont {Hall},\ and\ \citenamefont
  {Peacock}}]{Gow:2019pok}%
  \BibitemOpen
  \bibfield  {author} {\bibinfo {author} {\bibfnamefont {A.~D.}\ \bibnamefont
  {Gow}}, \bibinfo {author} {\bibfnamefont {C.~T.}\ \bibnamefont {Byrnes}},
  \bibinfo {author} {\bibfnamefont {A.}~\bibnamefont {Hall}}, \ and\ \bibinfo
  {author} {\bibfnamefont {J.~A.}\ \bibnamefont {Peacock}},\ }\href {\doibase
  10.1088/1475-7516/2020/01/031} {\bibfield  {journal} {\bibinfo  {journal}
  {JCAP}\ }\textbf {\bibinfo {volume} {2001}},\ \bibinfo {pages} {031}
  (\bibinfo {year} {2020})},\ \Eprint {http://arxiv.org/abs/1911.12685}
  {arXiv:1911.12685 [astro-ph.CO]} \BibitemShut {NoStop}%
\bibitem [{\citenamefont {De~Luca}\ \emph
  {et~al.}(2020{\natexlab{a}})\citenamefont {De~Luca}, \citenamefont
  {Franciolini},\ and\ \citenamefont {Riotto}}]{DeLuca:2020ioi}%
  \BibitemOpen
  \bibfield  {author} {\bibinfo {author} {\bibfnamefont {V.}~\bibnamefont
  {De~Luca}}, \bibinfo {author} {\bibfnamefont {G.}~\bibnamefont
  {Franciolini}}, \ and\ \bibinfo {author} {\bibfnamefont {A.}~\bibnamefont
  {Riotto}},\ }\href@noop {} {\  (\bibinfo {year} {2020}{\natexlab{a}})},\
  \Eprint {http://arxiv.org/abs/2001.04371} {arXiv:2001.04371 [astro-ph.CO]}
  \BibitemShut {NoStop}%
\bibitem [{\citenamefont {Young}\ and\ \citenamefont
  {Musso}(2020)}]{Young:2020xmk}%
  \BibitemOpen
  \bibfield  {author} {\bibinfo {author} {\bibfnamefont {S.}~\bibnamefont
  {Young}}\ and\ \bibinfo {author} {\bibfnamefont {M.}~\bibnamefont {Musso}},\
  }\href@noop {} {\  (\bibinfo {year} {2020})},\ \Eprint
  {http://arxiv.org/abs/2001.06469} {arXiv:2001.06469 [astro-ph.CO]}
  \BibitemShut {NoStop}%
\bibitem [{\citenamefont {Lin}\ \emph {et~al.}(2020)\citenamefont {Lin},
  \citenamefont {Gao}, \citenamefont {Gong}, \citenamefont {Lu}, \citenamefont
  {Zhang},\ and\ \citenamefont {Zhang}}]{Lin:2020goi}%
  \BibitemOpen
  \bibfield  {author} {\bibinfo {author} {\bibfnamefont {J.}~\bibnamefont
  {Lin}}, \bibinfo {author} {\bibfnamefont {Q.}~\bibnamefont {Gao}}, \bibinfo
  {author} {\bibfnamefont {Y.}~\bibnamefont {Gong}}, \bibinfo {author}
  {\bibfnamefont {Y.}~\bibnamefont {Lu}}, \bibinfo {author} {\bibfnamefont
  {C.}~\bibnamefont {Zhang}}, \ and\ \bibinfo {author} {\bibfnamefont
  {F.}~\bibnamefont {Zhang}},\ }\href@noop {} {\  (\bibinfo {year} {2020})},\
  \Eprint {http://arxiv.org/abs/2001.05909} {arXiv:2001.05909 [gr-qc]}
  \BibitemShut {NoStop}%
\bibitem [{\citenamefont {Liu}\ \emph {et~al.}(2020)\citenamefont {Liu},
  \citenamefont {Guo}, \citenamefont {Cai},\ and\ \citenamefont
  {Kim}}]{Liu:2020cds}%
  \BibitemOpen
  \bibfield  {author} {\bibinfo {author} {\bibfnamefont {L.}~\bibnamefont
  {Liu}}, \bibinfo {author} {\bibfnamefont {Z.-K.}\ \bibnamefont {Guo}},
  \bibinfo {author} {\bibfnamefont {R.-G.}\ \bibnamefont {Cai}}, \ and\
  \bibinfo {author} {\bibfnamefont {S.~P.}\ \bibnamefont {Kim}},\ }\href@noop
  {} {\  (\bibinfo {year} {2020})},\ \Eprint {http://arxiv.org/abs/2001.02984}
  {arXiv:2001.02984 [astro-ph.CO]} \BibitemShut {NoStop}%
\bibitem [{\citenamefont {Roncadelli}\ \emph {et~al.}(2009)\citenamefont
  {Roncadelli}, \citenamefont {Treves},\ and\ \citenamefont
  {Turolla}}]{Roncadelli:2009qj}%
  \BibitemOpen
  \bibfield  {author} {\bibinfo {author} {\bibfnamefont {M.}~\bibnamefont
  {Roncadelli}}, \bibinfo {author} {\bibfnamefont {A.}~\bibnamefont {Treves}},
  \ and\ \bibinfo {author} {\bibfnamefont {R.}~\bibnamefont {Turolla}},\
  }\href@noop {} {\  (\bibinfo {year} {2009})},\ \Eprint
  {http://arxiv.org/abs/0901.1093} {arXiv:0901.1093 [astro-ph.CO]} \BibitemShut
  {NoStop}%
\bibitem [{\citenamefont {Baibhav}\ \emph {et~al.}(2019)\citenamefont
  {Baibhav}, \citenamefont {Berti}, \citenamefont {Gerosa}, \citenamefont
  {Mapelli}, \citenamefont {Giacobbo}, \citenamefont {Bouffanais},\ and\
  \citenamefont {Di~Carlo}}]{Baibhav:2019gxm}%
  \BibitemOpen
  \bibfield  {author} {\bibinfo {author} {\bibfnamefont {V.}~\bibnamefont
  {Baibhav}}, \bibinfo {author} {\bibfnamefont {E.}~\bibnamefont {Berti}},
  \bibinfo {author} {\bibfnamefont {D.}~\bibnamefont {Gerosa}}, \bibinfo
  {author} {\bibfnamefont {M.}~\bibnamefont {Mapelli}}, \bibinfo {author}
  {\bibfnamefont {N.}~\bibnamefont {Giacobbo}}, \bibinfo {author}
  {\bibfnamefont {Y.}~\bibnamefont {Bouffanais}}, \ and\ \bibinfo {author}
  {\bibfnamefont {U.~N.}\ \bibnamefont {Di~Carlo}},\ }\href {\doibase
  10.1103/PhysRevD.100.064060} {\bibfield  {journal} {\bibinfo  {journal}
  {Phys. Rev.}\ }\textbf {\bibinfo {volume} {D100}},\ \bibinfo {pages} {064060}
  (\bibinfo {year} {2019})},\ \Eprint {http://arxiv.org/abs/1906.04197}
  {arXiv:1906.04197 [gr-qc]} \BibitemShut {NoStop}%
\bibitem [{\citenamefont {Sasaki}\ \emph {et~al.}(2018)\citenamefont {Sasaki},
  \citenamefont {Suyama}, \citenamefont {Tanaka},\ and\ \citenamefont
  {Yokoyama}}]{Sasaki:2018dmp}%
  \BibitemOpen
  \bibfield  {author} {\bibinfo {author} {\bibfnamefont {M.}~\bibnamefont
  {Sasaki}}, \bibinfo {author} {\bibfnamefont {T.}~\bibnamefont {Suyama}},
  \bibinfo {author} {\bibfnamefont {T.}~\bibnamefont {Tanaka}}, \ and\ \bibinfo
  {author} {\bibfnamefont {S.}~\bibnamefont {Yokoyama}},\ }\href {\doibase
  10.1088/1361-6382/aaa7b4} {\bibfield  {journal} {\bibinfo  {journal} {Class.
  Quant. Grav.}\ }\textbf {\bibinfo {volume} {35}},\ \bibinfo {pages} {063001}
  (\bibinfo {year} {2018})},\ \Eprint {http://arxiv.org/abs/1801.05235}
  {arXiv:1801.05235 [astro-ph.CO]} \BibitemShut {NoStop}%
\bibitem [{\citenamefont {Raidal}\ \emph {et~al.}(2019)\citenamefont {Raidal},
  \citenamefont {Spethmann}, \citenamefont {Vaskonen},\ and\ \citenamefont
  {Veermäe}}]{Raidal:2018bbj}%
  \BibitemOpen
  \bibfield  {author} {\bibinfo {author} {\bibfnamefont {M.}~\bibnamefont
  {Raidal}}, \bibinfo {author} {\bibfnamefont {C.}~\bibnamefont {Spethmann}},
  \bibinfo {author} {\bibfnamefont {V.}~\bibnamefont {Vaskonen}}, \ and\
  \bibinfo {author} {\bibfnamefont {H.}~\bibnamefont {Veermäe}},\ }\href
  {\doibase 10.1088/1475-7516/2019/02/018} {\bibfield  {journal} {\bibinfo
  {journal} {JCAP}\ }\textbf {\bibinfo {volume} {1902}},\ \bibinfo {pages}
  {018} (\bibinfo {year} {2019})},\ \Eprint {http://arxiv.org/abs/1812.01930}
  {arXiv:1812.01930 [astro-ph.CO]} \BibitemShut {NoStop}%
\bibitem [{\citenamefont {Albert}\ \emph {et~al.}(2019)\citenamefont {Albert}
  \emph {et~al.}}]{Albert:2019qxd}%
  \BibitemOpen
  \bibfield  {author} {\bibinfo {author} {\bibfnamefont {A.}~\bibnamefont
  {Albert}} \emph {et~al.} (\bibinfo {collaboration} {HAWC}),\ }\href@noop {}
  {\  (\bibinfo {year} {2019})},\ \Eprint {http://arxiv.org/abs/1911.04356}
  {arXiv:1911.04356 [astro-ph.HE]} \BibitemShut {NoStop}%
\bibitem [{\citenamefont {Hertzberg}\ \emph {et~al.}(2020)\citenamefont
  {Hertzberg}, \citenamefont {Schiappacasse},\ and\ \citenamefont
  {Yanagida}}]{Hertzberg:2020hsz}%
  \BibitemOpen
  \bibfield  {author} {\bibinfo {author} {\bibfnamefont {M.~P.}\ \bibnamefont
  {Hertzberg}}, \bibinfo {author} {\bibfnamefont {E.~D.}\ \bibnamefont
  {Schiappacasse}}, \ and\ \bibinfo {author} {\bibfnamefont {T.~T.}\
  \bibnamefont {Yanagida}},\ }\href@noop {} {\  (\bibinfo {year} {2020})},\
  \Eprint {http://arxiv.org/abs/2001.07476} {arXiv:2001.07476 [astro-ph.CO]}
  \BibitemShut {NoStop}%
\bibitem [{\citenamefont {De~Luca}\ \emph
  {et~al.}(2020{\natexlab{b}})\citenamefont {De~Luca}, \citenamefont
  {Franciolini}, \citenamefont {Pani},\ and\ \citenamefont
  {Riotto}}]{DeLuca:2020fpg}%
  \BibitemOpen
  \bibfield  {author} {\bibinfo {author} {\bibfnamefont {V.}~\bibnamefont
  {De~Luca}}, \bibinfo {author} {\bibfnamefont {G.}~\bibnamefont
  {Franciolini}}, \bibinfo {author} {\bibfnamefont {P.}~\bibnamefont {Pani}}, \
  and\ \bibinfo {author} {\bibfnamefont {A.}~\bibnamefont {Riotto}},\
  }\href@noop {} {\  (\bibinfo {year} {2020}{\natexlab{b}})},\ \Eprint
  {http://arxiv.org/abs/2003.12589} {arXiv:2003.12589 [astro-ph.CO]}
  \BibitemShut {NoStop}%
\bibitem [{\citenamefont {Carr}\ \emph {et~al.}(2020)\citenamefont {Carr},
  \citenamefont {Kohri}, \citenamefont {Sendouda},\ and\ \citenamefont
  {Yokoyama}}]{Carr:2020gox}%
  \BibitemOpen
  \bibfield  {author} {\bibinfo {author} {\bibfnamefont {B.}~\bibnamefont
  {Carr}}, \bibinfo {author} {\bibfnamefont {K.}~\bibnamefont {Kohri}},
  \bibinfo {author} {\bibfnamefont {Y.}~\bibnamefont {Sendouda}}, \ and\
  \bibinfo {author} {\bibfnamefont {J.}~\bibnamefont {Yokoyama}},\ }\href@noop
  {} {\  (\bibinfo {year} {2020})},\ \Eprint {http://arxiv.org/abs/2002.12778}
  {arXiv:2002.12778 [astro-ph.CO]} \BibitemShut {NoStop}%
\bibitem [{\citenamefont {Abbott}\ \emph
  {et~al.}(2019{\natexlab{b}})\citenamefont {Abbott} \emph
  {et~al.}}]{LIGOScientific:2018mvr}%
  \BibitemOpen
  \bibfield  {author} {\bibinfo {author} {\bibfnamefont {B.~P.}\ \bibnamefont
  {Abbott}} \emph {et~al.} (\bibinfo {collaboration} {LIGO Scientific,
  Virgo}),\ }\href {\doibase 10.1103/PhysRevX.9.031040} {\bibfield  {journal}
  {\bibinfo  {journal} {Phys. Rev.}\ }\textbf {\bibinfo {volume} {X9}},\
  \bibinfo {pages} {031040} (\bibinfo {year} {2019}{\natexlab{b}})},\ \Eprint
  {http://arxiv.org/abs/1811.12907} {arXiv:1811.12907 [astro-ph.HE]}
  \BibitemShut {NoStop}%
\bibitem [{\citenamefont {Abbott}\ \emph
  {et~al.}(2016{\natexlab{c}})\citenamefont {Abbott} \emph
  {et~al.}}]{TheLIGOScientific:2016pea}%
  \BibitemOpen
  \bibfield  {author} {\bibinfo {author} {\bibfnamefont {B.~P.}\ \bibnamefont
  {Abbott}} \emph {et~al.} (\bibinfo {collaboration} {LIGO Scientific,
  Virgo}),\ }\href {\doibase 10.1103/PhysRevX.6.041015,
  10.1103/PhysRevX.8.039903} {\bibfield  {journal} {\bibinfo  {journal} {Phys.
  Rev.}\ }\textbf {\bibinfo {volume} {X6}},\ \bibinfo {pages} {041015}
  (\bibinfo {year} {2016}{\natexlab{c}})},\ \bibinfo {note} {[erratum: Phys.
  Rev.X8,no.3,039903(2018)]},\ \Eprint {http://arxiv.org/abs/1606.04856}
  {arXiv:1606.04856 [gr-qc]} \BibitemShut {NoStop}%
\bibitem [{\citenamefont {Raaijmakers}\ \emph {et~al.}(2019)\citenamefont
  {Raaijmakers} \emph {et~al.}}]{Raaijmakers:2019qny}%
  \BibitemOpen
  \bibfield  {author} {\bibinfo {author} {\bibfnamefont {G.}~\bibnamefont
  {Raaijmakers}} \emph {et~al.},\ }\href {\doibase 10.3847/2041-8213/ab451a}
  {\bibfield  {journal} {\bibinfo  {journal} {Astrophys. J. Lett.}\ }\textbf
  {\bibinfo {volume} {887}},\ \bibinfo {pages} {L22} (\bibinfo {year}
  {2019})},\ \Eprint {http://arxiv.org/abs/1912.05703} {arXiv:1912.05703
  [astro-ph.HE]} \BibitemShut {NoStop}%
\bibitem [{\citenamefont {Riley}\ \emph {et~al.}(2019)\citenamefont {Riley}
  \emph {et~al.}}]{Riley:2019yda}%
  \BibitemOpen
  \bibfield  {author} {\bibinfo {author} {\bibfnamefont {T.~E.}\ \bibnamefont
  {Riley}} \emph {et~al.},\ }\href {\doibase 10.3847/2041-8213/ab481c}
  {\bibfield  {journal} {\bibinfo  {journal} {Astrophys. J. Lett.}\ }\textbf
  {\bibinfo {volume} {887}},\ \bibinfo {pages} {L21} (\bibinfo {year}
  {2019})},\ \Eprint {http://arxiv.org/abs/1912.05702} {arXiv:1912.05702
  [astro-ph.HE]} \BibitemShut {NoStop}%
\bibitem [{\citenamefont {Miller}\ \emph {et~al.}(2019)\citenamefont {Miller}
  \emph {et~al.}}]{Miller:2019cac}%
  \BibitemOpen
  \bibfield  {author} {\bibinfo {author} {\bibfnamefont {M.~C.}\ \bibnamefont
  {Miller}} \emph {et~al.},\ }\href {\doibase 10.3847/2041-8213/ab50c5}
  {\bibfield  {journal} {\bibinfo  {journal} {Astrophys. J. Lett.}\ }\textbf
  {\bibinfo {volume} {887}},\ \bibinfo {pages} {L24} (\bibinfo {year}
  {2019})},\ \Eprint {http://arxiv.org/abs/1912.05705} {arXiv:1912.05705
  [astro-ph.HE]} \BibitemShut {NoStop}%
\bibitem [{\citenamefont {Bogdanov}\ \emph
  {et~al.}(2019{\natexlab{a}})\citenamefont {Bogdanov} \emph
  {et~al.}}]{Bogdanov:2019ixe}%
  \BibitemOpen
  \bibfield  {author} {\bibinfo {author} {\bibfnamefont {S.}~\bibnamefont
  {Bogdanov}} \emph {et~al.},\ }\href {\doibase 10.3847/2041-8213/ab53eb}
  {\bibfield  {journal} {\bibinfo  {journal} {Astrophys. J. Lett.}\ }\textbf
  {\bibinfo {volume} {887}},\ \bibinfo {pages} {L25} (\bibinfo {year}
  {2019}{\natexlab{a}})},\ \Eprint {http://arxiv.org/abs/1912.05706}
  {arXiv:1912.05706 [astro-ph.HE]} \BibitemShut {NoStop}%
\bibitem [{\citenamefont {Bogdanov}\ \emph
  {et~al.}(2019{\natexlab{b}})\citenamefont {Bogdanov} \emph
  {et~al.}}]{Bogdanov:2019qjb}%
  \BibitemOpen
  \bibfield  {author} {\bibinfo {author} {\bibfnamefont {S.}~\bibnamefont
  {Bogdanov}} \emph {et~al.},\ }\href {\doibase 10.3847/2041-8213/ab5968}
  {\bibfield  {journal} {\bibinfo  {journal} {Astrophys. J.}\ }\textbf
  {\bibinfo {volume} {887}},\ \bibinfo {pages} {L26} (\bibinfo {year}
  {2019}{\natexlab{b}})},\ \Eprint {http://arxiv.org/abs/1912.05707}
  {arXiv:1912.05707 [astro-ph.HE]} \BibitemShut {NoStop}%
\bibitem [{\citenamefont {Hinderer}\ \emph {et~al.}(2018)\citenamefont
  {Hinderer}, \citenamefont {Rezzolla},\ and\ \citenamefont
  {Baiotti}}]{Hinderer:2018mrj}%
  \BibitemOpen
  \bibfield  {author} {\bibinfo {author} {\bibfnamefont {T.}~\bibnamefont
  {Hinderer}}, \bibinfo {author} {\bibfnamefont {L.}~\bibnamefont {Rezzolla}},
  \ and\ \bibinfo {author} {\bibfnamefont {L.}~\bibnamefont {Baiotti}},\
  }\enquote {\bibinfo {title} {{Gravitational Waves from Merging Binary
  Neutron-Star Systems}},}\ \ (\bibinfo {year} {2018})\ pp.\ \bibinfo {pages}
  {575--635}\BibitemShut {NoStop}%
\bibitem [{\citenamefont {Poisson}(2015)}]{Poisson:2014gka}%
  \BibitemOpen
  \bibfield  {author} {\bibinfo {author} {\bibfnamefont {E.}~\bibnamefont
  {Poisson}},\ }\href {\doibase 10.1103/PhysRevD.91.044004} {\bibfield
  {journal} {\bibinfo  {journal} {Phys. Rev.}\ }\textbf {\bibinfo {volume}
  {D91}},\ \bibinfo {pages} {044004} (\bibinfo {year} {2015})},\ \Eprint
  {http://arxiv.org/abs/1411.4711} {arXiv:1411.4711 [gr-qc]} \BibitemShut
  {NoStop}%
\bibitem [{\citenamefont {Taylor}\ and\ \citenamefont
  {Gerosa}(2018)}]{Taylor:2018iat}%
  \BibitemOpen
  \bibfield  {author} {\bibinfo {author} {\bibfnamefont {S.~R.}\ \bibnamefont
  {Taylor}}\ and\ \bibinfo {author} {\bibfnamefont {D.}~\bibnamefont
  {Gerosa}},\ }\href {\doibase 10.1103/PhysRevD.98.083017} {\bibfield
  {journal} {\bibinfo  {journal} {Phys. Rev.}\ }\textbf {\bibinfo {volume}
  {D98}},\ \bibinfo {pages} {083017} (\bibinfo {year} {2018})},\ \Eprint
  {http://arxiv.org/abs/1806.08365} {arXiv:1806.08365 [astro-ph.HE]}
  \BibitemShut {NoStop}%
\bibitem [{\citenamefont {Wong}\ and\ \citenamefont
  {Gerosa}(2019)}]{Wong:2019uni}%
  \BibitemOpen
  \bibfield  {author} {\bibinfo {author} {\bibfnamefont {K.~W.~K.}\
  \bibnamefont {Wong}}\ and\ \bibinfo {author} {\bibfnamefont {D.}~\bibnamefont
  {Gerosa}},\ }\href {\doibase 10.1103/PhysRevD.100.083015} {\bibfield
  {journal} {\bibinfo  {journal} {Phys. Rev.}\ }\textbf {\bibinfo {volume}
  {D100}},\ \bibinfo {pages} {083015} (\bibinfo {year} {2019})},\ \Eprint
  {http://arxiv.org/abs/1909.06373} {arXiv:1909.06373 [astro-ph.HE]}
  \BibitemShut {NoStop}%
\bibitem [{\citenamefont {Wong}\ \emph {et~al.}(2020)\citenamefont {Wong},
  \citenamefont {Contardo},\ and\ \citenamefont {Ho}}]{Wong:2020jdt}%
  \BibitemOpen
  \bibfield  {author} {\bibinfo {author} {\bibfnamefont {K.~W.~K.}\
  \bibnamefont {Wong}}, \bibinfo {author} {\bibfnamefont {G.}~\bibnamefont
  {Contardo}}, \ and\ \bibinfo {author} {\bibfnamefont {S.}~\bibnamefont
  {Ho}},\ }\href@noop {} {\  (\bibinfo {year} {2020})},\ \Eprint
  {http://arxiv.org/abs/2002.09491} {arXiv:2002.09491 [astro-ph.IM]}
  \BibitemShut {NoStop}%
\bibitem [{\citenamefont {{Pitkin}}\ \emph {et~al.}(2011)\citenamefont
  {{Pitkin}}, \citenamefont {{Reid}}, \citenamefont {{Rowan}},\ and\
  \citenamefont {{Hough}}}]{HVL}%
  \BibitemOpen
  \bibfield  {author} {\bibinfo {author} {\bibfnamefont {M.}~\bibnamefont
  {{Pitkin}}}, \bibinfo {author} {\bibfnamefont {S.}~\bibnamefont {{Reid}}},
  \bibinfo {author} {\bibfnamefont {S.}~\bibnamefont {{Rowan}}}, \ and\
  \bibinfo {author} {\bibfnamefont {J.}~\bibnamefont {{Hough}}},\ }\href
  {\doibase 10.12942/lrr-2011-5} {\bibfield  {journal} {\bibinfo  {journal}
  {Living Reviews in Relativity}\ }\textbf {\bibinfo {volume} {14}},\ \bibinfo
  {eid} {5} (\bibinfo {year} {2011})},\ \Eprint
  {http://arxiv.org/abs/1102.3355} {arXiv:1102.3355 [astro-ph.IM]} \BibitemShut
  {NoStop}%
\bibitem [{\citenamefont {Sathyaprakash}\ \emph {et~al.}(2019)\citenamefont
  {Sathyaprakash} \emph {et~al.}}]{Sathyaprakash:2019rom}%
  \BibitemOpen
  \bibfield  {author} {\bibinfo {author} {\bibfnamefont {B.~S.}\ \bibnamefont
  {Sathyaprakash}} \emph {et~al.},\ }\href@noop {} {\  (\bibinfo {year}
  {2019})},\ \Eprint {http://arxiv.org/abs/1903.09277} {arXiv:1903.09277
  [astro-ph.HE]} \BibitemShut {NoStop}%
\bibitem [{\citenamefont {{Ashton}}\ \emph
  {et~al.}(2019{\natexlab{a}})\citenamefont {{Ashton}}, \citenamefont
  {{H{\"u}bner}}, \citenamefont {{Lasky}}, \citenamefont {{Talbot}},
  \citenamefont {{Ackley}}, \citenamefont {{Biscoveanu}}, \citenamefont
  {{Chu}}, \citenamefont {{Divarkala}}, \citenamefont {{Easter}}, \citenamefont
  {{Goncharov}}, \citenamefont {{Hernandez Vivanco}}, \citenamefont {{Harms}},
  \citenamefont {{Lower}}, \citenamefont {{Meadors}}, \citenamefont
  {{Melchor}}, \citenamefont {{Payne}}, \citenamefont {{Pitkin}}, \citenamefont
  {{Powell}}, \citenamefont {{Sarin}}, \citenamefont {{Smith}},\ and\
  \citenamefont {{Thrane}}}]{bilby1}%
  \BibitemOpen
  \bibfield  {author} {\bibinfo {author} {\bibfnamefont {G.}~\bibnamefont
  {{Ashton}}}, \bibinfo {author} {\bibfnamefont {M.}~\bibnamefont
  {{H{\"u}bner}}}, \bibinfo {author} {\bibfnamefont {P.~D.}\ \bibnamefont
  {{Lasky}}}, \bibinfo {author} {\bibfnamefont {C.}~\bibnamefont {{Talbot}}},
  \bibinfo {author} {\bibfnamefont {K.}~\bibnamefont {{Ackley}}}, \bibinfo
  {author} {\bibfnamefont {S.}~\bibnamefont {{Biscoveanu}}}, \bibinfo {author}
  {\bibfnamefont {Q.}~\bibnamefont {{Chu}}}, \bibinfo {author} {\bibfnamefont
  {A.}~\bibnamefont {{Divarkala}}}, \bibinfo {author} {\bibfnamefont {P.~J.}\
  \bibnamefont {{Easter}}}, \bibinfo {author} {\bibfnamefont {B.}~\bibnamefont
  {{Goncharov}}}, \bibinfo {author} {\bibfnamefont {F.}~\bibnamefont
  {{Hernandez Vivanco}}}, \bibinfo {author} {\bibfnamefont {J.}~\bibnamefont
  {{Harms}}}, \bibinfo {author} {\bibfnamefont {M.~E.}\ \bibnamefont
  {{Lower}}}, \bibinfo {author} {\bibfnamefont {G.~D.}\ \bibnamefont
  {{Meadors}}}, \bibinfo {author} {\bibfnamefont {D.}~\bibnamefont
  {{Melchor}}}, \bibinfo {author} {\bibfnamefont {E.}~\bibnamefont {{Payne}}},
  \bibinfo {author} {\bibfnamefont {M.~D.}\ \bibnamefont {{Pitkin}}}, \bibinfo
  {author} {\bibfnamefont {J.}~\bibnamefont {{Powell}}}, \bibinfo {author}
  {\bibfnamefont {N.}~\bibnamefont {{Sarin}}}, \bibinfo {author} {\bibfnamefont
  {R.~J.~E.}\ \bibnamefont {{Smith}}}, \ and\ \bibinfo {author} {\bibfnamefont
  {E.}~\bibnamefont {{Thrane}}},\ }\href@noop {} {\enquote {\bibinfo {title}
  {{Bilby: Bayesian inference library}},}\ } (\bibinfo {year}
  {2019}{\natexlab{a}}),\ \Eprint {http://arxiv.org/abs/1901.011}
  {ascl:1901.011} \BibitemShut {NoStop}%
\bibitem [{\citenamefont {{Ashton}}\ \emph
  {et~al.}(2019{\natexlab{b}})\citenamefont {{Ashton}}, \citenamefont
  {{H{\"u}bner}}, \citenamefont {{Lasky}},\ and\ \citenamefont
  {{Talbot}}}]{bilby2}%
  \BibitemOpen
  \bibfield  {author} {\bibinfo {author} {\bibfnamefont {G.}~\bibnamefont
  {{Ashton}}}, \bibinfo {author} {\bibfnamefont {M.}~\bibnamefont
  {{H{\"u}bner}}}, \bibinfo {author} {\bibfnamefont {P.}~\bibnamefont
  {{Lasky}}}, \ and\ \bibinfo {author} {\bibfnamefont {C.}~\bibnamefont
  {{Talbot}}},\ }\href {\doibase 10.5281/zenodo.2602178} {\enquote {\bibinfo
  {title} {{Bilby: A User-Friendly Bayesian Inference Library}},}\ } (\bibinfo
  {year} {2019}{\natexlab{b}})\BibitemShut {NoStop}%
\bibitem [{\citenamefont {Ashton}\ \emph {et~al.}(2019)\citenamefont {Ashton}
  \emph {et~al.}}]{Ashton:2018jfp}%
  \BibitemOpen
  \bibfield  {author} {\bibinfo {author} {\bibfnamefont {G.}~\bibnamefont
  {Ashton}} \emph {et~al.},\ }\href {\doibase 10.3847/1538-4365/ab06fc}
  {\bibfield  {journal} {\bibinfo  {journal} {Astrophys. J. Suppl.}\ }\textbf
  {\bibinfo {volume} {241}},\ \bibinfo {pages} {27} (\bibinfo {year} {2019})},\
  \Eprint {http://arxiv.org/abs/1811.02042} {arXiv:1811.02042 [astro-ph.IM]}
  \BibitemShut {NoStop}%
\bibitem [{\citenamefont {Dietrich}\ \emph
  {et~al.}(2019{\natexlab{a}})\citenamefont {Dietrich} \emph
  {et~al.}}]{Dietrich:2018uni}%
  \BibitemOpen
  \bibfield  {author} {\bibinfo {author} {\bibfnamefont {T.}~\bibnamefont
  {Dietrich}} \emph {et~al.},\ }\href {\doibase 10.1103/PhysRevD.99.024029}
  {\bibfield  {journal} {\bibinfo  {journal} {Phys. Rev. D}\ }\textbf {\bibinfo
  {volume} {99}},\ \bibinfo {pages} {024029} (\bibinfo {year}
  {2019}{\natexlab{a}})},\ \Eprint {http://arxiv.org/abs/1804.02235}
  {arXiv:1804.02235 [gr-qc]} \BibitemShut {NoStop}%
\bibitem [{\citenamefont {Dietrich}\ \emph
  {et~al.}(2019{\natexlab{b}})\citenamefont {Dietrich}, \citenamefont
  {Samajdar}, \citenamefont {Khan}, \citenamefont {Johnson-McDaniel},
  \citenamefont {Dudi},\ and\ \citenamefont {Tichy}}]{Dietrich:2019kaq}%
  \BibitemOpen
  \bibfield  {author} {\bibinfo {author} {\bibfnamefont {T.}~\bibnamefont
  {Dietrich}}, \bibinfo {author} {\bibfnamefont {A.}~\bibnamefont {Samajdar}},
  \bibinfo {author} {\bibfnamefont {S.}~\bibnamefont {Khan}}, \bibinfo {author}
  {\bibfnamefont {N.~K.}\ \bibnamefont {Johnson-McDaniel}}, \bibinfo {author}
  {\bibfnamefont {R.}~\bibnamefont {Dudi}}, \ and\ \bibinfo {author}
  {\bibfnamefont {W.}~\bibnamefont {Tichy}},\ }\href {\doibase
  10.1103/PhysRevD.100.044003} {\bibfield  {journal} {\bibinfo  {journal}
  {Phys. Rev.}\ }\textbf {\bibinfo {volume} {D100}},\ \bibinfo {pages} {044003}
  (\bibinfo {year} {2019}{\natexlab{b}})},\ \Eprint
  {http://arxiv.org/abs/1905.06011} {arXiv:1905.06011 [gr-qc]} \BibitemShut
  {NoStop}%
\bibitem [{\citenamefont {Pannarale}\ \emph {et~al.}(2013)\citenamefont
  {Pannarale}, \citenamefont {Berti}, \citenamefont {Kyutoku},\ and\
  \citenamefont {Shibata}}]{Pannarale:2013uoa}%
  \BibitemOpen
  \bibfield  {author} {\bibinfo {author} {\bibfnamefont {F.}~\bibnamefont
  {Pannarale}}, \bibinfo {author} {\bibfnamefont {E.}~\bibnamefont {Berti}},
  \bibinfo {author} {\bibfnamefont {K.}~\bibnamefont {Kyutoku}}, \ and\
  \bibinfo {author} {\bibfnamefont {M.}~\bibnamefont {Shibata}},\ }\href
  {\doibase 10.1103/PhysRevD.88.084011} {\bibfield  {journal} {\bibinfo
  {journal} {Phys. Rev. D}\ }\textbf {\bibinfo {volume} {88}},\ \bibinfo
  {pages} {084011} (\bibinfo {year} {2013})},\ \Eprint
  {http://arxiv.org/abs/1307.5111} {arXiv:1307.5111 [gr-qc]} \BibitemShut
  {NoStop}%
\bibitem [{\citenamefont {Pannarale}\ \emph
  {et~al.}(2015{\natexlab{b}})\citenamefont {Pannarale}, \citenamefont {Berti},
  \citenamefont {Kyutoku}, \citenamefont {Lackey},\ and\ \citenamefont
  {Shibata}}]{Pannarale:2015jka}%
  \BibitemOpen
  \bibfield  {author} {\bibinfo {author} {\bibfnamefont {F.}~\bibnamefont
  {Pannarale}}, \bibinfo {author} {\bibfnamefont {E.}~\bibnamefont {Berti}},
  \bibinfo {author} {\bibfnamefont {K.}~\bibnamefont {Kyutoku}}, \bibinfo
  {author} {\bibfnamefont {B.~D.}\ \bibnamefont {Lackey}}, \ and\ \bibinfo
  {author} {\bibfnamefont {M.}~\bibnamefont {Shibata}},\ }\href {\doibase
  10.1103/PhysRevD.92.084050} {\bibfield  {journal} {\bibinfo  {journal} {Phys.
  Rev. D}\ }\textbf {\bibinfo {volume} {92}},\ \bibinfo {pages} {084050}
  (\bibinfo {year} {2015}{\natexlab{b}})},\ \Eprint
  {http://arxiv.org/abs/1509.00512} {arXiv:1509.00512 [gr-qc]} \BibitemShut
  {NoStop}%
\bibitem [{\citenamefont {Krüger}\ and\ \citenamefont
  {Foucart}(2020)}]{Kruger:2020gig}%
  \BibitemOpen
  \bibfield  {author} {\bibinfo {author} {\bibfnamefont {C.~J.}\ \bibnamefont
  {Krüger}}\ and\ \bibinfo {author} {\bibfnamefont {F.}~\bibnamefont
  {Foucart}},\ }\href {\doibase 10.1103/PhysRevD.101.103002} {\bibfield
  {journal} {\bibinfo  {journal} {Phys. Rev. D}\ }\textbf {\bibinfo {volume}
  {101}},\ \bibinfo {pages} {103002} (\bibinfo {year} {2020})},\ \Eprint
  {http://arxiv.org/abs/2002.07728} {arXiv:2002.07728 [astro-ph.HE]}
  \BibitemShut {NoStop}%
\bibitem [{\citenamefont {Thompson}\ \emph {et~al.}(2020)\citenamefont
  {Thompson}, \citenamefont {Fauchon-Jones}, \citenamefont {Khan},
  \citenamefont {Nitoglia}, \citenamefont {Pannarale}, \citenamefont
  {Dietrich},\ and\ \citenamefont {Hannam}}]{Thompson:2020nei}%
  \BibitemOpen
  \bibfield  {author} {\bibinfo {author} {\bibfnamefont {J.~E.}\ \bibnamefont
  {Thompson}}, \bibinfo {author} {\bibfnamefont {E.}~\bibnamefont
  {Fauchon-Jones}}, \bibinfo {author} {\bibfnamefont {S.}~\bibnamefont {Khan}},
  \bibinfo {author} {\bibfnamefont {E.}~\bibnamefont {Nitoglia}}, \bibinfo
  {author} {\bibfnamefont {F.}~\bibnamefont {Pannarale}}, \bibinfo {author}
  {\bibfnamefont {T.}~\bibnamefont {Dietrich}}, \ and\ \bibinfo {author}
  {\bibfnamefont {M.}~\bibnamefont {Hannam}},\ }\href@noop {} {\  (\bibinfo
  {year} {2020})},\ \Eprint {http://arxiv.org/abs/2002.08383} {arXiv:2002.08383
  [gr-qc]} \BibitemShut {NoStop}%
\bibitem [{\citenamefont {Ferrari}\ \emph {et~al.}(2010)\citenamefont
  {Ferrari}, \citenamefont {Gualtieri},\ and\ \citenamefont
  {Pannarale}}]{Ferrari:2009bw}%
  \BibitemOpen
  \bibfield  {author} {\bibinfo {author} {\bibfnamefont {V.}~\bibnamefont
  {Ferrari}}, \bibinfo {author} {\bibfnamefont {L.}~\bibnamefont {Gualtieri}},
  \ and\ \bibinfo {author} {\bibfnamefont {F.}~\bibnamefont {Pannarale}},\
  }\href {\doibase 10.1103/PhysRevD.81.064026} {\bibfield  {journal} {\bibinfo
  {journal} {Phys. Rev. D}\ }\textbf {\bibinfo {volume} {81}},\ \bibinfo
  {pages} {064026} (\bibinfo {year} {2010})},\ \Eprint
  {http://arxiv.org/abs/0912.3692} {arXiv:0912.3692 [gr-qc]} \BibitemShut
  {NoStop}%
\bibitem [{\citenamefont {Maselli}\ \emph {et~al.}(2013)\citenamefont
  {Maselli}, \citenamefont {Gualtieri},\ and\ \citenamefont
  {Ferrari}}]{Maselli:2013rza}%
  \BibitemOpen
  \bibfield  {author} {\bibinfo {author} {\bibfnamefont {A.}~\bibnamefont
  {Maselli}}, \bibinfo {author} {\bibfnamefont {L.}~\bibnamefont {Gualtieri}},
  \ and\ \bibinfo {author} {\bibfnamefont {V.}~\bibnamefont {Ferrari}},\ }\href
  {\doibase 10.1103/PhysRevD.88.104040} {\bibfield  {journal} {\bibinfo
  {journal} {Phys. Rev. D}\ }\textbf {\bibinfo {volume} {88}},\ \bibinfo
  {pages} {104040} (\bibinfo {year} {2013})},\ \Eprint
  {http://arxiv.org/abs/1310.5381} {arXiv:1310.5381 [gr-qc]} \BibitemShut
  {NoStop}%
\bibitem [{\citenamefont {Miller}\ and\ \citenamefont
  {Miller}(2014)}]{Miller:2014aaa}%
  \BibitemOpen
  \bibfield  {author} {\bibinfo {author} {\bibfnamefont {M.~C.}\ \bibnamefont
  {Miller}}\ and\ \bibinfo {author} {\bibfnamefont {J.~M.}\ \bibnamefont
  {Miller}},\ }\href {\doibase 10.1016/j.physrep.2014.09.003} {\bibfield
  {journal} {\bibinfo  {journal} {Phys. Rept.}\ }\textbf {\bibinfo {volume}
  {548}},\ \bibinfo {pages} {1} (\bibinfo {year} {2014})},\ \Eprint
  {http://arxiv.org/abs/1408.4145} {arXiv:1408.4145 [astro-ph.HE]} \BibitemShut
  {NoStop}%
\bibitem [{\citenamefont {Baibhav}\ \emph {et~al.}(2020)\citenamefont
  {Baibhav}, \citenamefont {Gerosa}, \citenamefont {Berti}, \citenamefont
  {Wong}, \citenamefont {Helfer},\ and\ \citenamefont
  {Mould}}]{Baibhav:2020xdf}%
  \BibitemOpen
  \bibfield  {author} {\bibinfo {author} {\bibfnamefont {V.}~\bibnamefont
  {Baibhav}}, \bibinfo {author} {\bibfnamefont {D.}~\bibnamefont {Gerosa}},
  \bibinfo {author} {\bibfnamefont {E.}~\bibnamefont {Berti}}, \bibinfo
  {author} {\bibfnamefont {K.~W.~K.}\ \bibnamefont {Wong}}, \bibinfo {author}
  {\bibfnamefont {T.}~\bibnamefont {Helfer}}, \ and\ \bibinfo {author}
  {\bibfnamefont {M.}~\bibnamefont {Mould}},\ }\href@noop {} {\  (\bibinfo
  {year} {2020})},\ \Eprint {http://arxiv.org/abs/2004.00650} {arXiv:2004.00650
  [astro-ph.HE]} \BibitemShut {NoStop}%
\bibitem [{\citenamefont {De~Luca}\ \emph
  {et~al.}(2020{\natexlab{c}})\citenamefont {De~Luca}, \citenamefont
  {Franciolini}, \citenamefont {Pani},\ and\ \citenamefont
  {Riotto}}]{DeLuca:2020bjf}%
  \BibitemOpen
  \bibfield  {author} {\bibinfo {author} {\bibfnamefont {V.}~\bibnamefont
  {De~Luca}}, \bibinfo {author} {\bibfnamefont {G.}~\bibnamefont
  {Franciolini}}, \bibinfo {author} {\bibfnamefont {P.}~\bibnamefont {Pani}}, \
  and\ \bibinfo {author} {\bibfnamefont {A.}~\bibnamefont {Riotto}},\
  }\href@noop {} {\  (\bibinfo {year} {2020}{\natexlab{c}})},\ \Eprint
  {http://arxiv.org/abs/2003.02778} {arXiv:2003.02778 [astro-ph.CO]}
  \BibitemShut {NoStop}%
\bibitem [{\citenamefont {Mora}\ and\ \citenamefont
  {Will}(2004)}]{Mora:2003wt}%
  \BibitemOpen
  \bibfield  {author} {\bibinfo {author} {\bibfnamefont {T.}~\bibnamefont
  {Mora}}\ and\ \bibinfo {author} {\bibfnamefont {C.~M.}\ \bibnamefont
  {Will}},\ }\href {\doibase 10.1103/PhysRevD.71.129901,
  10.1103/PhysRevD.69.104021} {\bibfield  {journal} {\bibinfo  {journal} {Phys.
  Rev.}\ }\textbf {\bibinfo {volume} {D69}},\ \bibinfo {pages} {104021}
  (\bibinfo {year} {2004})},\ \bibinfo {note} {[Erratum: Phys.
  Rev.D71,129901(2005)]},\ \Eprint {http://arxiv.org/abs/gr-qc/0312082}
  {arXiv:gr-qc/0312082 [gr-qc]} \BibitemShut {NoStop}%
\bibitem [{\citenamefont {Giddings}\ and\ \citenamefont
  {Mangano}(2008)}]{Giddings:2008gr}%
  \BibitemOpen
  \bibfield  {author} {\bibinfo {author} {\bibfnamefont {S.~B.}\ \bibnamefont
  {Giddings}}\ and\ \bibinfo {author} {\bibfnamefont {M.~L.}\ \bibnamefont
  {Mangano}},\ }\href {\doibase 10.1103/PhysRevD.78.035009} {\bibfield
  {journal} {\bibinfo  {journal} {Phys. Rev. D}\ }\textbf {\bibinfo {volume}
  {78}},\ \bibinfo {pages} {035009} (\bibinfo {year} {2008})},\ \Eprint
  {http://arxiv.org/abs/0806.3381} {arXiv:0806.3381 [hep-ph]} \BibitemShut
  {NoStop}%
\bibitem [{\citenamefont {East}\ and\ \citenamefont
  {Lehner}(2019)}]{East:2019dxt}%
  \BibitemOpen
  \bibfield  {author} {\bibinfo {author} {\bibfnamefont {W.~E.}\ \bibnamefont
  {East}}\ and\ \bibinfo {author} {\bibfnamefont {L.}~\bibnamefont {Lehner}},\
  }\href {\doibase 10.1103/PhysRevD.100.124026} {\bibfield  {journal} {\bibinfo
   {journal} {Phys. Rev. D}\ }\textbf {\bibinfo {volume} {100}},\ \bibinfo
  {pages} {124026} (\bibinfo {year} {2019})},\ \Eprint
  {http://arxiv.org/abs/1909.07968} {arXiv:1909.07968 [gr-qc]} \BibitemShut
  {NoStop}%
\bibitem [{\citenamefont {Gilks}\ \emph {et~al.}(1996)\citenamefont {Gilks},
  \citenamefont {Richardson},\ and\ \citenamefont
  {Spiegelhalter}}]{Gilks:1996}%
  \BibitemOpen
  \bibfield  {author} {\bibinfo {author} {\bibfnamefont {W.~R.}\ \bibnamefont
  {Gilks}}, \bibinfo {author} {\bibfnamefont {S.}~\bibnamefont {Richardson}}, \
  and\ \bibinfo {author} {\bibfnamefont {D.~J.}\ \bibnamefont
  {Spiegelhalter}},\ }\href@noop {} {\emph {\bibinfo {title} {{Markov Chain
  Monte Carlo in Practice}}}}\ (\bibinfo  {publisher} {Chapman \& Hall},\
  \bibinfo {address} {London, UK},\ \bibinfo {year} {1996})\BibitemShut
  {NoStop}%
\bibitem [{\citenamefont {Belczynski}\ \emph {et~al.}(2012)\citenamefont
  {Belczynski}, \citenamefont {Wiktorowicz}, \citenamefont {Fryer},
  \citenamefont {Holz},\ and\ \citenamefont {Kalogera}}]{Belczynski:2011bn}%
  \BibitemOpen
  \bibfield  {author} {\bibinfo {author} {\bibfnamefont {K.}~\bibnamefont
  {Belczynski}}, \bibinfo {author} {\bibfnamefont {G.}~\bibnamefont
  {Wiktorowicz}}, \bibinfo {author} {\bibfnamefont {C.}~\bibnamefont {Fryer}},
  \bibinfo {author} {\bibfnamefont {D.}~\bibnamefont {Holz}}, \ and\ \bibinfo
  {author} {\bibfnamefont {V.}~\bibnamefont {Kalogera}},\ }\href {\doibase
  10.1088/0004-637X/757/1/91} {\bibfield  {journal} {\bibinfo  {journal}
  {Astrophys. J.}\ }\textbf {\bibinfo {volume} {757}},\ \bibinfo {pages} {91}
  (\bibinfo {year} {2012})},\ \Eprint {http://arxiv.org/abs/1110.1635}
  {arXiv:1110.1635 [astro-ph.GA]} \BibitemShut {NoStop}%
\bibitem [{\citenamefont {{Nakamura}}(1983)}]{1983PThPh..70.1144N}%
  \BibitemOpen
  \bibfield  {author} {\bibinfo {author} {\bibfnamefont {T.}~\bibnamefont
  {{Nakamura}}},\ }\href {\doibase 10.1143/PTP.70.1144} {\bibfield  {journal}
  {\bibinfo  {journal} {Progress of Theoretical Physics}\ }\textbf {\bibinfo
  {volume} {70}},\ \bibinfo {pages} {1144} (\bibinfo {year}
  {1983})}\BibitemShut {NoStop}%
\bibitem [{\citenamefont {Vietri}\ and\ \citenamefont
  {Stella}(1999)}]{Vietri:1999kj}%
  \BibitemOpen
  \bibfield  {author} {\bibinfo {author} {\bibfnamefont {M.}~\bibnamefont
  {Vietri}}\ and\ \bibinfo {author} {\bibfnamefont {L.}~\bibnamefont
  {Stella}},\ }\href {\doibase 10.1086/312386} {\bibfield  {journal} {\bibinfo
  {journal} {Astrophys. J.}\ }\textbf {\bibinfo {volume} {527}},\ \bibinfo
  {pages} {L43} (\bibinfo {year} {1999})},\ \Eprint
  {http://arxiv.org/abs/astro-ph/9910008} {arXiv:astro-ph/9910008 [astro-ph]}
  \BibitemShut {NoStop}%
\bibitem [{\citenamefont {MacFadyen}\ \emph {et~al.}(2005)\citenamefont
  {MacFadyen}, \citenamefont {Ramirez-Ruiz},\ and\ \citenamefont
  {Zhang}}]{MacFadyen:2005xm}%
  \BibitemOpen
  \bibfield  {author} {\bibinfo {author} {\bibfnamefont {A.~I.}\ \bibnamefont
  {MacFadyen}}, \bibinfo {author} {\bibfnamefont {E.}~\bibnamefont
  {Ramirez-Ruiz}}, \ and\ \bibinfo {author} {\bibfnamefont {W.}~\bibnamefont
  {Zhang}},\ }\href@noop {} {\  (\bibinfo {year} {2005})},\ \Eprint
  {http://arxiv.org/abs/astro-ph/0510192} {arXiv:astro-ph/0510192 [astro-ph]}
  \BibitemShut {NoStop}%
\bibitem [{\citenamefont {Dermer}\ and\ \citenamefont
  {Atoyan}(2006)}]{Dermer:2006pw}%
  \BibitemOpen
  \bibfield  {author} {\bibinfo {author} {\bibfnamefont {C.~D.}\ \bibnamefont
  {Dermer}}\ and\ \bibinfo {author} {\bibfnamefont {A.}~\bibnamefont
  {Atoyan}},\ }\href {\doibase 10.1086/504895} {\bibfield  {journal} {\bibinfo
  {journal} {Astrophys. J.}\ }\textbf {\bibinfo {volume} {643}},\ \bibinfo
  {pages} {L13} (\bibinfo {year} {2006})},\ \Eprint
  {http://arxiv.org/abs/astro-ph/0601142} {arXiv:astro-ph/0601142 [astro-ph]}
  \BibitemShut {NoStop}%
\bibitem [{\citenamefont {Fishbach}\ \emph {et~al.}(2017)\citenamefont
  {Fishbach}, \citenamefont {Holz},\ and\ \citenamefont
  {Farr}}]{fishbach2017ligo}%
  \BibitemOpen
  \bibfield  {author} {\bibinfo {author} {\bibfnamefont {M.}~\bibnamefont
  {Fishbach}}, \bibinfo {author} {\bibfnamefont {D.~E.}\ \bibnamefont {Holz}},
  \ and\ \bibinfo {author} {\bibfnamefont {B.}~\bibnamefont {Farr}},\ }\href
  {\doibase 10.3847/2041-8213/aa7045} {\bibfield  {journal} {\bibinfo
  {journal} {Astrophys. J.}\ }\textbf {\bibinfo {volume} {840}},\ \bibinfo
  {pages} {L24} (\bibinfo {year} {2017})},\ \Eprint
  {http://arxiv.org/abs/1703.06869} {arXiv:1703.06869 [astro-ph.HE]}
  \BibitemShut {NoStop}%
\bibitem [{\citenamefont {Gerosa}\ and\ \citenamefont
  {Berti}(2017)}]{gerosa2017merging}%
  \BibitemOpen
  \bibfield  {author} {\bibinfo {author} {\bibfnamefont {D.}~\bibnamefont
  {Gerosa}}\ and\ \bibinfo {author} {\bibfnamefont {E.}~\bibnamefont {Berti}},\
  }\href {\doibase 10.1103/PhysRevD.95.124046} {\bibfield  {journal} {\bibinfo
  {journal} {Phys. Rev.}\ }\textbf {\bibinfo {volume} {D95}},\ \bibinfo {pages}
  {124046} (\bibinfo {year} {2017})},\ \Eprint
  {http://arxiv.org/abs/1703.06223} {arXiv:1703.06223 [gr-qc]} \BibitemShut
  {NoStop}%
\bibitem [{\citenamefont {Antonini}\ and\ \citenamefont
  {Rasio}(2016)}]{antonini2016merging}%
  \BibitemOpen
  \bibfield  {author} {\bibinfo {author} {\bibfnamefont {F.}~\bibnamefont
  {Antonini}}\ and\ \bibinfo {author} {\bibfnamefont {F.~A.}\ \bibnamefont
  {Rasio}},\ }\href {\doibase 10.3847/0004-637X/831/2/187} {\bibfield
  {journal} {\bibinfo  {journal} {Astrophys. J.}\ }\textbf {\bibinfo {volume}
  {831}},\ \bibinfo {pages} {187} (\bibinfo {year} {2016})},\ \Eprint
  {http://arxiv.org/abs/1606.04889} {arXiv:1606.04889 [astro-ph.HE]}
  \BibitemShut {NoStop}%
\bibitem [{\citenamefont {Gupta}\ \emph {et~al.}(2019)\citenamefont {Gupta},
  \citenamefont {Gerosa}, \citenamefont {Arun}, \citenamefont {Berti},
  \citenamefont {Farr},\ and\ \citenamefont {Sathyaprakash}}]{Gupta:2019nwj}%
  \BibitemOpen
  \bibfield  {author} {\bibinfo {author} {\bibfnamefont {A.}~\bibnamefont
  {Gupta}}, \bibinfo {author} {\bibfnamefont {D.}~\bibnamefont {Gerosa}},
  \bibinfo {author} {\bibfnamefont {K.~G.}\ \bibnamefont {Arun}}, \bibinfo
  {author} {\bibfnamefont {E.}~\bibnamefont {Berti}}, \bibinfo {author}
  {\bibfnamefont {W.}~\bibnamefont {Farr}}, \ and\ \bibinfo {author}
  {\bibfnamefont {B.~S.}\ \bibnamefont {Sathyaprakash}},\ }\href@noop {} {\
  (\bibinfo {year} {2019})},\ \Eprint {http://arxiv.org/abs/1909.05804}
  {arXiv:1909.05804 [gr-qc]} \BibitemShut {NoStop}%
\bibitem [{\citenamefont {Ye}\ \emph {et~al.}(2020)\citenamefont {Ye},
  \citenamefont {Fong}, \citenamefont {Kremer}, \citenamefont {Rodriguez},
  \citenamefont {Chatterjee}, \citenamefont {Fragione},\ and\ \citenamefont
  {Rasio}}]{Ye:2019xvf}%
  \BibitemOpen
  \bibfield  {author} {\bibinfo {author} {\bibfnamefont {C.~S.}\ \bibnamefont
  {Ye}}, \bibinfo {author} {\bibfnamefont {W.-f.}\ \bibnamefont {Fong}},
  \bibinfo {author} {\bibfnamefont {K.}~\bibnamefont {Kremer}}, \bibinfo
  {author} {\bibfnamefont {C.~L.}\ \bibnamefont {Rodriguez}}, \bibinfo {author}
  {\bibfnamefont {S.}~\bibnamefont {Chatterjee}}, \bibinfo {author}
  {\bibfnamefont {G.}~\bibnamefont {Fragione}}, \ and\ \bibinfo {author}
  {\bibfnamefont {F.~A.}\ \bibnamefont {Rasio}},\ }\href {\doibase
  10.3847/2041-8213/ab5dc5} {\bibfield  {journal} {\bibinfo  {journal}
  {Astrophys. J.}\ }\textbf {\bibinfo {volume} {888}},\ \bibinfo {pages} {L10}
  (\bibinfo {year} {2020})},\ \Eprint {http://arxiv.org/abs/1910.10740}
  {arXiv:1910.10740 [astro-ph.HE]} \BibitemShut {NoStop}%
\bibitem [{\citenamefont {Fragione}\ \emph {et~al.}(2020)\citenamefont
  {Fragione}, \citenamefont {Loeb},\ and\ \citenamefont
  {Rasio}}]{Fragione:2020aki}%
  \BibitemOpen
  \bibfield  {author} {\bibinfo {author} {\bibfnamefont {G.}~\bibnamefont
  {Fragione}}, \bibinfo {author} {\bibfnamefont {A.}~\bibnamefont {Loeb}}, \
  and\ \bibinfo {author} {\bibfnamefont {F.~A.}\ \bibnamefont {Rasio}},\
  }\href@noop {} {\  (\bibinfo {year} {2020})},\ \Eprint
  {http://arxiv.org/abs/2002.11278} {arXiv:2002.11278 [astro-ph.GA]}
  \BibitemShut {NoStop}%
\bibitem [{\citenamefont {East}\ \emph {et~al.}(2012)\citenamefont {East},
  \citenamefont {Pretorius},\ and\ \citenamefont {Stephens}}]{East:2011xa}%
  \BibitemOpen
  \bibfield  {author} {\bibinfo {author} {\bibfnamefont {W.~E.}\ \bibnamefont
  {East}}, \bibinfo {author} {\bibfnamefont {F.}~\bibnamefont {Pretorius}}, \
  and\ \bibinfo {author} {\bibfnamefont {B.~C.}\ \bibnamefont {Stephens}},\
  }\href {\doibase 10.1103/PhysRevD.85.124009} {\bibfield  {journal} {\bibinfo
  {journal} {Phys. Rev. D}\ }\textbf {\bibinfo {volume} {85}},\ \bibinfo
  {pages} {124009} (\bibinfo {year} {2012})},\ \Eprint
  {http://arxiv.org/abs/1111.3055} {arXiv:1111.3055 [astro-ph.HE]} \BibitemShut
  {NoStop}%
\bibitem [{\citenamefont {Capela}\ \emph {et~al.}(2013)\citenamefont {Capela},
  \citenamefont {Pshirkov},\ and\ \citenamefont {Tinyakov}}]{Capela:2013yf}%
  \BibitemOpen
  \bibfield  {author} {\bibinfo {author} {\bibfnamefont {F.}~\bibnamefont
  {Capela}}, \bibinfo {author} {\bibfnamefont {M.}~\bibnamefont {Pshirkov}}, \
  and\ \bibinfo {author} {\bibfnamefont {P.}~\bibnamefont {Tinyakov}},\ }\href
  {\doibase 10.1103/PhysRevD.87.123524} {\bibfield  {journal} {\bibinfo
  {journal} {Phys. Rev.}\ }\textbf {\bibinfo {volume} {D87}},\ \bibinfo {pages}
  {123524} (\bibinfo {year} {2013})},\ \Eprint {http://arxiv.org/abs/1301.4984}
  {arXiv:1301.4984 [astro-ph.CO]} \BibitemShut {NoStop}%
\bibitem [{\citenamefont {Fuller}\ \emph {et~al.}(2017)\citenamefont {Fuller},
  \citenamefont {Kusenko},\ and\ \citenamefont {Takhistov}}]{Fuller:2017uyd}%
  \BibitemOpen
  \bibfield  {author} {\bibinfo {author} {\bibfnamefont {G.~M.}\ \bibnamefont
  {Fuller}}, \bibinfo {author} {\bibfnamefont {A.}~\bibnamefont {Kusenko}}, \
  and\ \bibinfo {author} {\bibfnamefont {V.}~\bibnamefont {Takhistov}},\ }\href
  {\doibase 10.1103/PhysRevLett.119.061101} {\bibfield  {journal} {\bibinfo
  {journal} {Phys. Rev. Lett.}\ }\textbf {\bibinfo {volume} {119}},\ \bibinfo
  {pages} {061101} (\bibinfo {year} {2017})},\ \Eprint
  {http://arxiv.org/abs/1704.01129} {arXiv:1704.01129 [astro-ph.HE]}
  \BibitemShut {NoStop}%
\bibitem [{\citenamefont {Goldman}\ and\ \citenamefont
  {Nussinov}(1989)}]{Goldman:1989nd}%
  \BibitemOpen
  \bibfield  {author} {\bibinfo {author} {\bibfnamefont {I.}~\bibnamefont
  {Goldman}}\ and\ \bibinfo {author} {\bibfnamefont {S.}~\bibnamefont
  {Nussinov}},\ }\href {\doibase 10.1103/PhysRevD.40.3221} {\bibfield
  {journal} {\bibinfo  {journal} {Phys. Rev.}\ }\textbf {\bibinfo {volume}
  {D40}},\ \bibinfo {pages} {3221} (\bibinfo {year} {1989})}\BibitemShut
  {NoStop}%
\bibitem [{\citenamefont {Bramante}\ and\ \citenamefont
  {Linden}(2014)}]{Bramante:2014zca}%
  \BibitemOpen
  \bibfield  {author} {\bibinfo {author} {\bibfnamefont {J.}~\bibnamefont
  {Bramante}}\ and\ \bibinfo {author} {\bibfnamefont {T.}~\bibnamefont
  {Linden}},\ }\href {\doibase 10.1103/PhysRevLett.113.191301} {\bibfield
  {journal} {\bibinfo  {journal} {Phys. Rev. Lett.}\ }\textbf {\bibinfo
  {volume} {113}},\ \bibinfo {pages} {191301} (\bibinfo {year} {2014})},\
  \Eprint {http://arxiv.org/abs/1405.1031} {arXiv:1405.1031 [astro-ph.HE]}
  \BibitemShut {NoStop}%
\bibitem [{\citenamefont {Bramante}\ \emph {et~al.}(2018)\citenamefont
  {Bramante}, \citenamefont {Linden},\ and\ \citenamefont
  {Tsai}}]{Bramante:2017ulk}%
  \BibitemOpen
  \bibfield  {author} {\bibinfo {author} {\bibfnamefont {J.}~\bibnamefont
  {Bramante}}, \bibinfo {author} {\bibfnamefont {T.}~\bibnamefont {Linden}}, \
  and\ \bibinfo {author} {\bibfnamefont {Y.-D.}\ \bibnamefont {Tsai}},\ }\href
  {\doibase 10.1103/PhysRevD.97.055016} {\bibfield  {journal} {\bibinfo
  {journal} {Phys. Rev.}\ }\textbf {\bibinfo {volume} {D97}},\ \bibinfo {pages}
  {055016} (\bibinfo {year} {2018})},\ \Eprint
  {http://arxiv.org/abs/1706.00001} {arXiv:1706.00001 [hep-ph]} \BibitemShut
  {NoStop}%
\bibitem [{\citenamefont {Vallisneri}(2000)}]{Vallisneri:1999nq}%
  \BibitemOpen
  \bibfield  {author} {\bibinfo {author} {\bibfnamefont {M.}~\bibnamefont
  {Vallisneri}},\ }\href {\doibase 10.1103/PhysRevLett.84.3519} {\bibfield
  {journal} {\bibinfo  {journal} {Phys. Rev. Lett.}\ }\textbf {\bibinfo
  {volume} {84}},\ \bibinfo {pages} {3519} (\bibinfo {year} {2000})},\ \Eprint
  {http://arxiv.org/abs/gr-qc/9912026} {arXiv:gr-qc/9912026 [gr-qc]}
  \BibitemShut {NoStop}%
\end{thebibliography}%

\end{document}